\documentclass[12pt,a4paper,fleqn]{article}

\usepackage{lscape,exscale,amsthm}
\usepackage[intlimits]{amsmath}
\usepackage{rawfonts}
\usepackage{latexsym}
\usepackage[cp850]{inputenc}
\usepackage{epsfig}
\usepackage{bbm,enumerate}

\RequirePackage[OT1]{fontenc}
\RequirePackage{amsthm,amsmath}
\usepackage{natbib}
\RequirePackage[colorlinks,citecolor=blue,urlcolor=blue]{hyperref}

\newcommand{\bol}[1]{\mbox{\boldmath$#1$}}

\newcommand{\bSigma}{\bol{\Sigma}}
\newcommand{\tbS}{\widetilde{\mathbf{S}}}

\newcommand{\bm}{\bol{\mu}}

\newcommand{\bx}{\mathbf{X}}
\newcommand{\bX}{\mathbf{x}}
\newcommand{\bQ}{\mathbf{Q}}

\newcommand{\by}{\mathbf{Y}}

\newcommand{\bi}{\mathbf{1}}

\newcommand{\bI}{\mathbf{I}}

\newcommand{\bxi}{\boldsymbol{\xi}}

\newcommand{\bS}{\mathbf{S}}
\newcommand{\bby}{\bar{\mathbf{y}}}
\newcommand{\bbx}{\bar{\mathbf{x}}}

\newcommand{\bTheta}{\mathbf{\Theta}}
\newcommand{\btheta}{\boldsymbol{\theta}}
\newcommand{\tbtheta}{\tilde{\boldsymbol{\theta}}}

\usepackage{stackrel}

\numberwithin{equation}{section}
\theoremstyle{plain}
\newtheorem{theorem}{Theorem}[section]

\newtheorem{lemma}{Lemma}[section]
\newtheorem{corollary}{Corollary}[section]

\usepackage{color}

\usepackage{ulem}

\begin{document}

\begin{center}
\vspace*{2cm} \noindent {\bf \large Consistent Estimation of the High-Dimensional Efficient Frontier}\\
\vspace{1cm} \noindent {\sc  Taras Bodnar$^{a}$, Nikolaus Hautsch$^{b}$, Yarema Okhrin$^{c}$ and Nestor Parolya$^{d,}$\footnote{ Corresponding author. E-mail address: N.Parolya@tudelft.nl}}\\
\vspace{1cm}
{\it \footnotesize  $^a$
Department of Management and Engineering, Link\"{o}ping University, SE-581 83 Link\"{o}ping, Sweden}\\
{\it \footnotesize  $^b$
Department of Statistics and Operations Research, University of Vienna, A-1090 Vienna, Austria} \\
{\it \footnotesize  $^c$
Chair of Statistics and Data Science, Augsburg University, 86159 Augsburg, Germany} \\
{\it \footnotesize  $^d$
Delft Institute of Applied Mathematics, Delft University of Technology, 2628CD Delft, The Netherlands} 
\end{center}

\begin{abstract}
In this paper, we analyze the asymptotic behavior of the main characteristics of the mean-variance efficient frontier employing random matrix theory. Our particular interest covers the case when the dimension $p$ and the sample size $n$ tend to infinity simultaneously and their ratio $p/n$ tends to a positive constant $c\in(0,1)$.
We neither impose any distributional nor structural assumptions on the asset
returns. For the developed theoretical framework, some regularity conditions, like the existence of the $4$th moments, are needed.

It is shown that two out of three quantities of interest are biased and
overestimated by their sample counterparts under the high-dimensional asymptotic
regime. This becomes evident based on the asymptotic deterministic equivalents of the sample plug-in estimators. Using them we construct consistent estimators of
the three characteristics of the efficient frontier. It it shown that the
additive and/or the multiplicative biases of the sample estimates are solely
functions of the concentration ratio $c$. Furthermore, the asymptotic normality of the considered estimators of the parameters of the efficient frontier is proved.
Verifying the theoretical results based on an extensive simulation study we show that the proposed estimator for the
efficient frontier is a valuable alternative to the sample estimator for high
dimensional data. Finally, we present an empirical application, where we estimate the efficient frontier based on the stocks included in S\&P 500 index.
\end{abstract}

\vspace{0.7cm}


\section{Introduction}

The efficient frontier is a key object of modern portfolio theory as derived by  \cite{Markowitz1952}.  \cite{Merton1972} shows that the efficient frontier is the upper part of the parabola in the mean-variance space defined by three characteristics. In particular, using Merton's notation, the efficient frontier is given by (cf.  \cite{Merton1972})
\begin{align}\label{Merton}
V&=\dfrac{a-2bR+cR^2}{ac-b^2},
\end{align}
where the constants $a=\bm^\prime\bSigma^{-1}\bm$, $b=\bi_p^\prime\bSigma^{-1}\bm$, and $c=\bi_p^\prime\bSigma^{-1}\bi_p$ fully determine the location of the vertex and the slope coefficient of the parabola in the mean-variance space. The vector $\bm$ denotes the $p$-dimensional mean vector of asset returns and the $p\times p$ matrix $\bSigma$ is the corresponding positive definite covariance matrix. The vector $\bi_p$ is the $p$-dimensional unity vector. 

The three parameters $a$, $b$, and $c$ do not possess an appropriate interpretation in the financial literature. For that reason, we rewrite (\ref{Merton}) as
\begin{equation}\label{Bodnar}
(R-R_{GMV})^2=s(V-V_{GMV}),
\end{equation}
where
\begin{equation}\label{rgmv}
R_{GMV}=\dfrac{\bi_p^\prime\bSigma^{-1}\bm}{\bi_p^\prime\bSigma^{-1}\bi_p}
\end{equation}
is the expected return and
\begin{equation}\label{vgmv}
V_{GMV}=\dfrac{1}{\bi_p^\prime\bSigma^{-1}\bi_p}
\end{equation}
is the variance of the global minimum variance (GMV) portfolio. The parameter
\begin{equation}\label{slope}
s=\bm^\prime\bQ\bm~~\text{with}~~\bQ=\bSigma^{-1}-\dfrac{\bSigma^{-1}\bi_p\bi_p^\prime\bSigma^{-1}}{\bi_p^\prime\bSigma^{-1}\bi_p},
\end{equation}
denotes the slope parameter of the parabola in the mean-variance space. The parameters $\{R_{GMV}, V_{GMV}, s\}$ are the objects of main interest in this paper.

Although the calculation of the set of parameters ${R_{GMV}, V_{GMV}, s}$ is straightforward when $\bm$ and $\bSigma$ are known, their computation becomes a challenging task otherwise. Since the mean vector $\bm$ and the covariance matrix $\bSigma$ are unknown in practice, they must be estimated from return data before the efficient frontier is constructed. The usual estimation technique is based on sample estimators or so-called plug-in estimators (see, e.g. \cite{BodnarSchmid2008, BodnarSchmid2009}, \cite{Kan2008}) which replace $\bm$ and $\bSigma$ in (\ref{rgmv}), (\ref{vgmv}) and (\ref{slope}) by their sample counterparts. This estimation method yields the sample efficient frontier, specified by the sample estimators $\hat{R}_{GMV}$, $\hat{V}_{GMV}$, and $\hat{s}$ of $R_{GMV}$, $V_{GMV}$, and $s$, respectively.

The asymptotic behavior of the sample efficient frontier for finite dimension $p$ under the assumption of normality is investigated in \cite{Jobson1980} and \cite{jobson1991}, while \cite{BodnarSchmid2008} and \cite{Kan2008} study the finite sample distributional properties of the sample parameters $\{\hat{R}_{GMV}, \hat{V}_{GMV}, \hat{s}\}$ of the efficient frontier. However, \cite{Basak2005} and \cite{Siegel2007} show that the sample efficient frontier overestimates the true location of the efficient frontier in the mean-variance space. To correct this overestimation,  \cite{Kan2008} derives improved estimates for the parameters of the efficient frontier. At the same time, \cite{Bodnar2010} constructs an unbiased estimator of the efficient frontier. Finally, Bayesian estimators of the efficient frontier were derived in \cite{bauder2019bayesian} and \cite{bauder2021bayesian}.

The finite-sample techniques work well only under the assumption of normally distributed asset returns,  which is not supported by in most applications. Typically, asset returns are skewed and heavy-tailed (e.g., \cite{Fama1965},  \cite{Markowitz1991}, \cite{adcock2014mean}, and \cite{adcock2015skewed}). \cite{BodnarGupta2009} and \cite{Gupta2013}  construct the efficient frontier under elliptical models and provide inference procedures for its main parameters.

More challenging problems arise by considering the asymptotic properties of the sample efficient frontier when the dimension of the portfolio $p$ increases together with the sample size $n$. It is noted that the sample estimators work well only if the number of assets $p$ is fixed and significantly smaller than the number of observations $n$. This case is often used in statistics and is refered to the case of standard asymptotics (see, \cite{LeCam2000}). In this regime, the traditional sample estimators are consistent for the main parameters of the efficient frontier. As a result, for a small fixed dimension $p\in\{2,3, 5\}$ with a large $n$ the sample estimator can be used. However, it is unclear what to do if the number of assets in the portfolio are (very) large comparable to $n$. Here, we are in a situation when both the number of assets $p$ and the sample size $n$ tend to infinity. This double asymptotic has an interpretation when $p$ and $n$ are of comparable size, i.e., when $p/n$ tends to a concentration ratio $c>0$. This regime of asymptotics is known as a high-dimensional asymptotic or ``Kolmogorov'' asymptotic (see, e.g., \cite{Buehlmann2011}, \cite{CaiShen2011}). In this setting, the sample estimators behave in an unpredictable way and are far from the optimal ones. In general, the larger the concentration ratio $c$ becomes, the worse is the performance of the sample estimators. This is a well-known problem in statistics called ``the curse of dimensionality'' (see, e.g., \cite{bellman1961}).

Under the presence of financial high-frequency (intraday) data, the curse of dimensionality can be reduced by employing in-fill asymptotics, however, requires to handle the impact of market microstructure noise (and potentially ill-conditioned estimates). One of the first applications of high-frequency data to estimate huge-dimensional covariance matrices is \cite{Hautsch2012} who propose a blocking and regularization approach. \cite{Hautsch2015} further extend this idea and show its statistical and economic usefulness in a global minimum variance framework. 

To handle the curse of dimensionality for the efficient frontier, we employ  results from asymptotic random matrix theory. Random matrix theory is a fast growing branch of probability theory with many applications in statistics and finance. It studies the asymptotic behavior of the eigenvalues of random matrices under general asymptotics (see, e.g., \cite{Anderson2010},  \cite{Bai2010}). The asymptotic behavior of the functionals of the sample covariance matrices is analyzed in \cite{Marcenko1967},  \cite{Yin1986},  \cite{Girko1994,Girko1996a,Girko1996b},  \cite{Silverstein1995}, \cite{Bai2007},  \cite{Bai2010},  \cite{Rubio2011}, \cite{Bodnar2014,Bodnar2015}, among others. \cite{Marcenko1967} formally derive the behavior of the limiting spectral measure of the sample covariance matrix. It depends on the corresponding spectral measure of the population covariance matrix, which is unknown.  \cite{Silverstein1995} proves the validity of the Ma$\breve{\text{c}}$enko and Pastur (MP) equation under more general assumptions and show the strong convergence of the spectral measure of the sample covariance matrix. Recently, high-dimensional optimal portfolio theory has attracted the attention of researchers and practitioners of the financial sector (see, e.g., \cite{jagannathan2003risk, el2010high, glombek2014, bodnar2018estimation, BodnarDmytriv2019, ao2019approaching, cai2020high, bodnar2021tests, ding2021high, bodnar2022recent, bodnar2022optimal, bodnar2022sampling, bodnarparolyathorsen2024, kan2024optimal, lassance2024combination}).

We use the above-mentioned theoretical results to find the asymptotic behavior of the sample efficient frontier $\{\hat{R}_{GMV}, \hat{V}_{GMV}, \hat{s}\}$ and to prove that $\hat{R}_{GMV}$ is indeed a consistent estimator for $R_{GMV}$, while $\hat{V}_{GMV}$ and $\hat{s}$ are highly biased and inconsistent in high dimensions. We show that the additive and multiplicative biases are solely functions of the concentration ratio $c$, thus we can construct consistent estimators handling these biases. No distributional neither structural assumptions are imposed on the asset returns. Theoretically, we only need the existence of the $4$th moments. Moreover, under additional assumptions imposed on the distribution of the asset returns we prove that the consistent estimators of the three parameters of the efficient frontier are mutually independent and asymptotically normally distributed under the high-dimensional asymptotic regime. Our findings confirm that the Bayesian estimator of the efficient frontier, as derived in \cite{bauder2021bayesian}, is indeed a consistent estimator from a frequentist perspective.

The rest of the paper is organized as follows. In Section 2, we provide the main assumptions and notations used in the paper and formulate the main theoretical result on the asymptotic behavior of the efficient frontier. At the end of Section 2, consistent estimators of the three parameters of the efficient frontier are presented, whereas the results on the asymptotic normality are summarized in Section 2.1. In Section 3, we provide an extensive simulation study to verify the rate of convergence as well as the performance of the obtained estimators under high-dimensional asymptotics. Here, the performance of the derived estimator is compared with existent estimation techniques for high-dimensional data. The results of the empirical study are presented in Section 4. Section 5 provides concluding remarks. The proofs of the theoretical results are presented in the appendix (Section 6).

\section{Consistent estimation of the efficient frontier under large dimensional asymptotics}

In this section, we present our main result on the asymptotics of the main characteristics of the efficient frontier.
%

Let the $p\times n$ matrix $\by_n$ be the observation matrix of the $p$-dimensional vector of the asset returns taken at time points $1,\ldots,n$. The vector of the expected asset returns is denoted by $\bm_n$, while $\bSigma_n$ denotes the covariance matrix.\footnote{Under large-dimensional asymptotics, both the dimension $p$ and the sample size $n$ tend to infinity. Therefore, it is natural to assume, without loss of generality, that the dimension $p\equiv p(n)$ is the function of the sample size $n$. Consequently, the mean vector and the covariance matrix are indexed with $n$ as well.} We assume that the observation matrix is equal in distribution to
\begin{equation}\label{obs}
 \by_n\overset{d.}{=}\bSigma_n^{\frac{1}{2}}\bx_n+\bm_n\bi_n^\prime,
 \end{equation}
where the symbol $'\overset{d.}{=}~'$ denotes the equality in distribution and the $p\times n$ matrix $\bx_n$ contains independent and identically distributed (i.i.d.) real random variables with zero mean and unit variance. Note that only the matrix $\by_n$ is observable. We know neither $\bx_{n}$, $\bSigma_n$ nor $\bm_n$.

Note that the observation matrix $\by_n$ has dependent rows, which is assured by the covariance matrix $\bSigma_n$, but independent columns. The assumption of the independence of the samples can be weakened to allow for  dependent elements of $\bx_n$ by controlling the growth of the number of dependent entries but not their joint distribution (see \cite{Friesen2013}). Here, we stick to the assumption of independence of the random samples to simplify the proof of the theoretical results.

The sample mean vector of the asset returns is given by
\begin{equation}\label{sample_m}
 \bby_n=\dfrac{1}{n}\by_n\bi_n\,,
\end{equation}
whereas the sample covariance matrix is defined as
\begin{equation}\label{samplecov}
 \bS_n=\dfrac{1}{n}\left(\by_n-\bby_n\bi_n^\prime\right)\left(\by_n-\bby_n\bi_n^\prime\right)^{\prime}=\dfrac{1}{n}\by_n\by_n^\prime-\bby_n\bby_n^\prime\,.
\end{equation}

Using \eqref{sample_m} and \eqref{samplecov} we obtain the sample estimators for the three parameters of the efficient frontier given by
\begin{equation}\label{sample_rgmv_vgmv}
\hat{R}_{GMV}=\dfrac{\bi_p^\prime\bS_n^{-1}\bby_n}{\bi_p^\prime\bS_n^{-1}\bi_p},\, \quad \hat{V}_{GMV}=\dfrac{1}{\bi_p^\prime\bS_n^{-1}\bi_p},
\end{equation}
and
\begin{equation}\label{sample_slope}
\hat{s}=\bby_n^\prime\widehat{\bQ}\bby_n~~\text{with}~~\widehat{\bQ}=\bS_n^{-1}-\dfrac{\bS_n^{-1}\bi_p\bi_p^\prime\bS_n^{-1}}{\bi_p^\prime\bS_n^{-1}\bi_p}\,.
\end{equation}

Our theoretical findings are based on the following assumptions:
\begin{description}
\item[(A1)] The population covariance matrix $\bSigma_n$ is a nonrandom $p$-dimensional positive definite matrix for all dimensions $p$.

\item[(A2)] The elements of the matrix $\bx_n$ have uniformly bounded $4+\varepsilon,~\varepsilon>0$ moments.

\item[(A3)] It exists $M_l,M_u \in (0,+\infty)$ and $q \in [0,+\infty)$ such that
\begin{equation*}
M_lp^q \le \bi_p^\prime \bSigma_n^{-1} \bi_p, \bm_n^\prime \bSigma_n^{-1} \bm_n \le  M_u p^q \,.
\end{equation*}
\end{description}

All of these regularity assumptions are very general and fit many practical situations. The assumption (A1) together with (\ref{obs}) is typical for financial and statistical problems and does not impose strong restrictions. The assumption (A2) is a technical one. Our simulation study shows that this assumption can be relaxed for practical purposes. The assumption (A3) requires that the quantities used in the calculation of $R_{GMV}$, $V_{GMV}$, and $s$ are of the same order. This assumption is very general and imposes no additional constraints neither on the mean vector $\bm_n$ (such that its Euclidean norm is bounded) nor on the covariance matrix $\bSigma_n$ (such that its eigenvalues lie in the compact interval). The last point allows us to assume a factor model for the data matrix $\by_n$ which implies that the largest eigenvalue of $\bSigma_n$ is of order $p$ (c.f. \cite{Fan2008,Fan2012,Fan2013}). Finally, assumption (A3) ensures that $\bi_p^\prime \bSigma_n^{-1} \bm_n$ is at most of order $p^{-q}$ which follows directly from the Cauchy-Schwarz inequality.

The main result about the strong convergence of the sample estimators for the three parameters of the efficient frontier is presented in Theorem \ref{th1}.

\vspace{0.2cm}
\begin{theorem}\label{th1}
Let $q \ge 1$. Then, under the assumptions (A1)-(A3) it holds that
\begin{align}
&\Biggl|\hat{R}_{GMV}-R_{GMV}\Biggr|\stackrel{a.s.}{\longrightarrow} 0~~\text{for}~p/n\rightarrow c\in(0, 1)~\text{as}~n\rightarrow\infty,\label{Rgmv}\\
&p^q\Biggl|\hat{V}_{GMV}-(1-c)V_{GMV}\Biggr|\stackrel{a.s.}{\longrightarrow}0~~\text{for}~p/n\rightarrow c\in(0, 1)~\text{as}~n\rightarrow\infty,\label{Vgmv}\\
&p^{-q}\Biggl|\hat{s}-\dfrac{1}{1-c}s\Biggr|\stackrel{a.s.}{\longrightarrow} 0~~\text{for}~p/n\rightarrow c\in(0, 1)~\text{as}~n\rightarrow\infty\label{s}\,.
\end{align}
\end{theorem}

\vspace{0.2cm}
The proof of Theorem \ref{th1} is given in the appendix (Section 6). Theorem \ref{th1} establishes the asymptotic behavior of the main characteristics of the efficient frontier as both the dimension and the sample size grow to the infinity simultaneously. It claims that the sample estimator of the expected return of the GMV portfolio $\hat{R}_{GMV}$ is indeed a consistent estimator under the high-dimensional asymptotics regime. In contrast, the sample estimators for the other parameters of the efficient frontier are not consistent. The estimator for the sample variance of the GMV portfolio, $\hat{V}_{GMV}$, possesses a multiplicative bias $\dfrac{1}{1-c}$, whereas the sample estimator of the slope parameter {$s$ has a multiplicative bias $(1-c)$}. Consequently, if the concentration ratio $c$ is close to one, both the estimators produce a high bias implying inconsistency.

Fortunately, these biases can be easily handled using Theorem \ref{th1}. Hence, consistent estimators $\hat{R}_c$, $\hat{V}_c$ and $\hat{s}_c$ for the three parameters of the efficient frontier are presented in Corollary \ref{cor1}. This  result follows directly from Theorem \ref{th1}.

\vspace{0.2cm}
\begin{corollary}\label{cor1}
Under assumptions (A1)-(A3), consistent estimators of $R_{GMV}$, $V_{GMV}$ and $s$ are given by
\begin{eqnarray}
&\hat{R}_{c}=\hat{R}_{GMV} \,,\\
&\hat{V}_{c}=\dfrac{1}{1-p/n}\hat{V}_{GMV} \,,\\
&\hat{s}_{c}=(1-p/n)\hat{s}\,.
\end{eqnarray}
\end{corollary}

\vspace{0.2cm}
Assuming that the asset returns are normally distributed and the number of assets is fixed, i.e., $c=0$, \cite{Bodnar2010} derive unbiased estimators for the three parameters of the efficient frontier given by
\begin{equation}\label{suu}
\hat{R}_{u}=\hat{R}_{GMV},~ \hat{V}_{u}=\dfrac{n-1}{n-p}\hat{V}_{GMV},~~ \mbox{and} ~~ \hat{s}_{u}=\dfrac{n-p-1}{n-1}\hat{s}-\dfrac{p-1}{n}\,,
\end{equation}
which are asymptotically equivalent to $\hat{R}_{c}$, $\hat{V}_{c}$, and $\hat{s}_c$.

Returning to the equivalent Merton notation of the efficient frontier, we present in Corollary \ref{cor2} the corresponding consistent estimators of the parameters $a$, $b$, and $c$ given in (\ref{Merton}).

\vspace{0.2cm}
\begin{corollary}\label{cor2}
 Under the assumptions (A1)-(A3) the consistent estimators of the Merton's constants $a$, $b$ and $c$ are given by
\begin{eqnarray}
&\hat{a}_c=(1-p/n)\hat{a}\,,\\
&\hat{b}_{c}=(1-p/n)\hat{b}\,,\\
&\hat{c}_{c}=(1-p/n)\hat{c}\,,
\end{eqnarray}
where $\hat{a}=\bby_n^\prime\bS_n^{-1}\bby_n$, $\hat{b}=\bi_p^\prime\bS_n^{-1}\bby_n$, and $\hat{c}=\bi_p^\prime\bS_n^{-1}\bi_p$ are the corresponding sample counterparts.
\end{corollary}

\vspace{0.2cm}
The proof of Corollary \ref{cor2} follows directly from the proof of Theorem \ref{th1} and Corollary \ref{cor1}.
Both corollaries provide equivalent results. Nevertheless, the parametrization discussed in Corollary \ref{cor1} is more tractable. It is easy to deduce that under the standard asymptotics $p/n\rightarrow0$ the consistent estimators of the parameters of the efficient frontier coincide with their sample counterparts.

The obtained results make the practical implementation of the Markowitz portfolio analysis feasible. They ensure the consistency of the estimators in the case of high-dimensional portfolios if the concentration ratio $c=p/n$ does not exceed $1$. In the case $c>1$, the sample covariance matrix is singular. Thus, it is more complicated to find consistent estimators for the parameters of the efficient frontier. In this case, the naive choice would be to replace the inverse sample covariance matrix by its pseudo-inverse, but the consistency of such replacement is not obvious. We leave this question for future research.

\subsection{Asymptotic normality}

Below we prove that the consistent estimators of the three parameters of the efficient frontier are asymptotically normally distributed under the high-dimensional asymptotic regime. These results are formulated as Theorem \ref{th2}. Note that only for the proof of this theorem we impose an additional assumption on the distribution of the entries of $\bx_n$, which are assumed to be standard normally distributed.

\vspace{0.2cm}
\begin{theorem}\label{th2}
Let the assumptions (A1)-(A3) are fulfilled. If $q=0$ and $x_{ij}\sim \mathcal{N}(0,1)$, then it holds that
\[
\sqrt{n}\left(
          \begin{array}{c}
            \hat{R}_{c}-R_{GMV} \\
            \hat{V}_{c}-V_{GMV} \\
            \hat{s}_{c}-s-\frac{p}{n} \\
          \end{array}
        \right)
\stackrel{d.}{\longrightarrow}\mathcal{N}
\left(\mathbf{0},\left(
  \begin{array}{ccc}
    \left(1+\frac{s+c}{1-c}\right)V_{GMV} & 0 & 0 \\
    0 & \dfrac{2V_{GMV}^2}{1-c} & 0 \\
    0 & 0 & \sigma_s^2 \\
  \end{array}
\right)\right),
\]
where
\begin{equation}\label{sig_s2}
\sigma_s^2=2\left(c+2s\right)+2\dfrac{\left(c+s\right)^2}{1-c}
\end{equation}
$\text{for}~ p/n\longrightarrow c \in (0, +\infty) ~~ \text{as} ~n\rightarrow\infty$.
\end{theorem}

The findings of Theorem \ref{th2} show that the consistent estimators for the three parameters of the efficient frontier are independent under high-dimensional asymptotics. We observe that their asymptotic variances increase as $c$ approaches $1$. On the other side, interesting results are obtained in the special case $c=0$. Here, it holds that
\[
\sqrt{n}\left(
          \begin{array}{c}
            \hat{R}_{c}-R_{GMV} \\
            \hat{V}_{c}-V_{GMV} \\
            \hat{s}_{c}-s \\
          \end{array}
        \right)
\stackrel{d.}{\longrightarrow}\mathcal{N}
\left(\mathbf{0},\left(
  \begin{array}{ccc}
    \left(1+s\right)V_{GMV} & 0 & 0 \\
    0 & 2V_{GMV}^2 & 0 \\
    0 & 0 & 4s+2s^2 \\
  \end{array}
\right)\right)\,,
\]
which coincides with the findings obtained for finite $p$ under the standard asymptotic regime in \cite{BodnarSchmid2009}. Comparing the above asymptotic distribution with those presented in Theorem \ref{th2}, we conclude that ignoring the high-dimensional effect in the calculation of the asymptotic distribution yields a considerable underestimation of the uncertainties of $\hat{V}_c$ and $\hat{s}_c$, especially for values of $c$ closed to $1$. Finally, we also observe that an additional additive bias is present in the sample estimator for $s$ when the quadratic forms $\bi_p^\prime \bSigma_n^{-1} \bi_p$ and $\bm_n^\prime \bSigma_n^{-1} \bm_n$ are bounded, i.e., in case of $q=0$.

Finally, application of Slutzky's Lemma (cf., Theorem 1.5 in \cite{dasgupta2008asymptotic}) leads to the marginal confidence intervals for $R_{GMV}$, $V_{GMV}$, and s given by
\begin{eqnarray*}
I_{1-\alpha}(R_{GMV})&=&\left[\hat{R}_{c}-\dfrac{z_{1-\alpha/2}}{\sqrt{n}}\sqrt{1+\frac{\hat{s}_c+c}{1-c}}\sqrt{\hat{V}_{c}},
\hat{R}_{c}+\dfrac{z_{1-\alpha/2}}{\sqrt{n}}\sqrt{1+\frac{\hat{s}_c+c}{1-c}}\sqrt{\hat{V}_{c}} \right] \label{I_R_GMV}\,,\\
I_{1-\alpha}(V_{GMV})&=&\left[\hat{V}_{c}-\dfrac{z_{1-\alpha/2}}{\sqrt{n}}\sqrt{\frac{2}{1-c}}\hat{V}_{c},
\hat{V}_{c}+\dfrac{z_{1-\alpha/2}}{\sqrt{n}}\sqrt{\frac{2}{1-c}}\hat{V}_{c} \right] \label{I_V_GMV}\,,\\
I_{1-\alpha}(s)&=&\left[\hat{s}_{c}-\dfrac{z_{1-\alpha/2}}{\sqrt{n}}\sqrt{2\left(c+2\hat{s}_c\right)+2\dfrac{\left(c+\hat{s}_c\right)^2}{1-c}},
\hat{s}_{c}+\dfrac{z_{1-\alpha/2}}{\sqrt{n}}\sqrt{2\left(c+2\hat{s}_c\right)+2\dfrac{\left(c+\hat{s}_c\right)^2}{1-c}} \right] \label{I_s}\,,
\end{eqnarray*}
where $z_{\beta}$ denotes the $\beta$-quantile of the standard normal distribution.

\section{Finite sample performance}
\subsection{Estimation of $R_{GMV}$, $V_{GMV}$, and $s$}

In this section, we examine the rate of convergence and the performance of the
derived consistent estimators given in Corollary \ref{cor1}. In order to analyze the
large sample performance of the estimators, we use a quadratic loss
function. Without loss of generality, in our simulation study we take the
covariance matrix $\bSigma_n$ as given with the same proportion of the
eigenvalues over all dimensions, namely $20\%$ of eigenvalues are equal to
$0.5$,
$40\%$ to $1$ and $40\%$ to $5$, respectively. This guarantees to keep
the spectrum of the covariance matrix unchanged for all $p$. The elements of the mean vector
$\bm_n$ are generated from the uniform distribution, i.e., $\mu^{i}_n\sim\text{Unif}[-0.2,0.2]$ for
$i\in\{1,\ldots,n\}$.

Three scenarios are considered in the simulation study, namely

\begin{enumerate}[\text{Scenario} 1:]
\item The entries of $\bx_n$ are generated independently from $\mathcal{N}(0,1)$.
\item The entries of $\bx_n$ are generated independently from $t_3(0,1/3)$ ($t$-distribution with $3$ degrees of freedom and scale parameter $1/3$). This choice of the scale parameter ensures that the covariance matrix of each column of $\by_n$ is equal to $\bSigma_n$.
\item The columns of $\by_n$ are assumed to follow a CCC-GARCH(1,1) process with correlation matrix $\mathbf{\Omega}_n$ computed from $\bSigma_n$ and conditional variance equations expressed as
\[h_{i,t}=\alpha_{0,i}+\alpha_{1,i} (Y^{i, t-1}_n-\mu^{i}_n)^2+\beta_{1,i} h_{i,t-1} \quad \text{for} \quad i=1,...,p ~\text{and}~ t=1,...,n\]
with $\alpha_{0,i}>0$ and $\alpha_{1,i},\beta_{1,i} \in [0,1)$ such that $\alpha_{1,i}+\beta_{1,i}<1$ for all $i=1,...,p$. The parameters $\alpha_{1,i}$ are generated from $\text{Unif}[0.0,0.1]$, while the parameters $\beta_{1,i}$ are simulated from $\text{Unif}[0.8,0.89]$. Finally, the values $\alpha_{0,i}$ are calculated using $\mathbf{\Omega}_n$, $\alpha_{1,i}$ and $\beta_{1,i}$ ensuring that unconditional covariance matrix of the simulated CCC-GARCH(1,1) process is equal to $\bSigma_n$.
\end{enumerate}

The model in Scenario 1 satisfies the conditions (A1)-(A3) and is used to study the speed of the convergence of the obtained theoretical results. The stochastic models from Scenarios 2 and 3 are designed in such a way, that at least one of the assumptions (A1)-(A3) is violated. In particular, in case of Scenario 2, the moments of order $(4+\varepsilon)$ do not exist. For the model in Scenario 3, we additionally assume dependence between the columns of matrix $\by_n$. These two scenarios are used in the simulation study in order to investigate the robustness of the obtained theoretical results to the violation of the imposed assumptions. The results of the simulation study are based on $1000$ independent repetitions performed for $c \in \{0.5,0.9\}$.

\begin{figure}[h!!]
\caption{\footnotesize The quadratic loss of $\hat{R}_c$ for the normal
distribution (above), for the $t$-distribution with $3$ degrees of freedom (in the middle), and for the CCC-GARCH process (below).
}
\begin{tabular}{cc}
&\\
$c=0.5$, Normal distribution&$c=0.9$, Normal distribution\\
\includegraphics[scale=0.28]{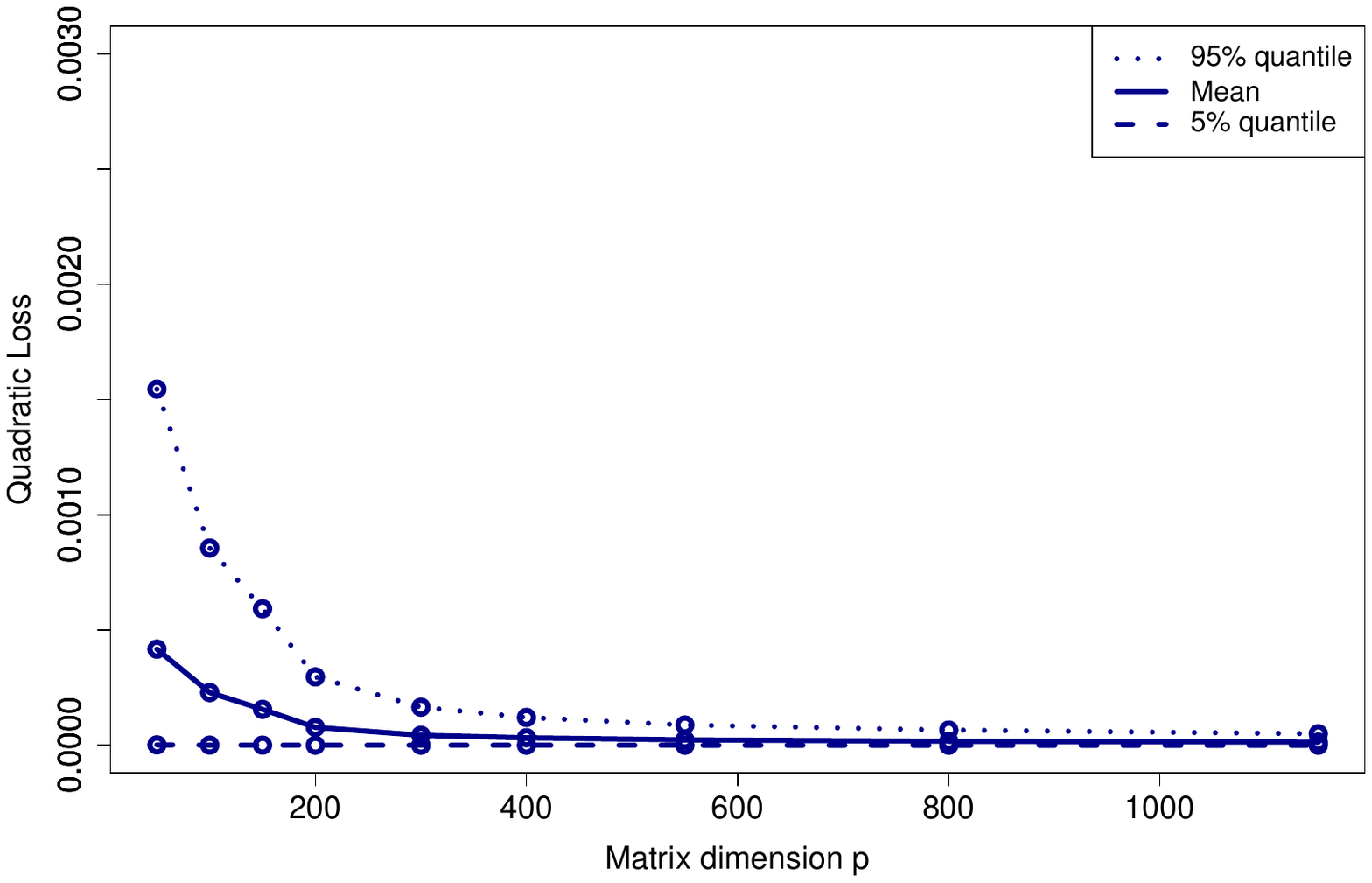}&\includegraphics[scale=0.28]{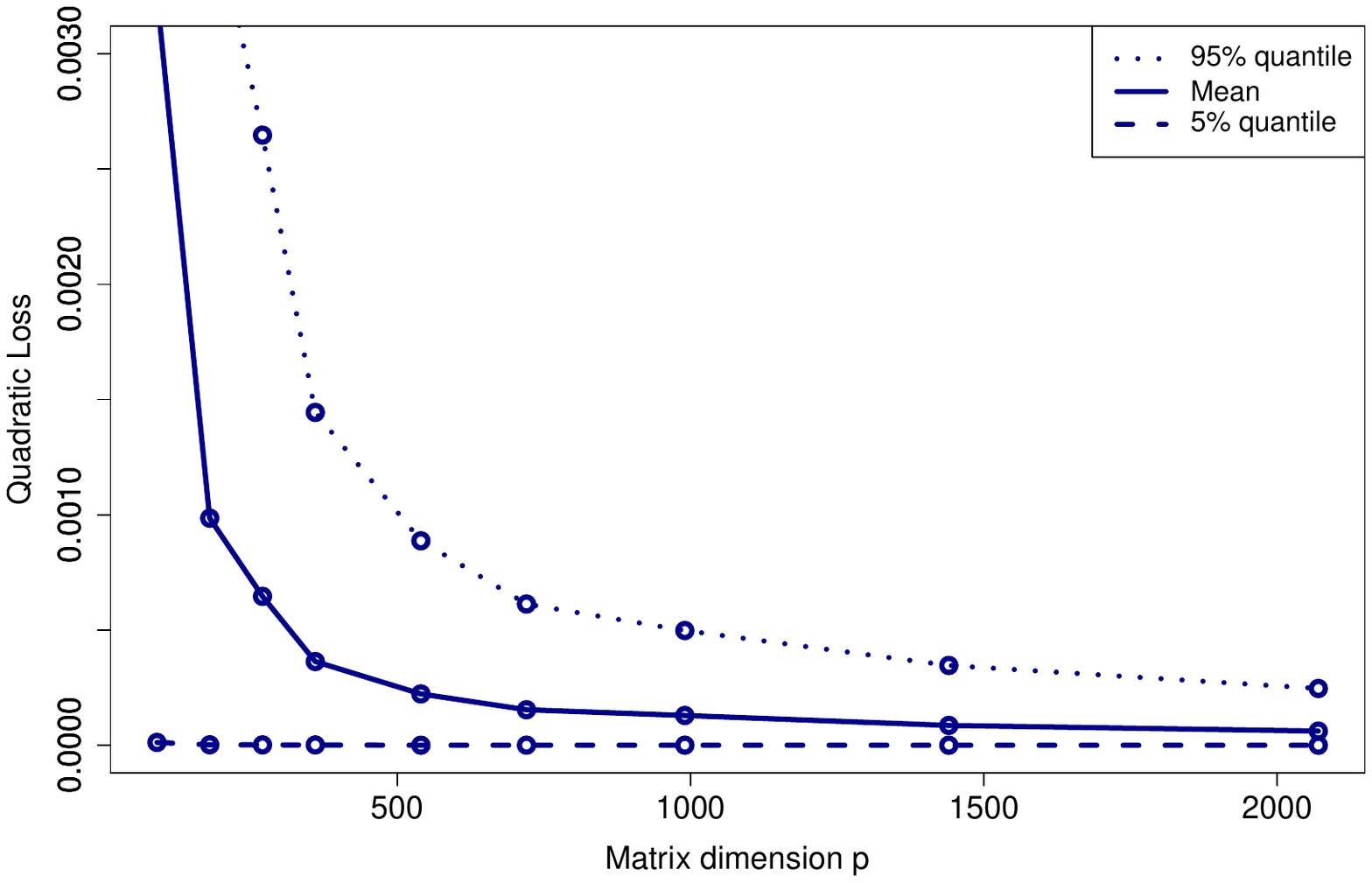}\\
$c=0.5$, $t$ distribution&$c=0.9$, $t$ distribution\\
\includegraphics[scale=0.28]{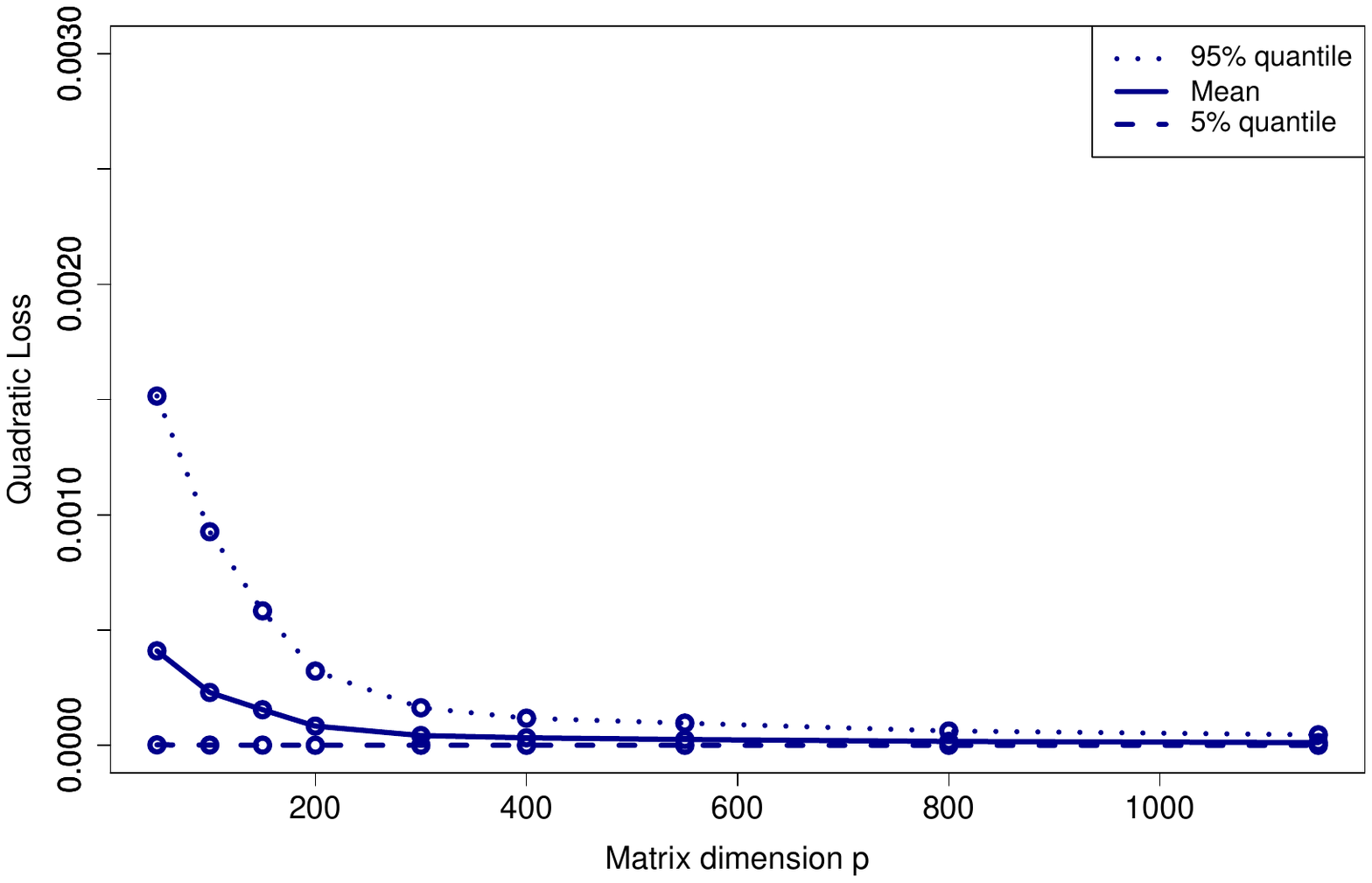}&\includegraphics[scale=0.28]{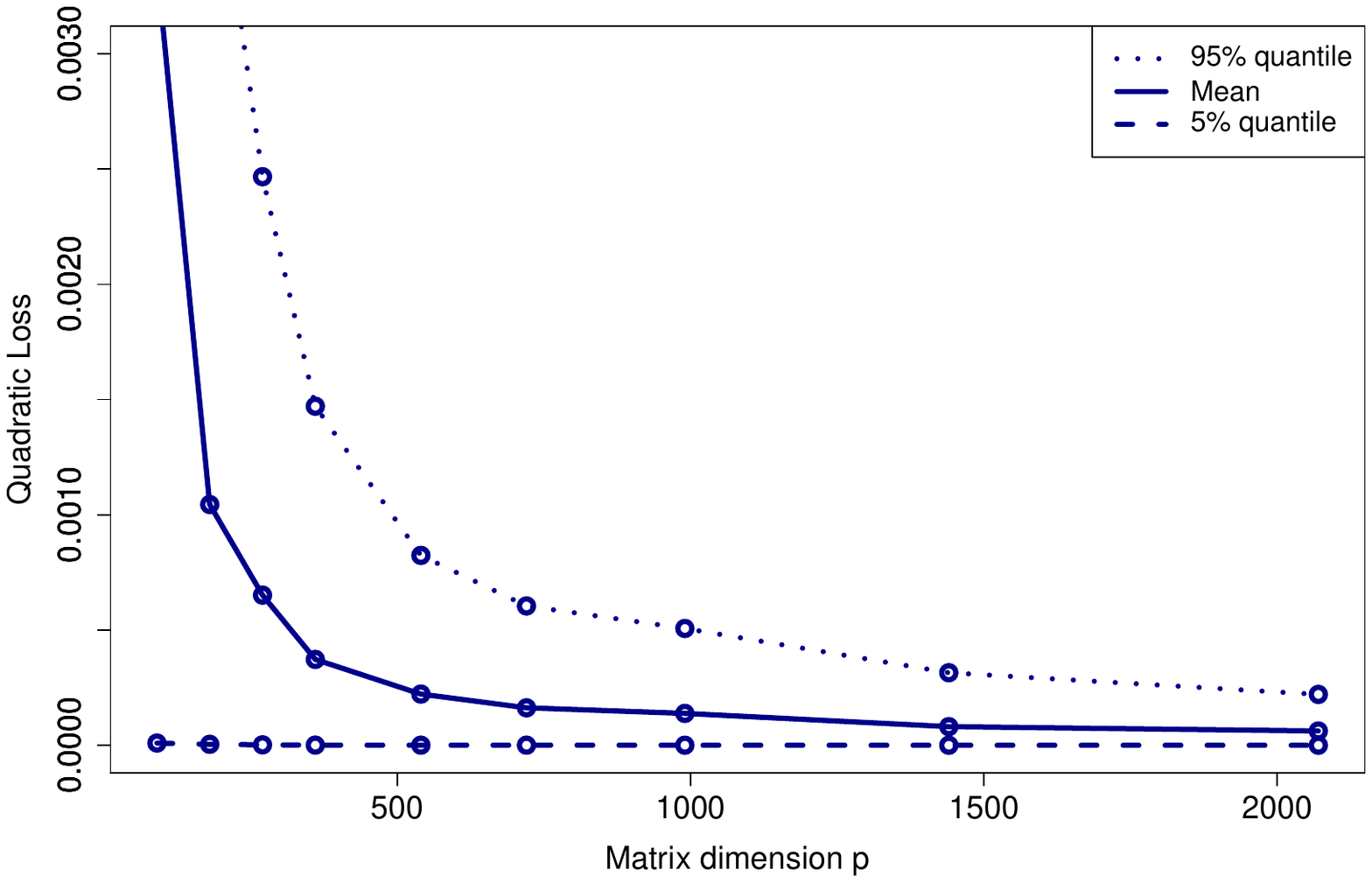}\\
$c=0.5$, CCC-GARCH &$c=0.9$, CCC-GARCH\\
\includegraphics[scale=0.28]{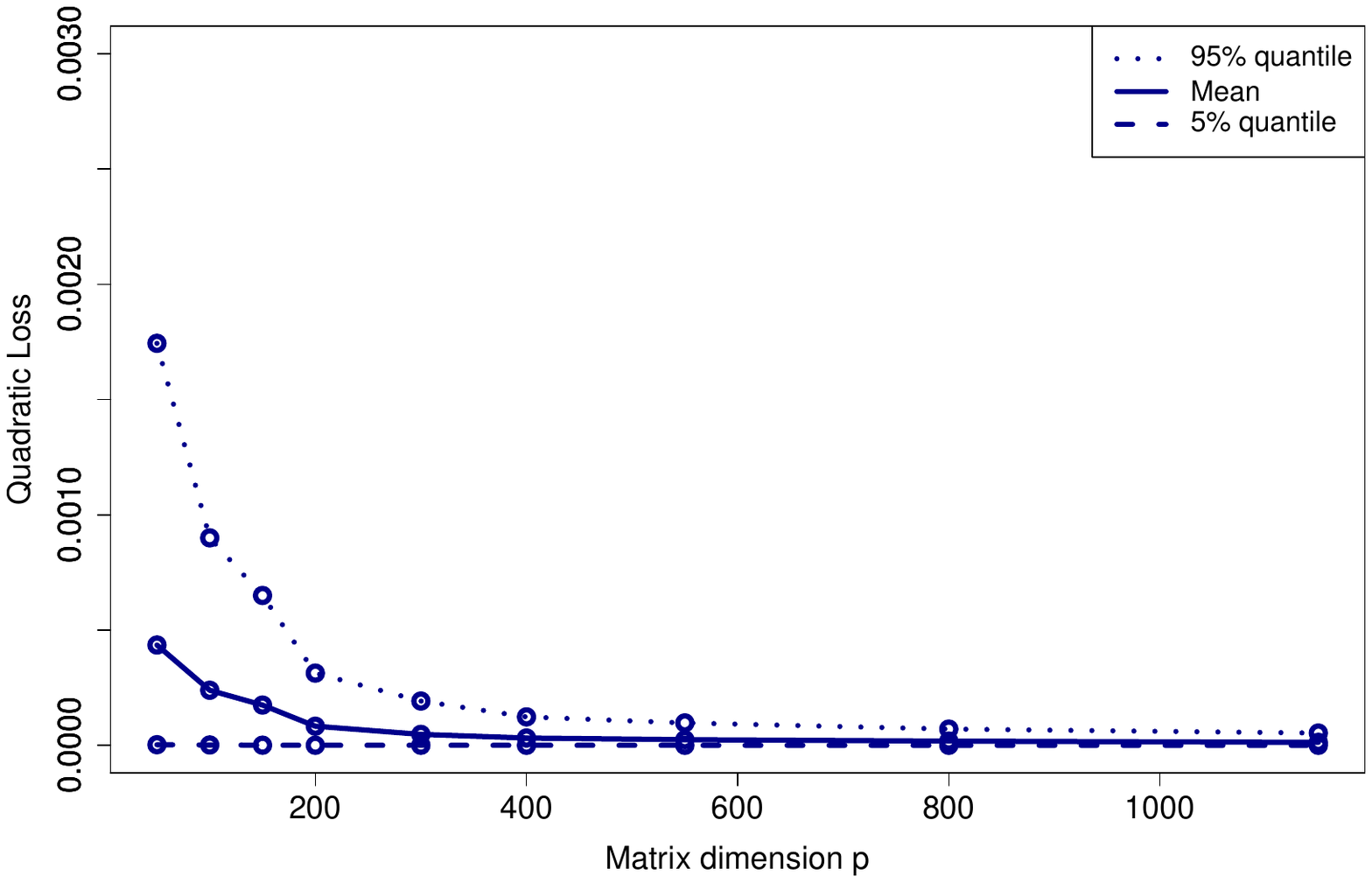}&\includegraphics[scale=0.28]{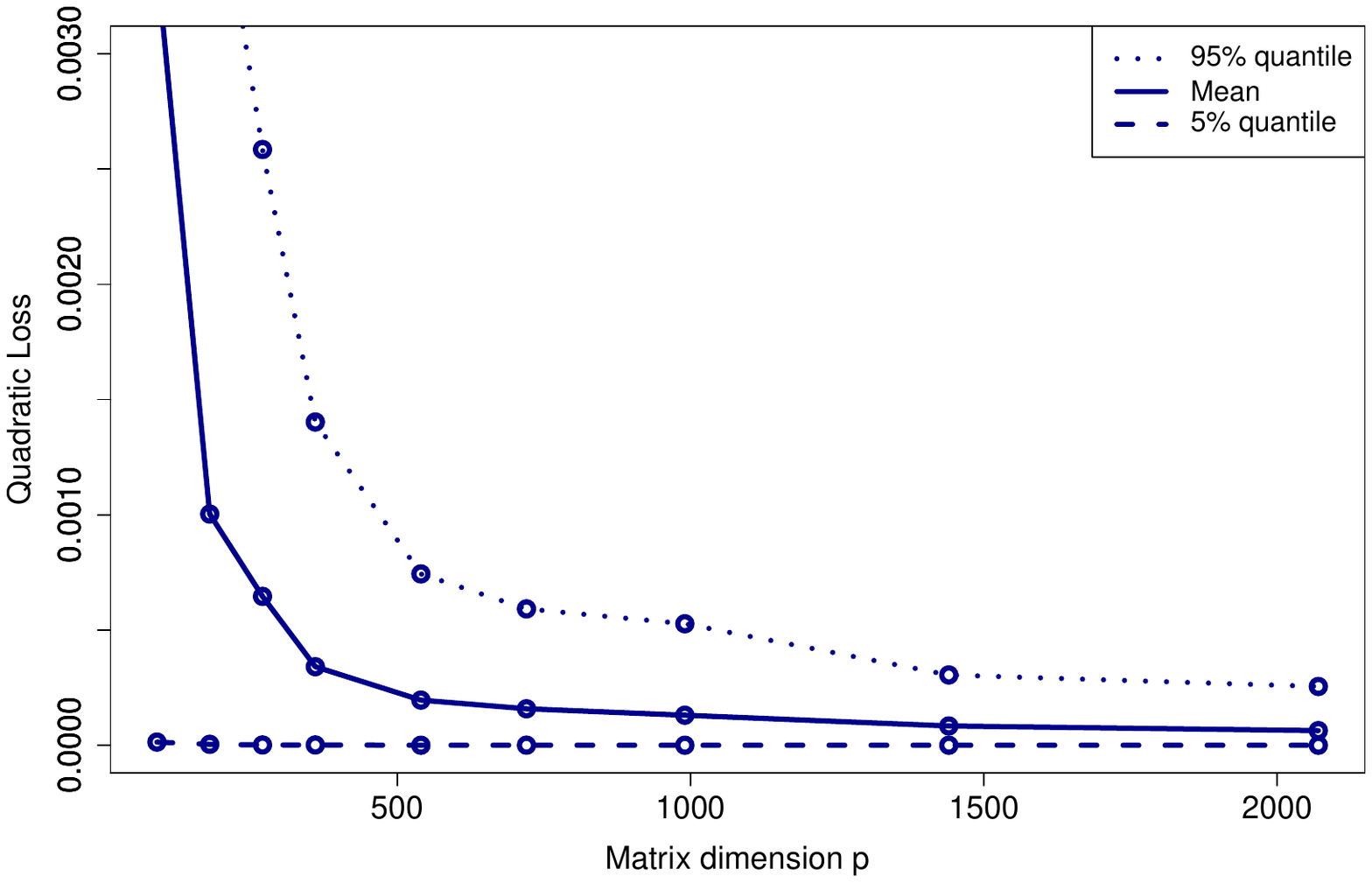}\\
\end{tabular}
\label{sim_loss_R}
\end{figure}

In Figure \ref{sim_loss_R}, we present the results for the estimator $\hat{R}_c$. The solid line denotes the average quadratic loss while the dotted lines present the $0.95$ and $0.05$ quantiles of the quadratic loss. In the top of the figure, the results for the normal distribution are presented, whereas findings for the $t$-distribution and the CCC-GARCH model are given in the middle and in the bottom, correspondingly. In case of the normal distribution, we observe a fast convergence speed of the estimator for moderate values of $c=0.5$. The variance of the quadratic loss is small and vanishes as the dimension increases. In the case of very noisy data, namely $c=0.9$, the estimator $\hat{R}_c$ also reveals a fast convergence rate in the average quadratic loss. However, as it can be seen from the figure, this estimator is quite noisy. Similar results are observed for the $t$-distribution and the CCC-GARCH process, where only the convergence rate is little slower than for the normal distribution.

\begin{figure}[h!!]
\caption{\footnotesize The quadratic loss of $\hat{V}_c$ for the normal
distribution (above), for the $t$-distribution with $3$ degrees of freedom (in the middle), and for the CCC-GARCH process (below).
}
\begin{tabular}{cc}
&\\
$c=0.5$, Normal distribution&$c=0.9$, Normal distribution\\
\includegraphics[scale=0.28]{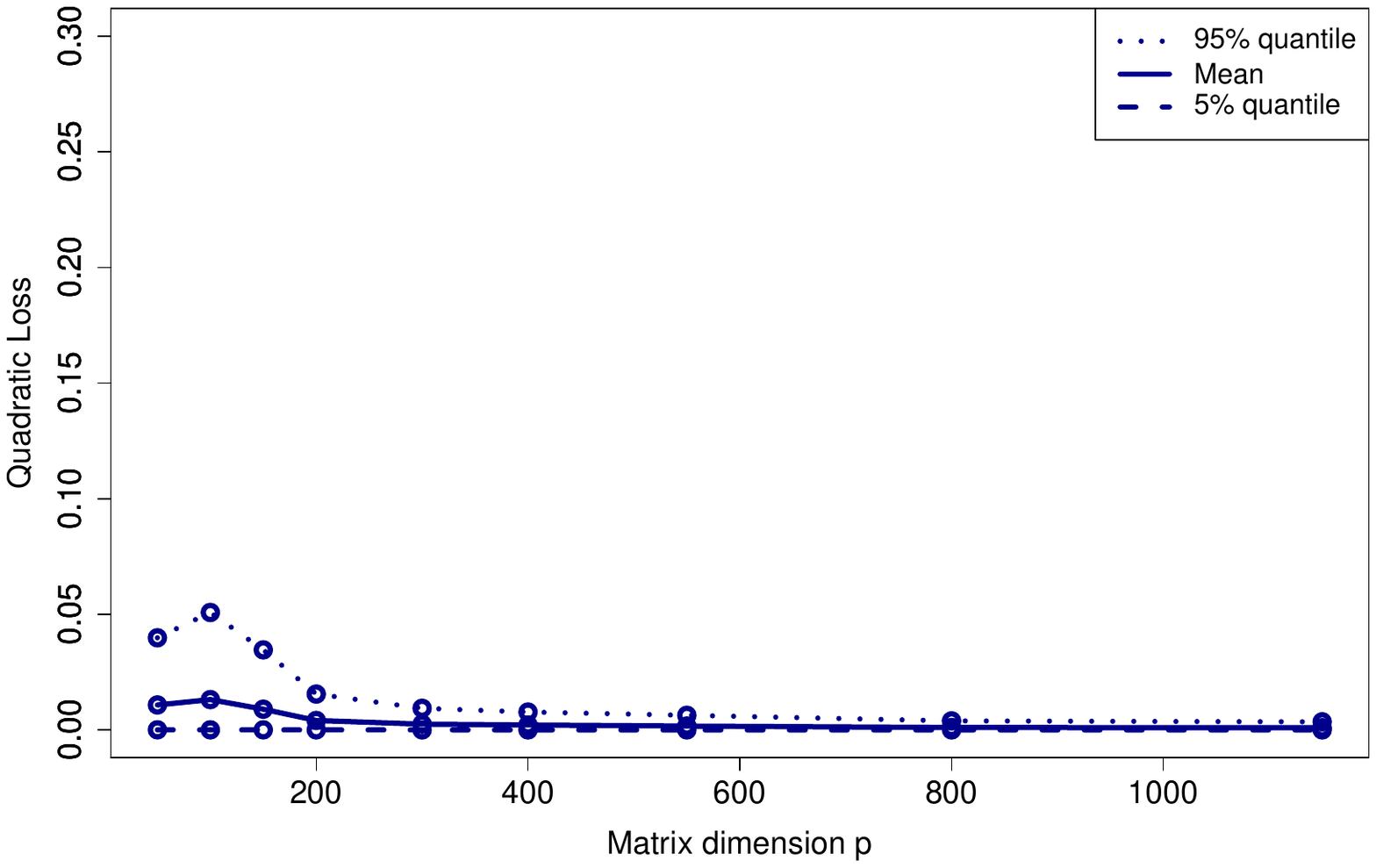}&\includegraphics[scale=0.28]{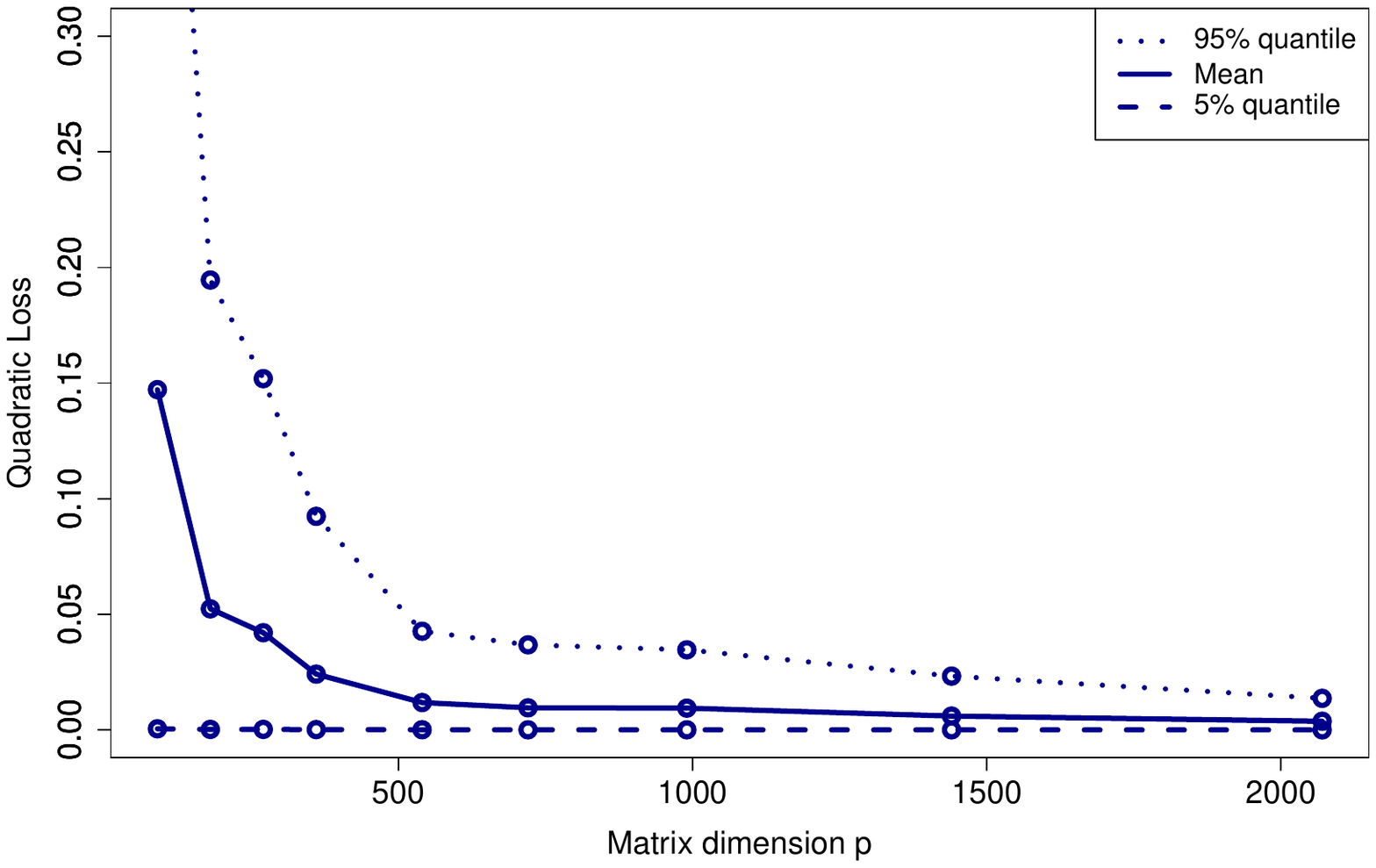}\\
$c=0.5$, $t$ distribution&$c=0.9$, $t$ distribution\\
\includegraphics[scale=0.28]{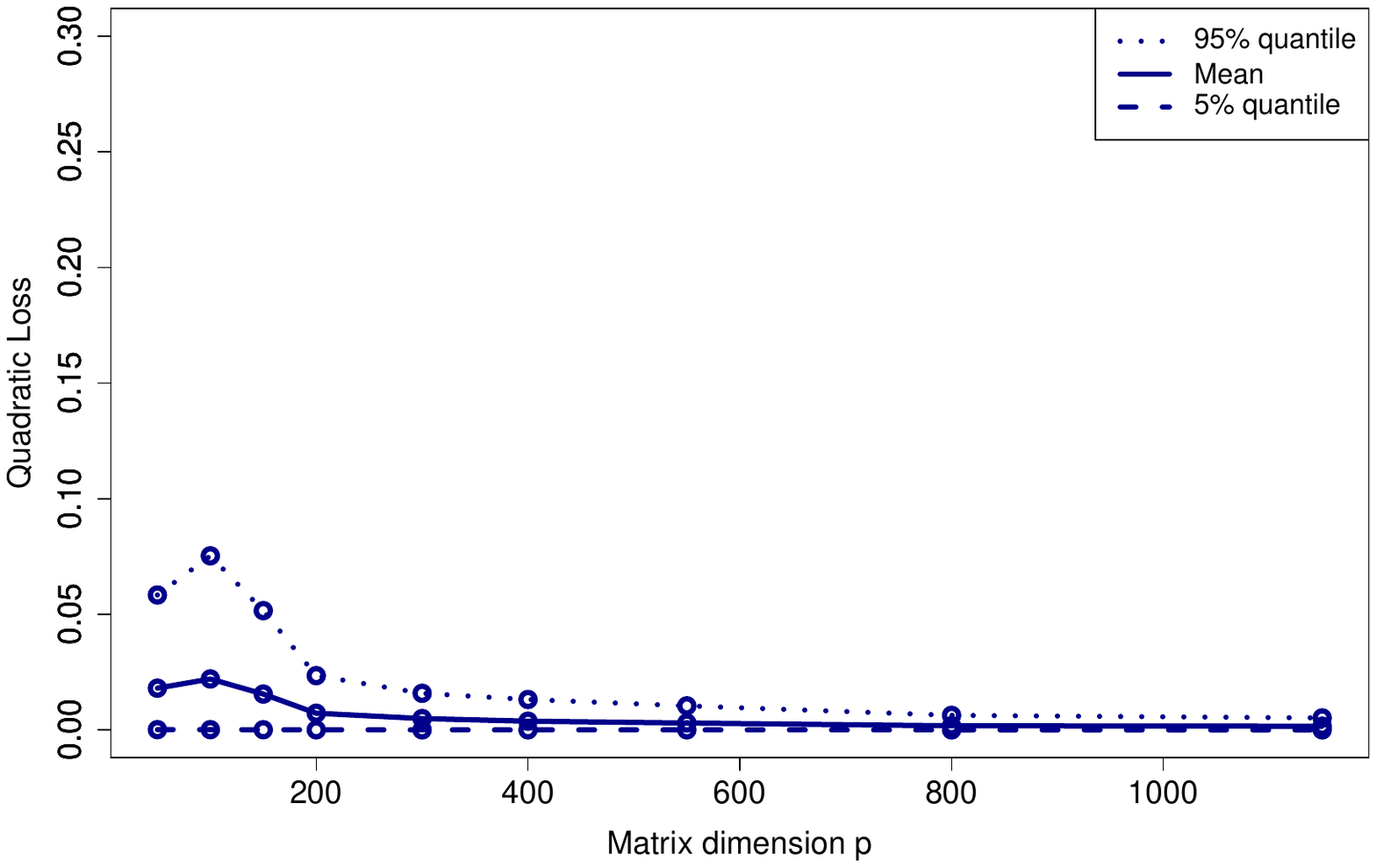}&\includegraphics[scale=0.28]{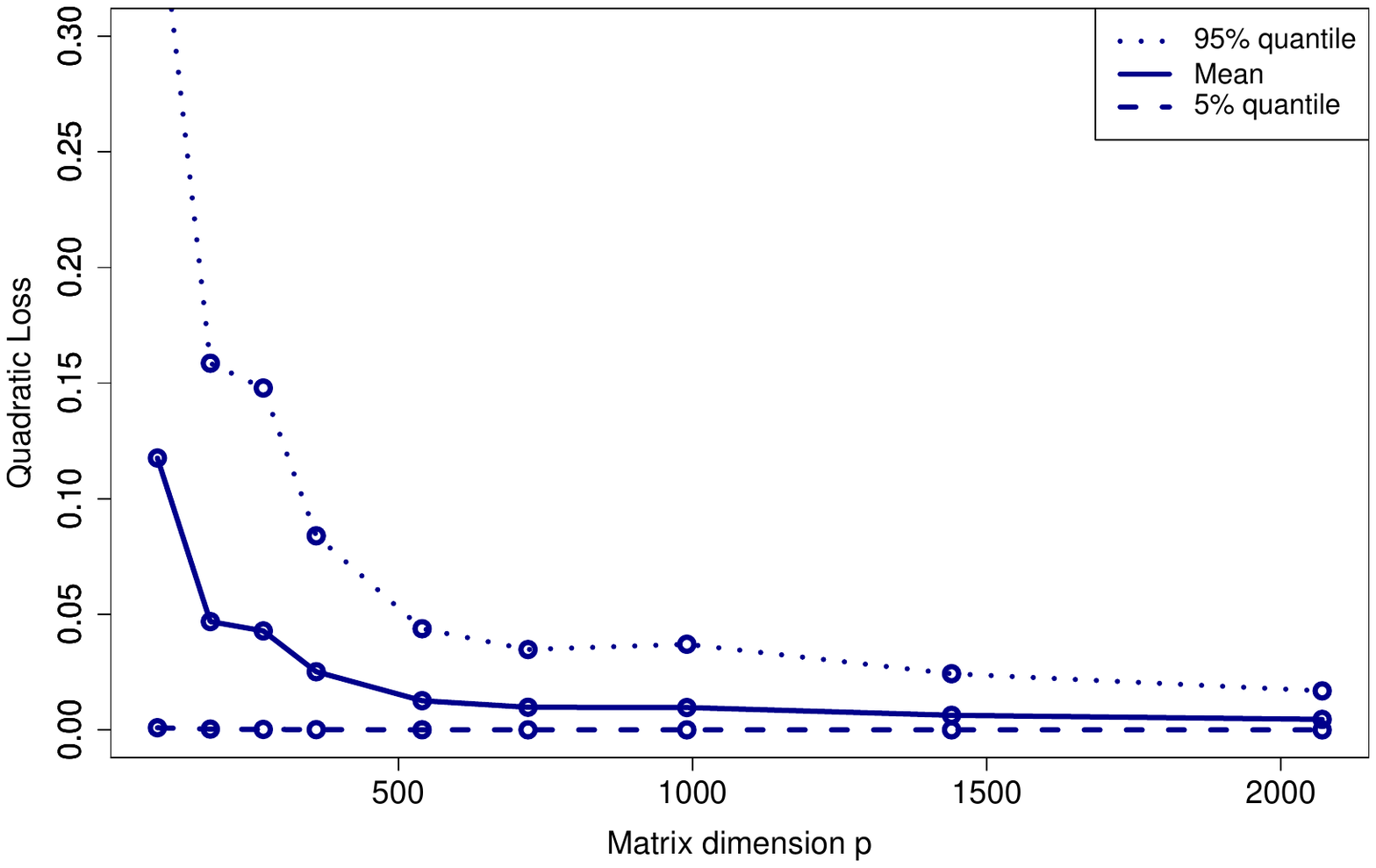}\\
$c=0.5$, CCC-GARCH &$c=0.9$, CCC-GARCH\\
\includegraphics[scale=0.28]{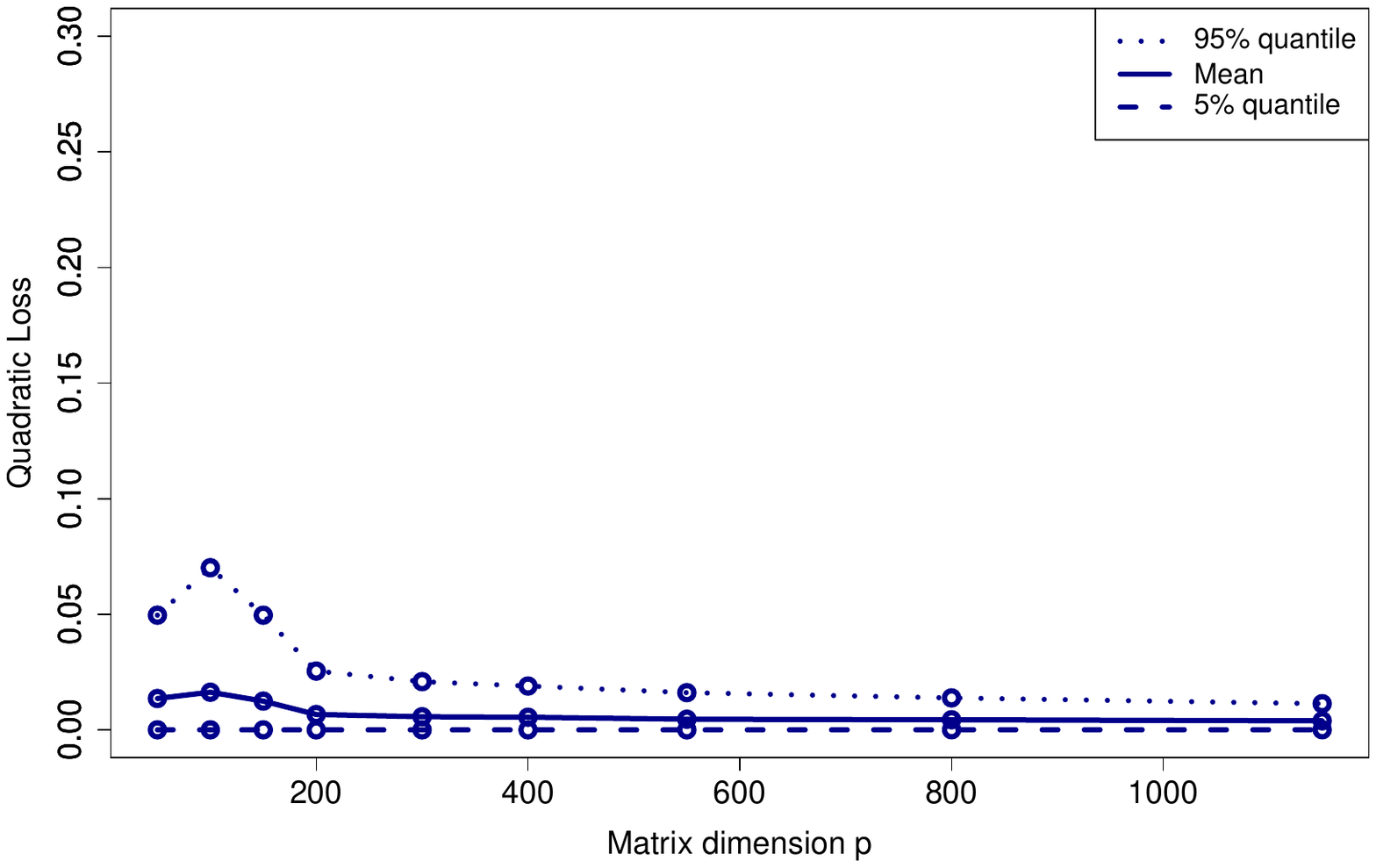}&\includegraphics[scale=0.28]{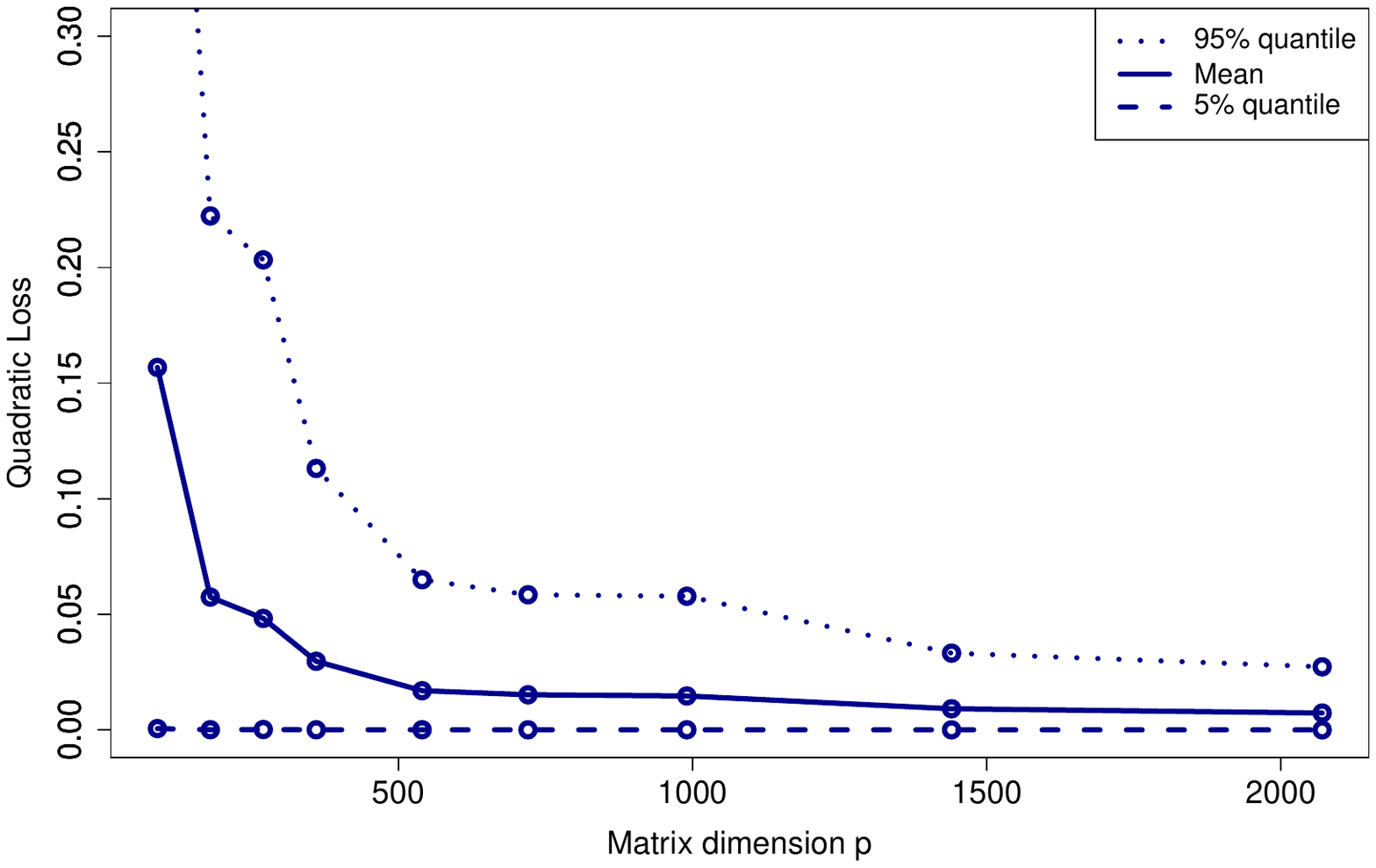}\\
\end{tabular}
\label{sim_loss_V}
\end{figure}

Figure \ref{sim_loss_V} contains the simulation results for the estimator $\hat{V}_c$, which is a consistent estimator for $V_{GMV}$. We observe a similar behavior independently of the scenario used to simulate the data. As expected, better results are presented in the case of the normal distribution, whereas the worst ones correspond to the CCC-GARCH process, although the difference is not large. The convergence rate is relatively small for $p<200$ in the case of $c=0.5$ and for $p<500$ in the case of $c=0.9$. If $p$ increases the quadratic loss tends to zero.

\begin{figure}[h!!]
\caption{\footnotesize The quadratic loss of $\hat{s}_c$ for the normal
distribution (above), for the $t$-distribution with $3$ degrees of freedom (in the middle), and for the CCC-GARCH process (below).
}
\begin{tabular}{cc}
&\\
$c=0.5$, Normal distribution&$c=0.9$, Normal distribution\\
\includegraphics[scale=0.28]{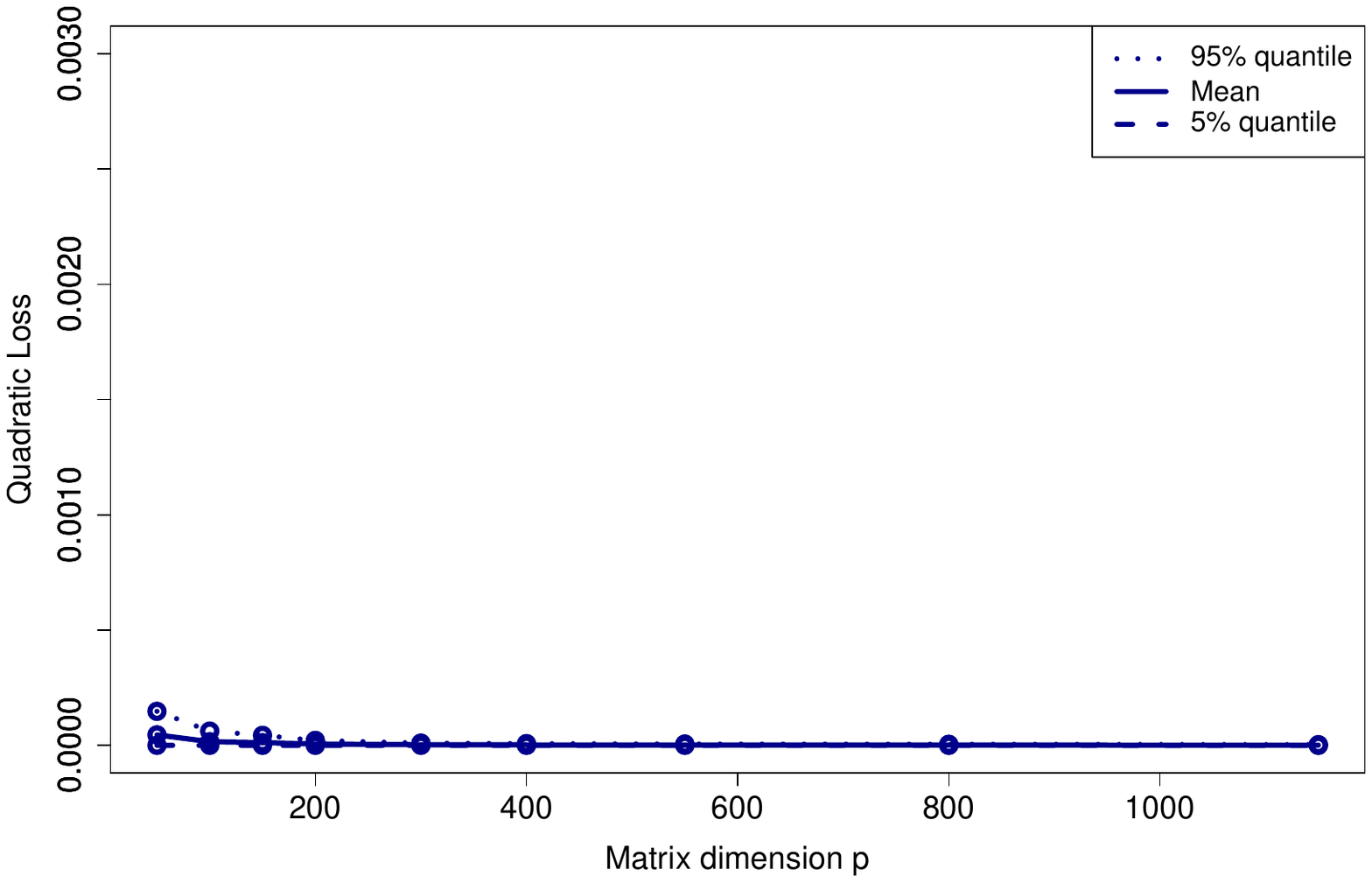}&\includegraphics[scale=0.28]{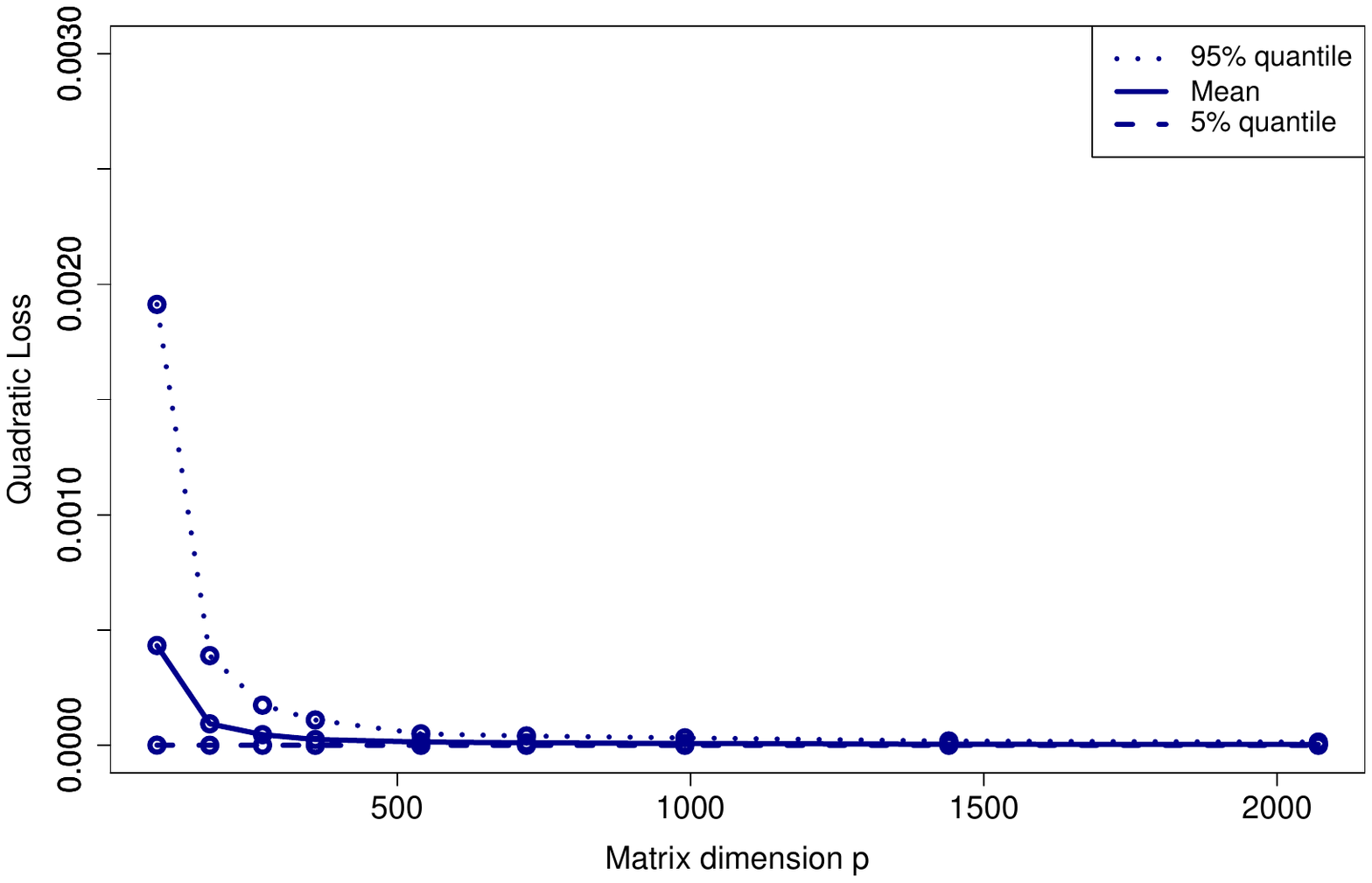}\\
$c=0.5$, $t$ distribution&$c=0.9$, $t$ distribution\\
\includegraphics[scale=0.28]{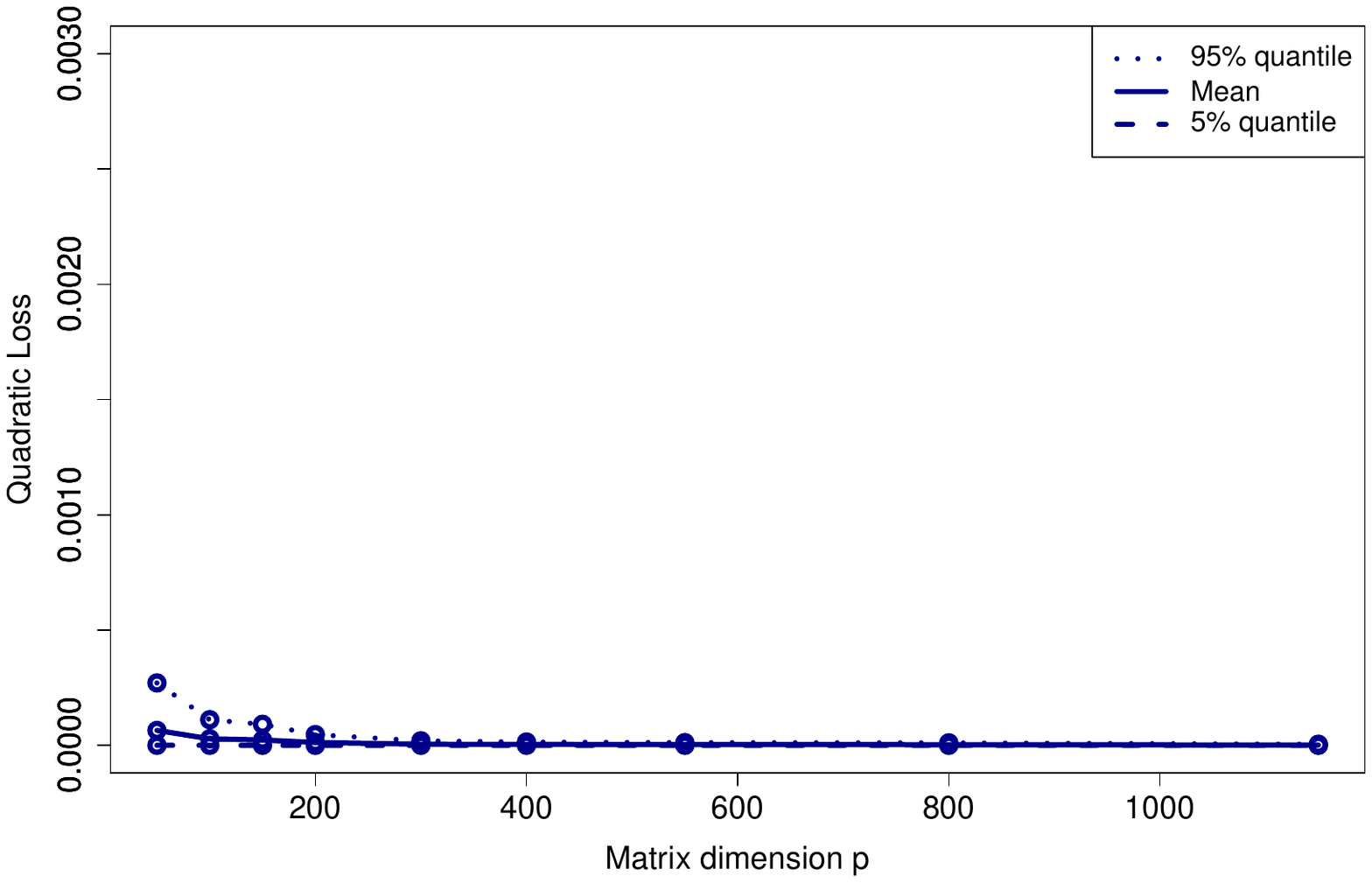}&\includegraphics[scale=0.28]{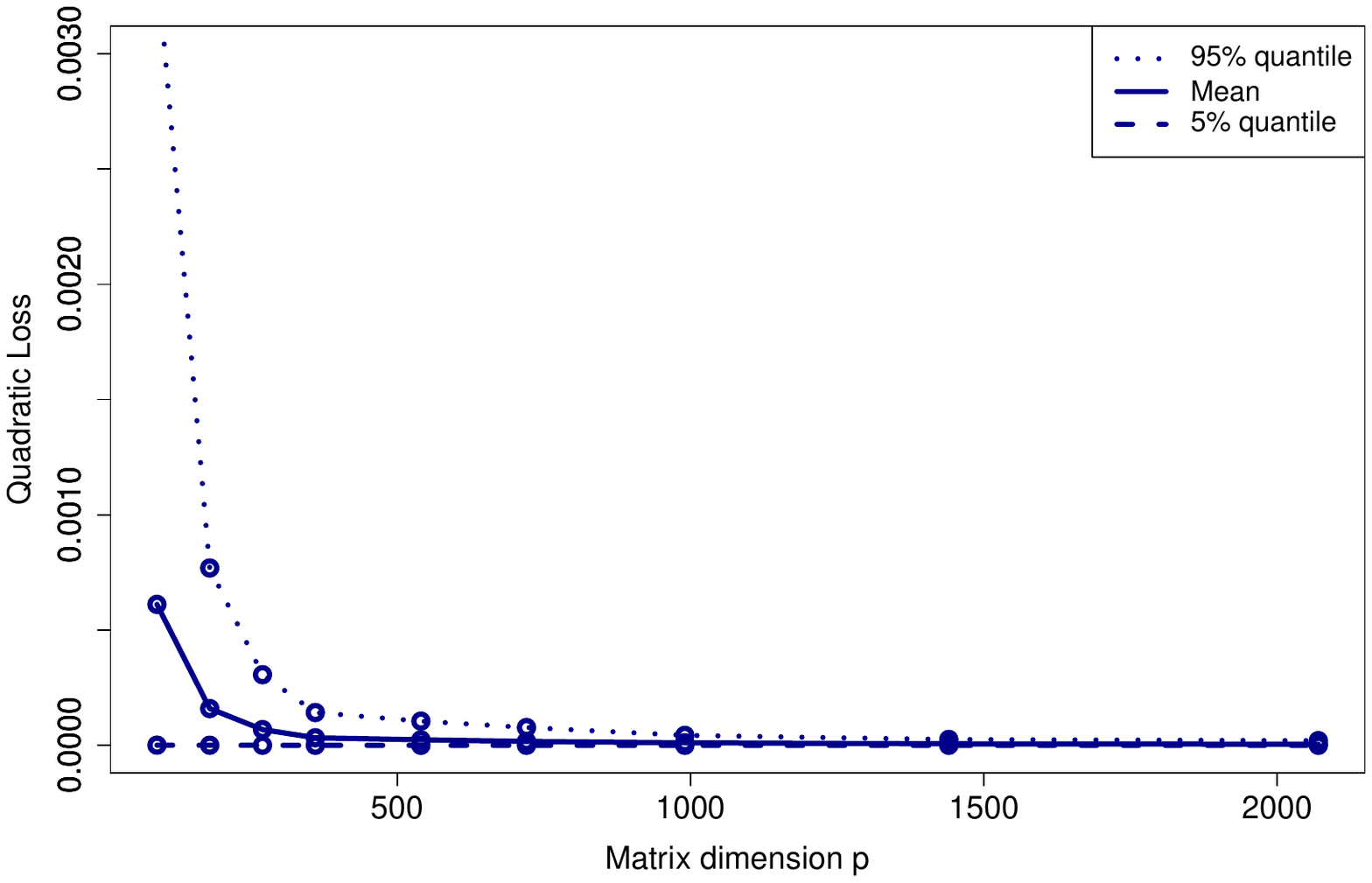}\\
$c=0.5$, CCC-GARCH &$c=0.9$, CCC-GARCH\\
\includegraphics[scale=0.28]{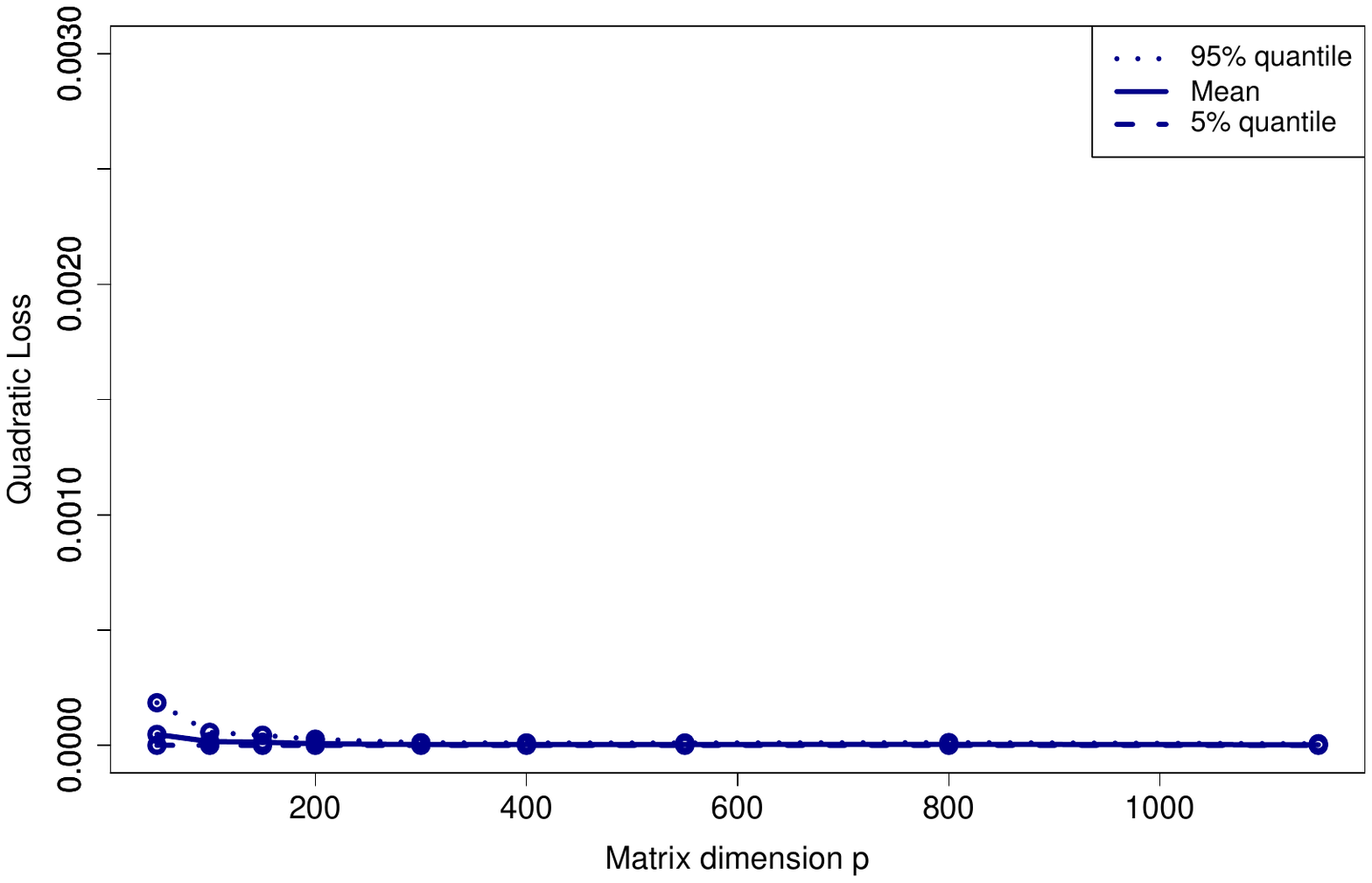}&\includegraphics[scale=0.28]{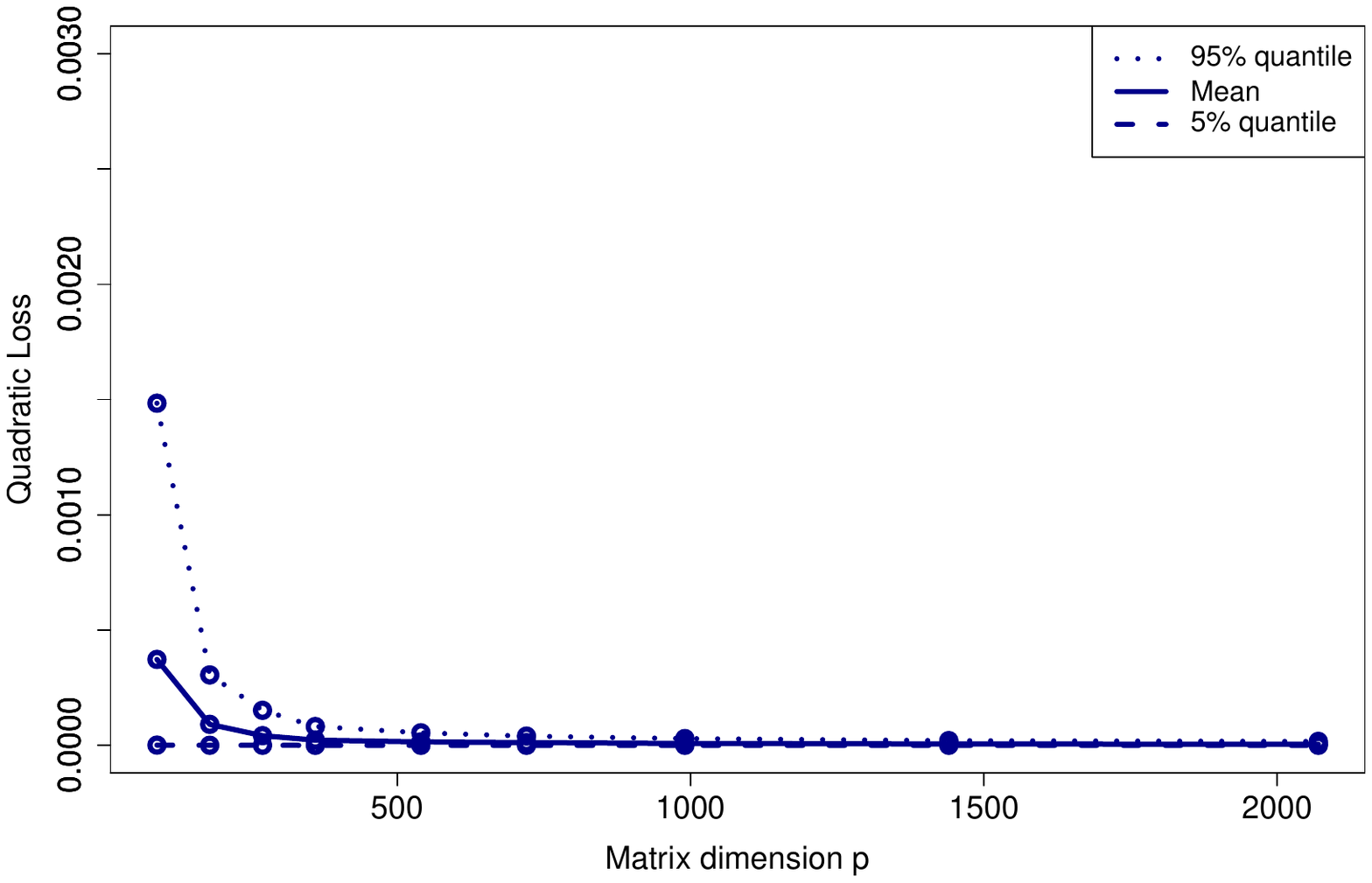}\\
\end{tabular}
\label{sim_loss_s}
\end{figure}

The results obtained for $\hat{s}_c$ in Figure \ref{sim_loss_s} are stronger as those obtained for $\hat{R}_c$ and $\hat{V}_c$. In the case of $c=0.5$, the quadratic loss function tends to zero already for small values of $p$, while for $c=0.9$ larger values of $p$ are needed. Interestingly, the best results are present in the case of the CCC-GARCH model, where the convergence rate is even larger than in case of the normal distribution.

\begin{figure}[h!!]
\caption{\footnotesize Histogram and the asymptotic density (see Theorem \ref{th2}) of $\sqrt{n}(\hat{R}_c-R_{GMV})$ for the normal
distribution (above), for the $t$-distribution with $3$ degrees of freedom (in the middle), and for the CCC-GARCH process (below). We put $n=1000$.
}
\begin{tabular}{cc}
&\\
$c=0.5$, Normal distribution&$c=0.9$, Normal distribution\\
\includegraphics[scale=0.28]{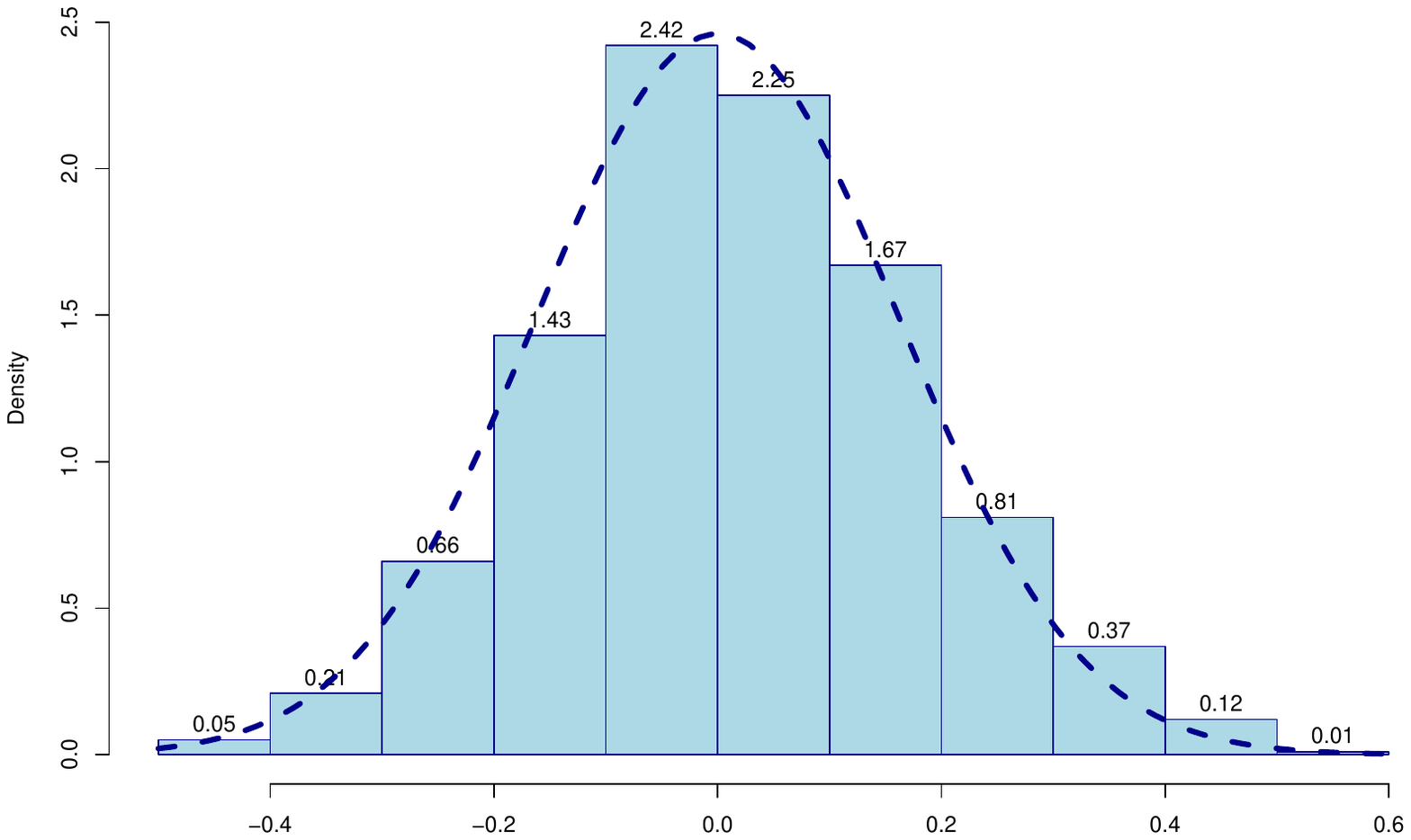}&\includegraphics[scale=0.28]{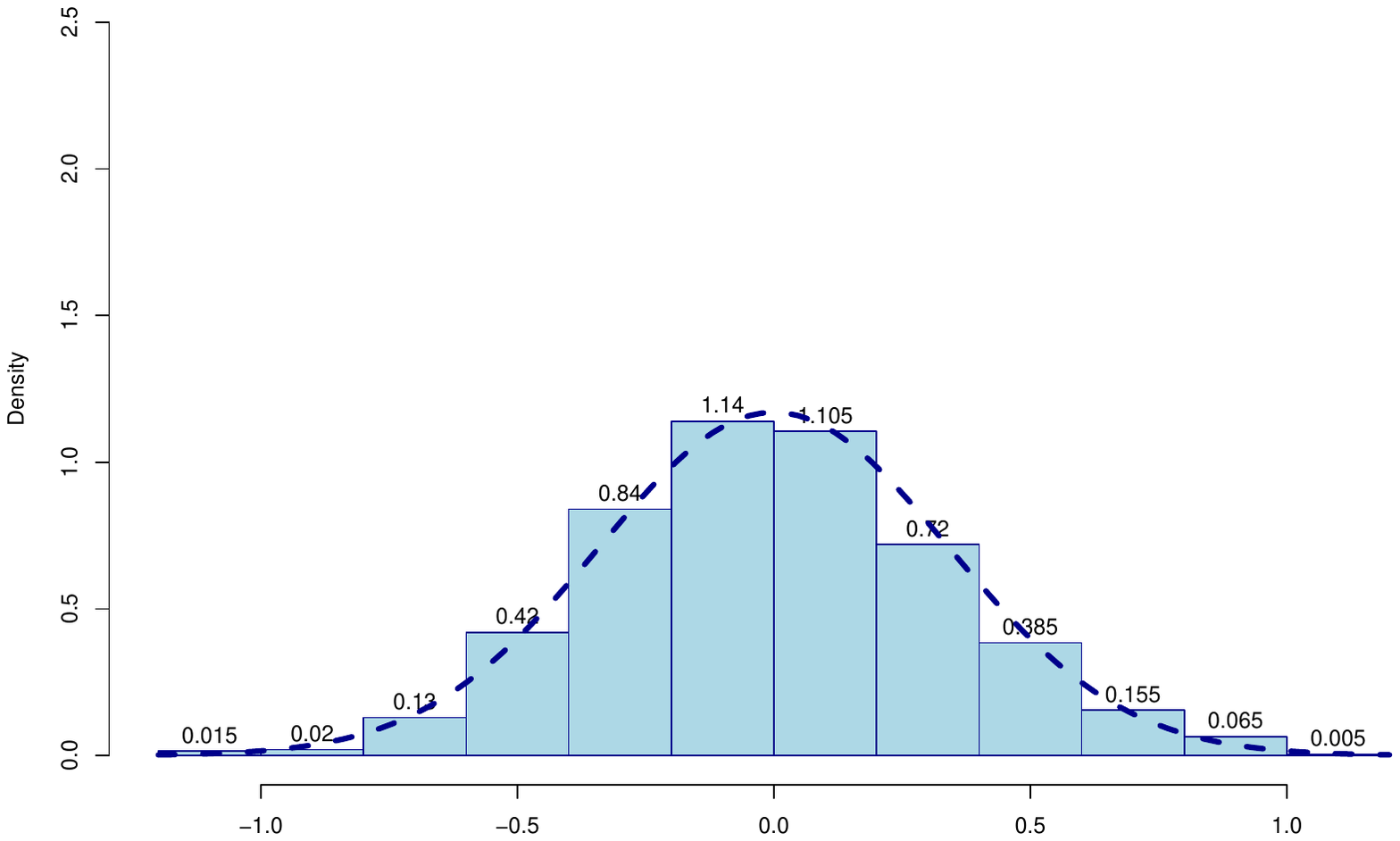}\\
$c=0.5$, $t$ distribution&$c=0.9$, $t$ distribution\\
\includegraphics[scale=0.28]{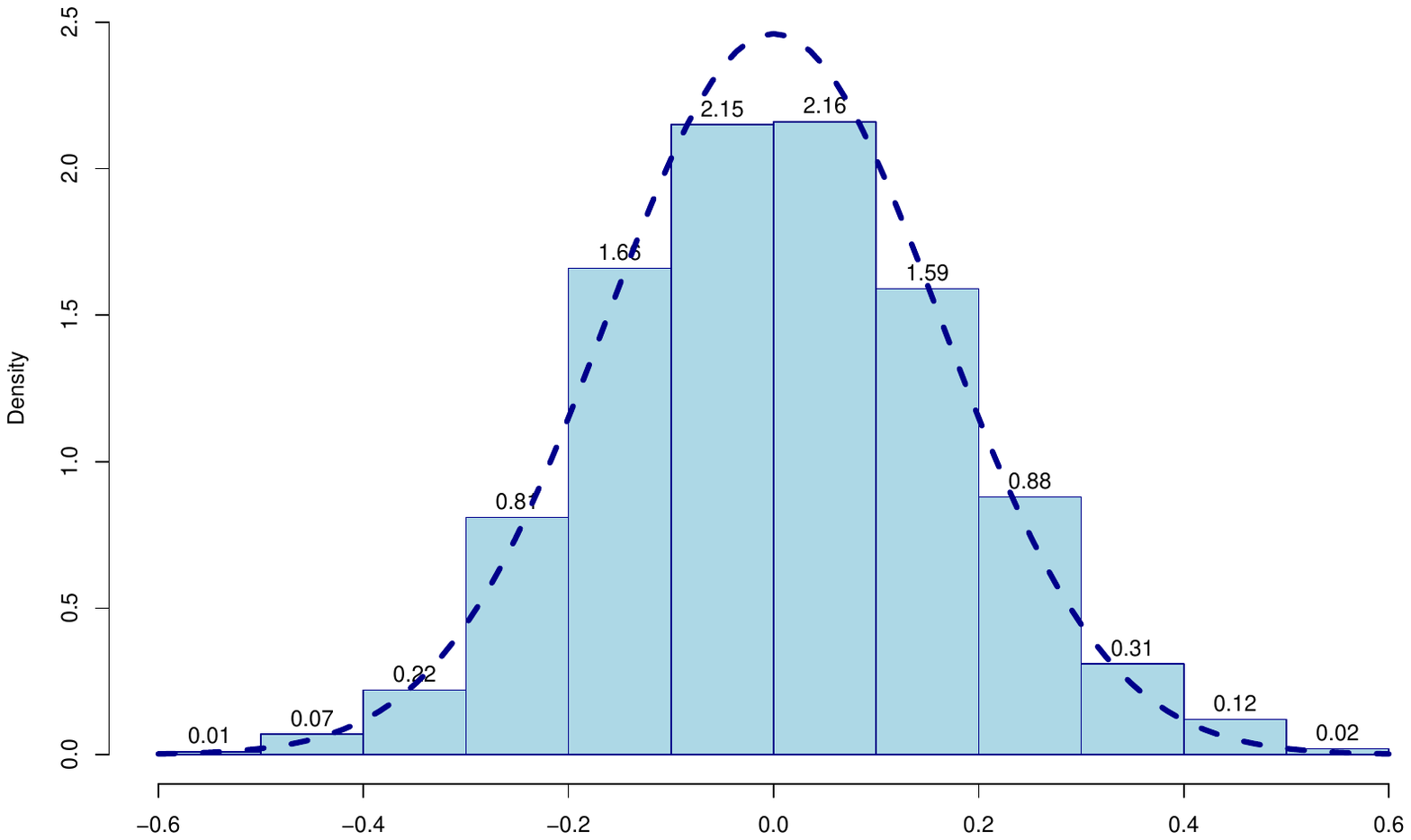}&\includegraphics[scale=0.28]{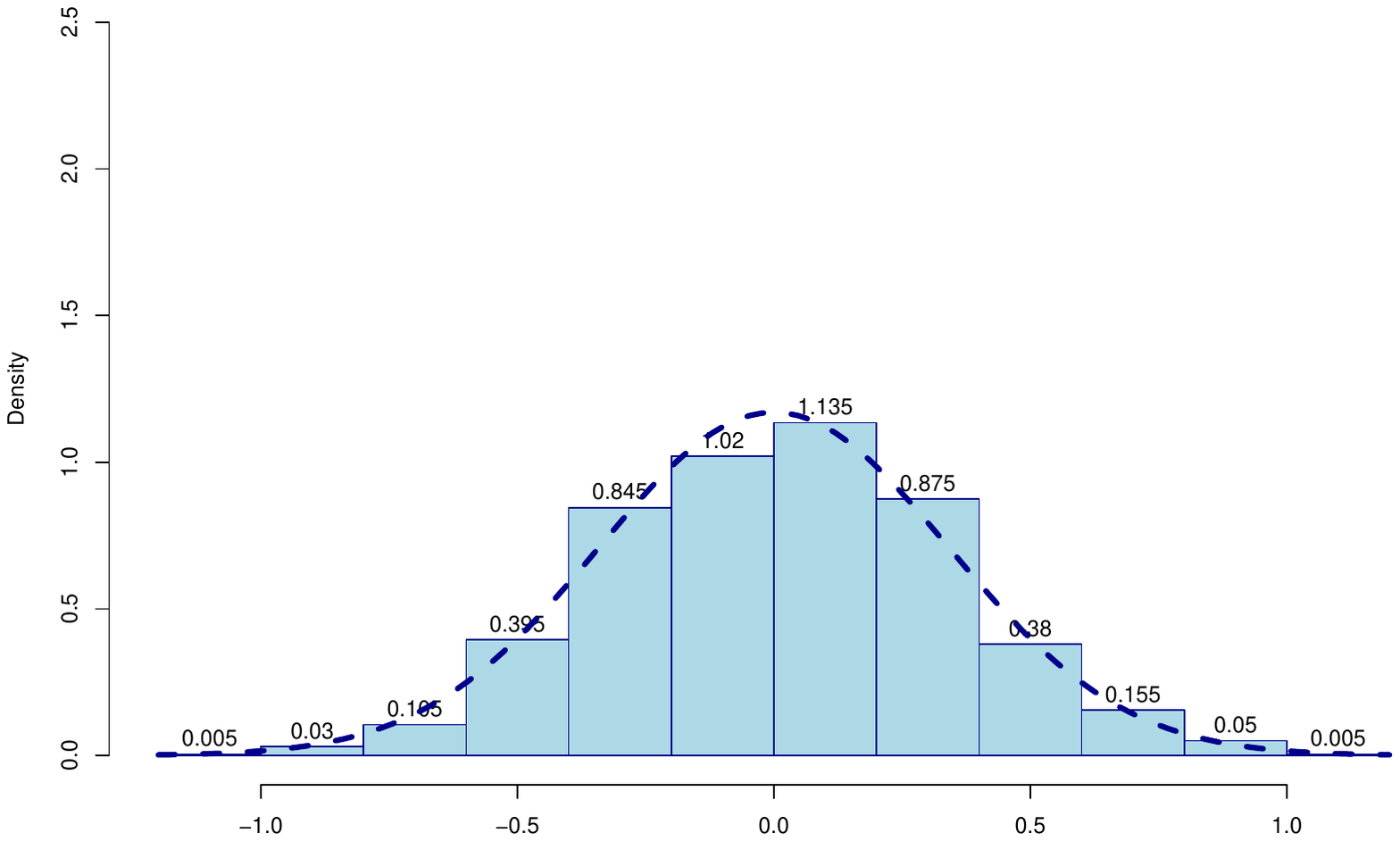}\\
$c=0.5$, CCC-GARCH &$c=0.9$, CCC-GARCH\\
\includegraphics[scale=0.28]{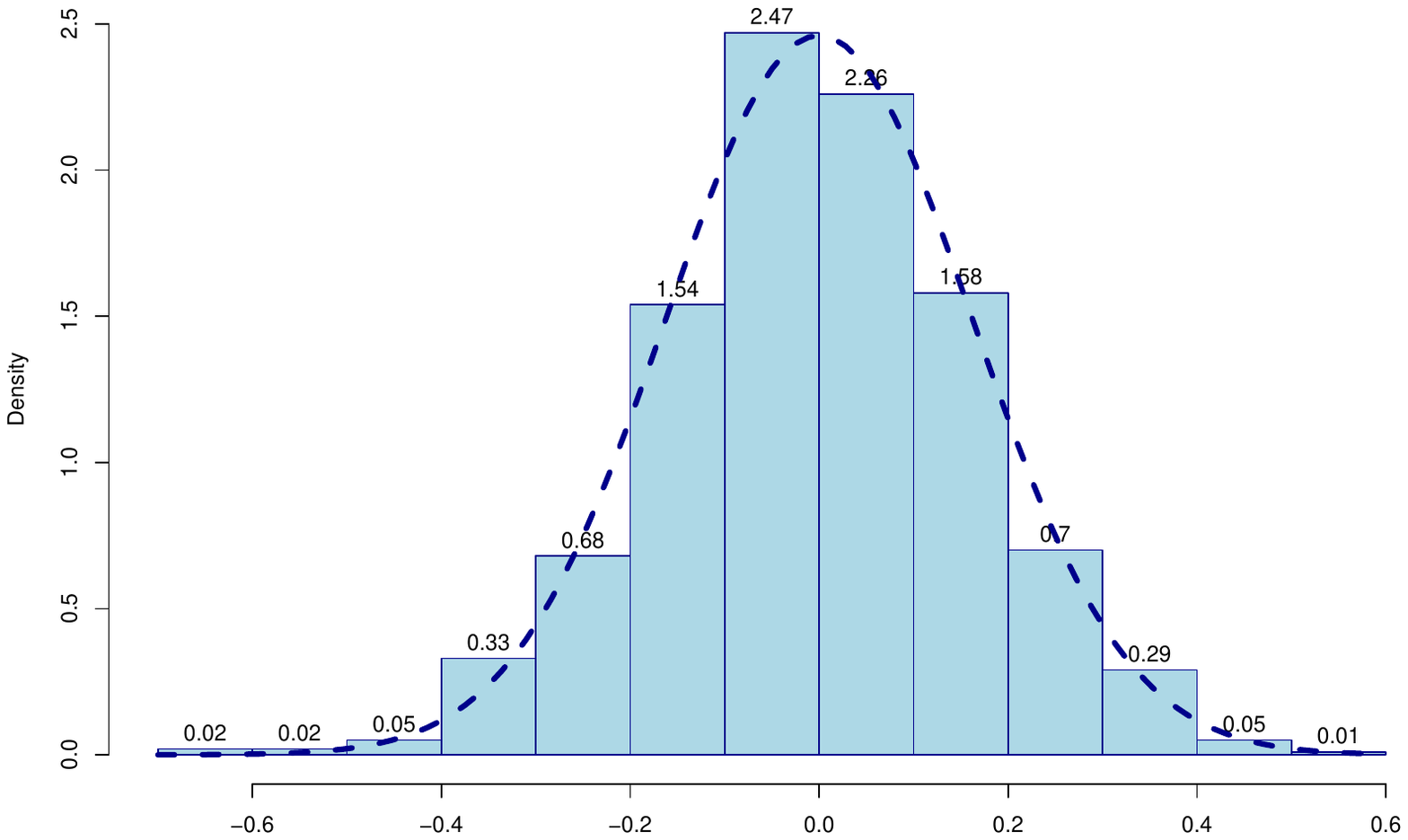}&\includegraphics[scale=0.28]{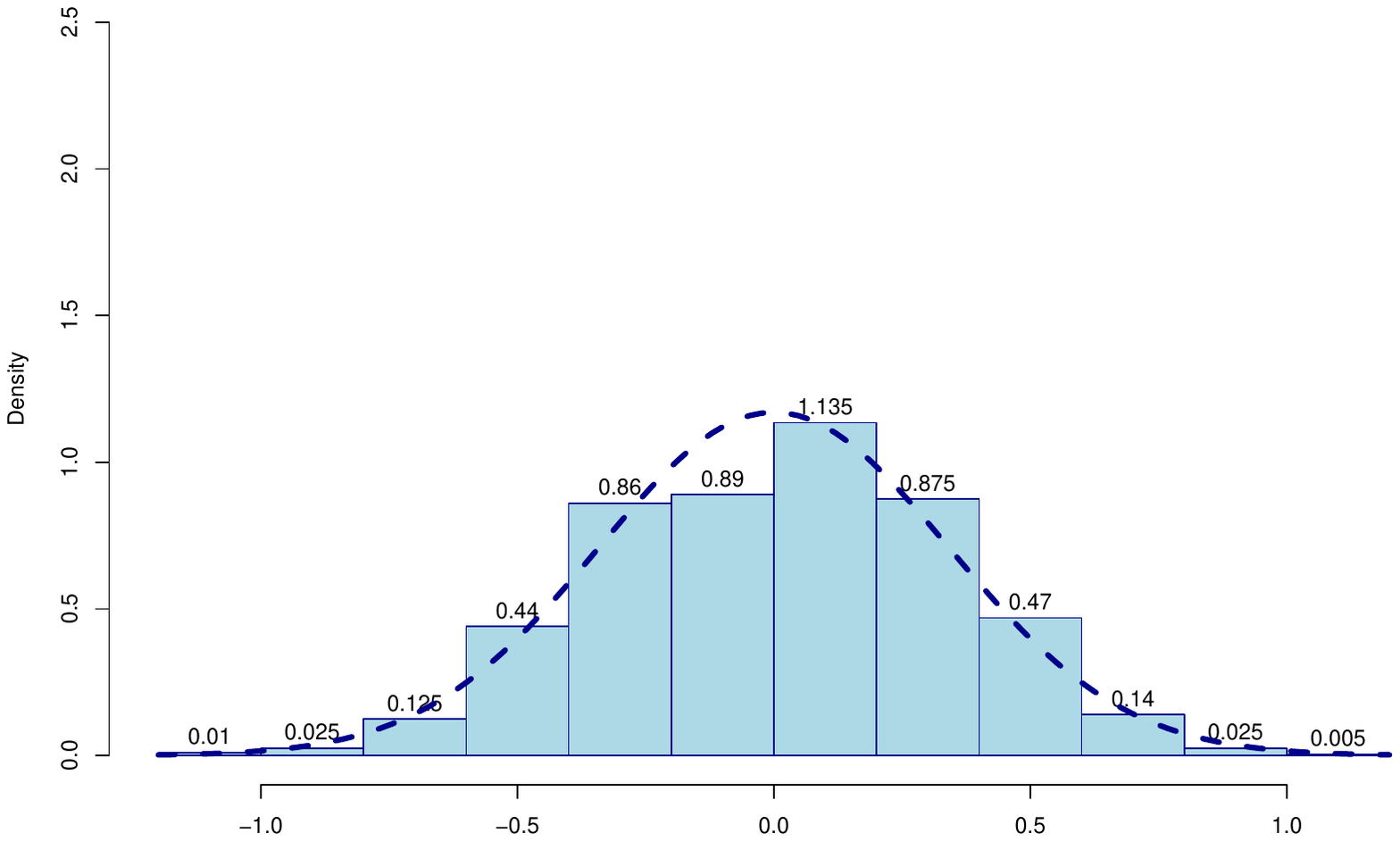}\\
\end{tabular}
\label{sim_hist_R}
\end{figure}

\begin{figure}[h!!]
\caption{\footnotesize Histogram and the asymptotic density (see Theorem \ref{th2}) of $\sqrt{n}(\hat{V}_c-V_{GMV})$ for the normal
distribution (above), for the $t$-distribution with $3$ degrees of freedom (in the middle), and for the CCC-GARCH process (below). We put $n=1000$.
}
\begin{tabular}{cc}
&\\
$c=0.5$, Normal distribution&$c=0.9$, Normal distribution\\
\includegraphics[scale=0.28]{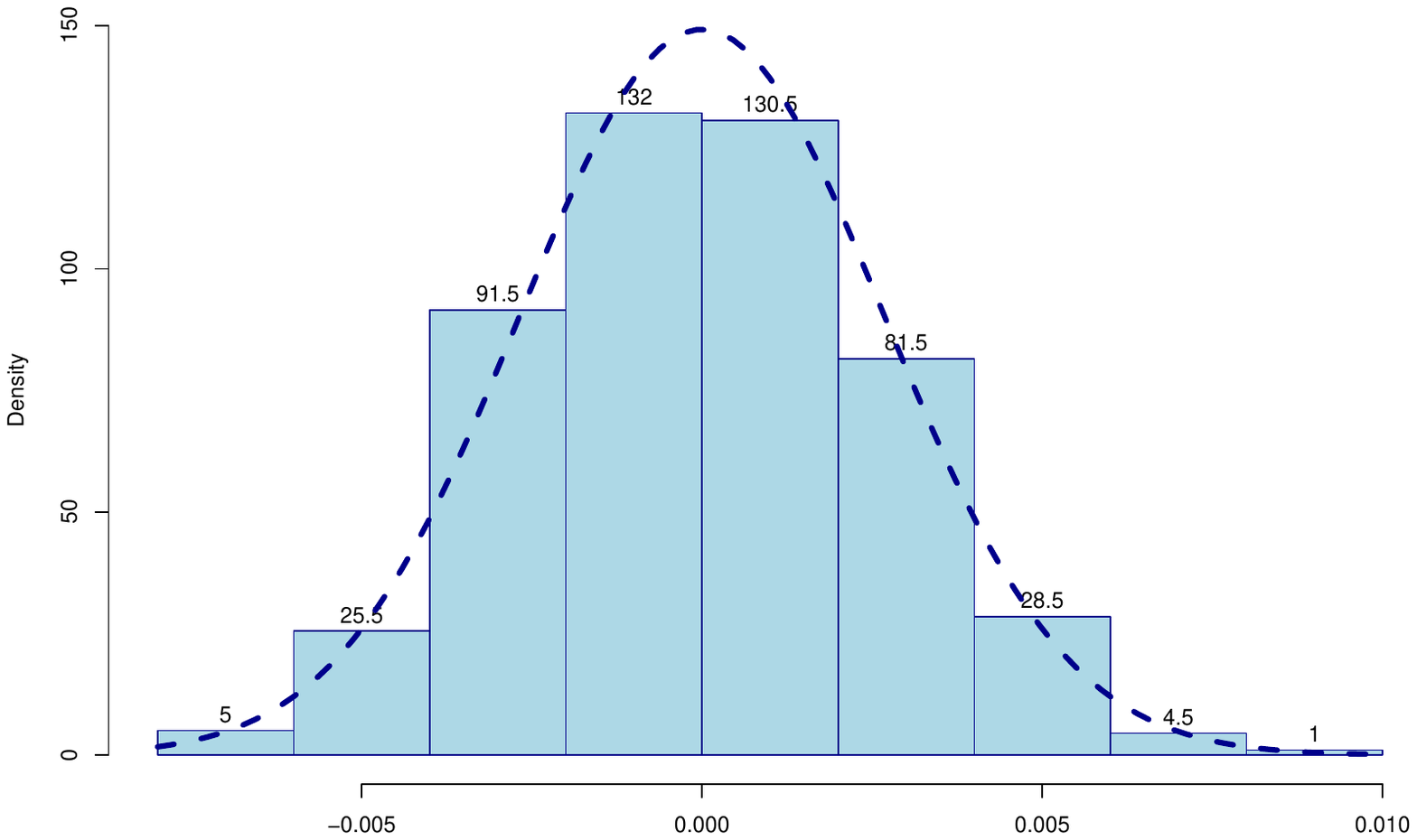}&\includegraphics[scale=0.28]{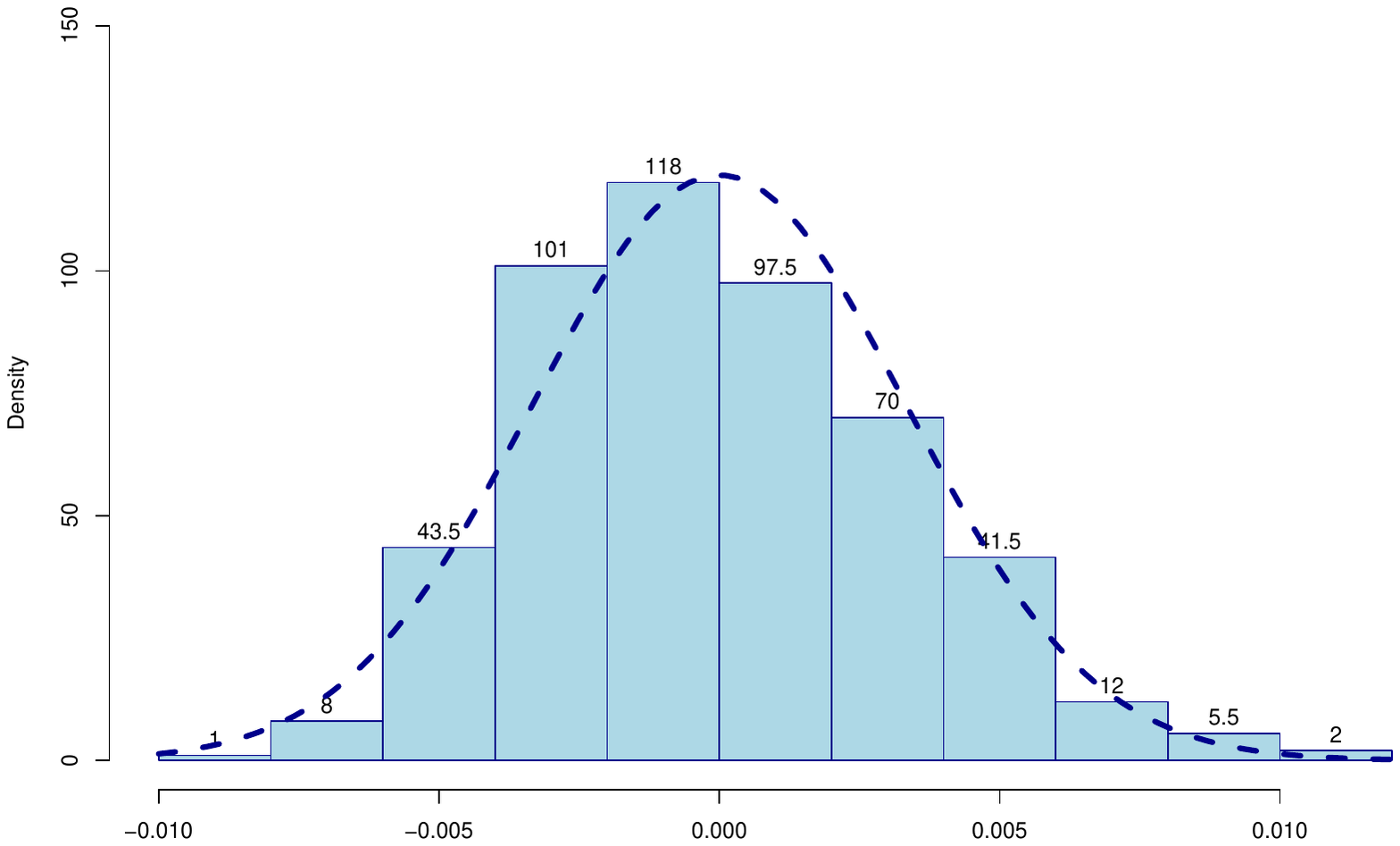}\\
$c=0.5$, $t$ distribution&$c=0.9$, $t$ distribution\\
\includegraphics[scale=0.28]{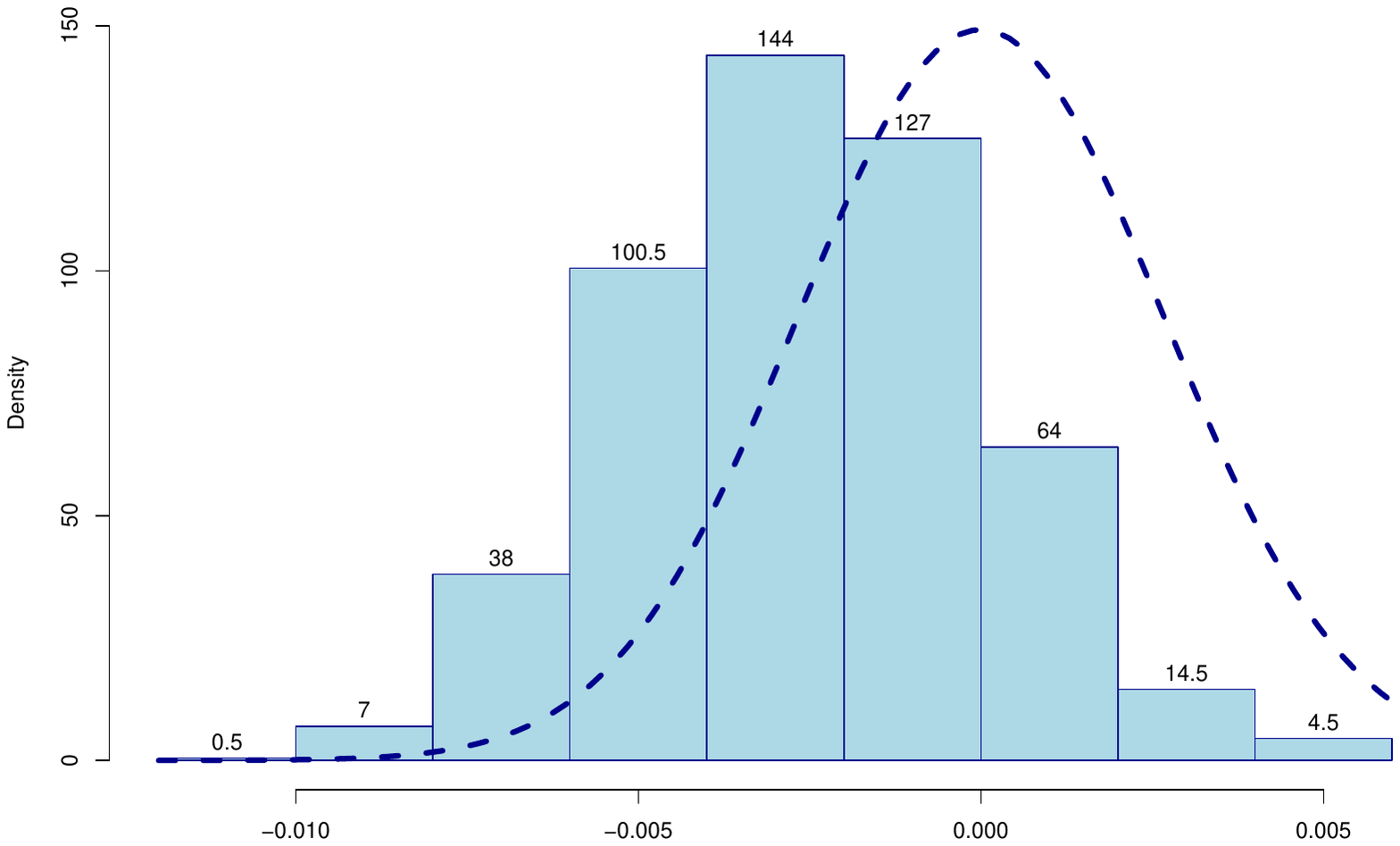}&\includegraphics[scale=0.28]{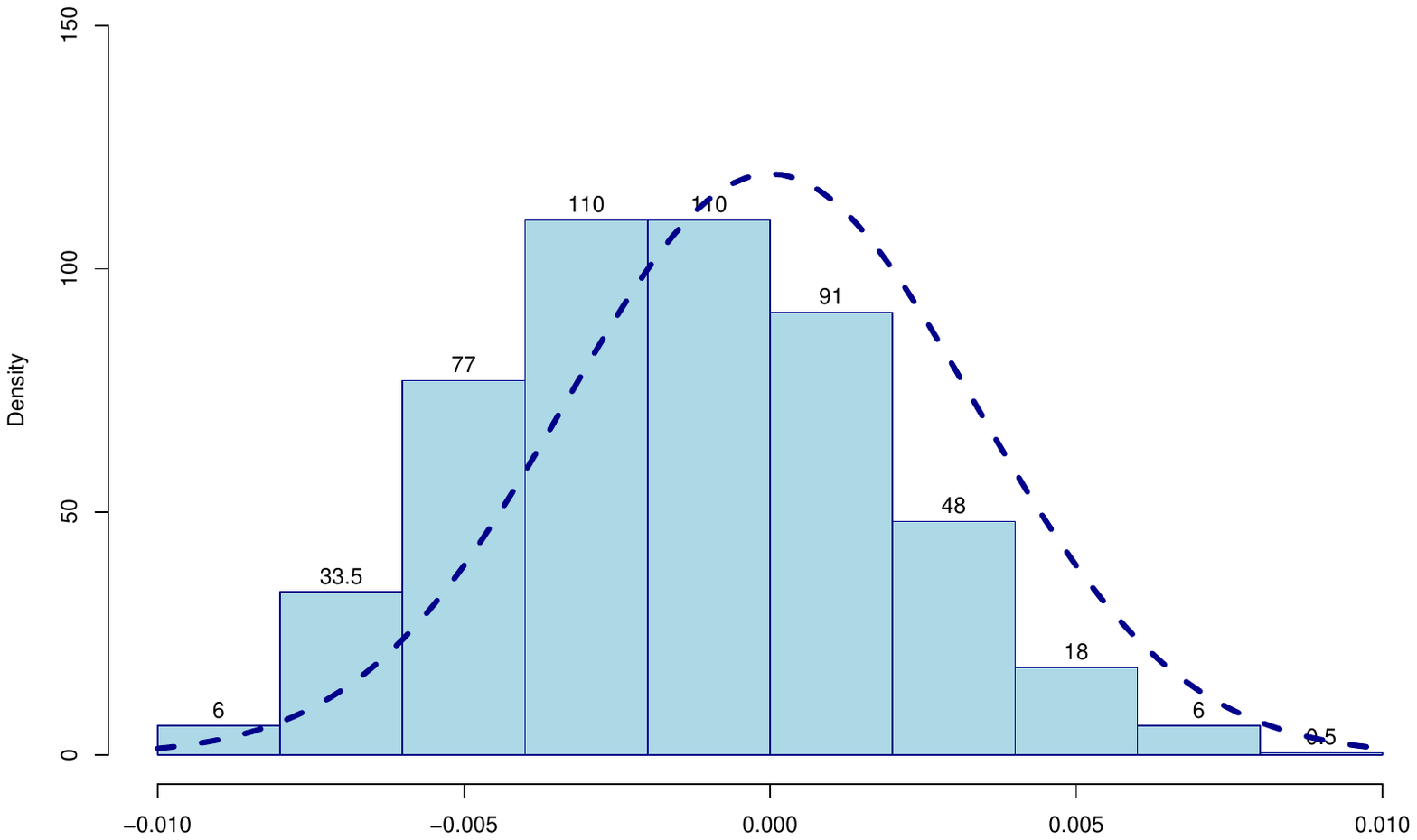}\\
$c=0.5$, CCC-GARCH &$c=0.9$, CCC-GARCH\\
\includegraphics[scale=0.28]{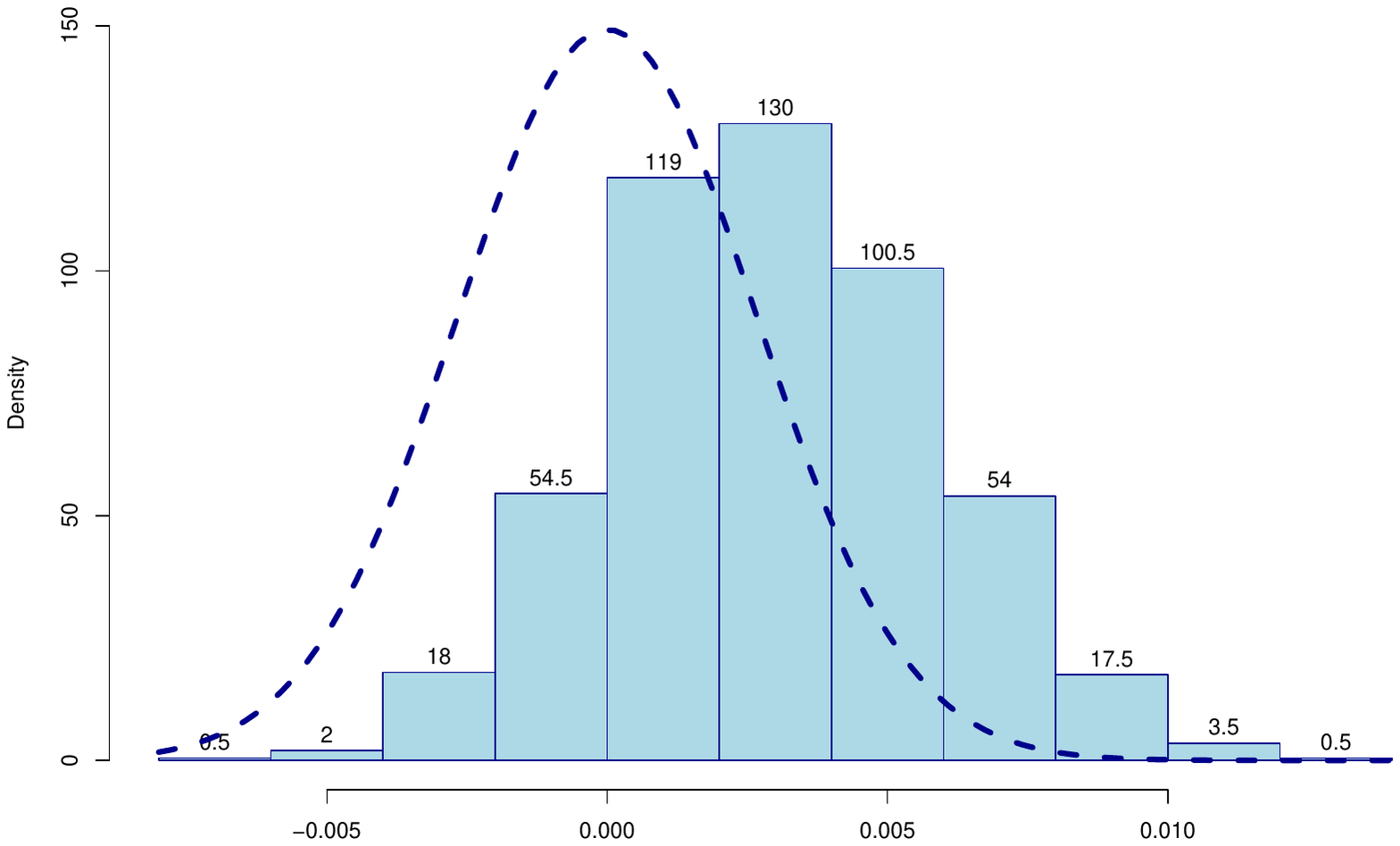}&\includegraphics[scale=0.28]{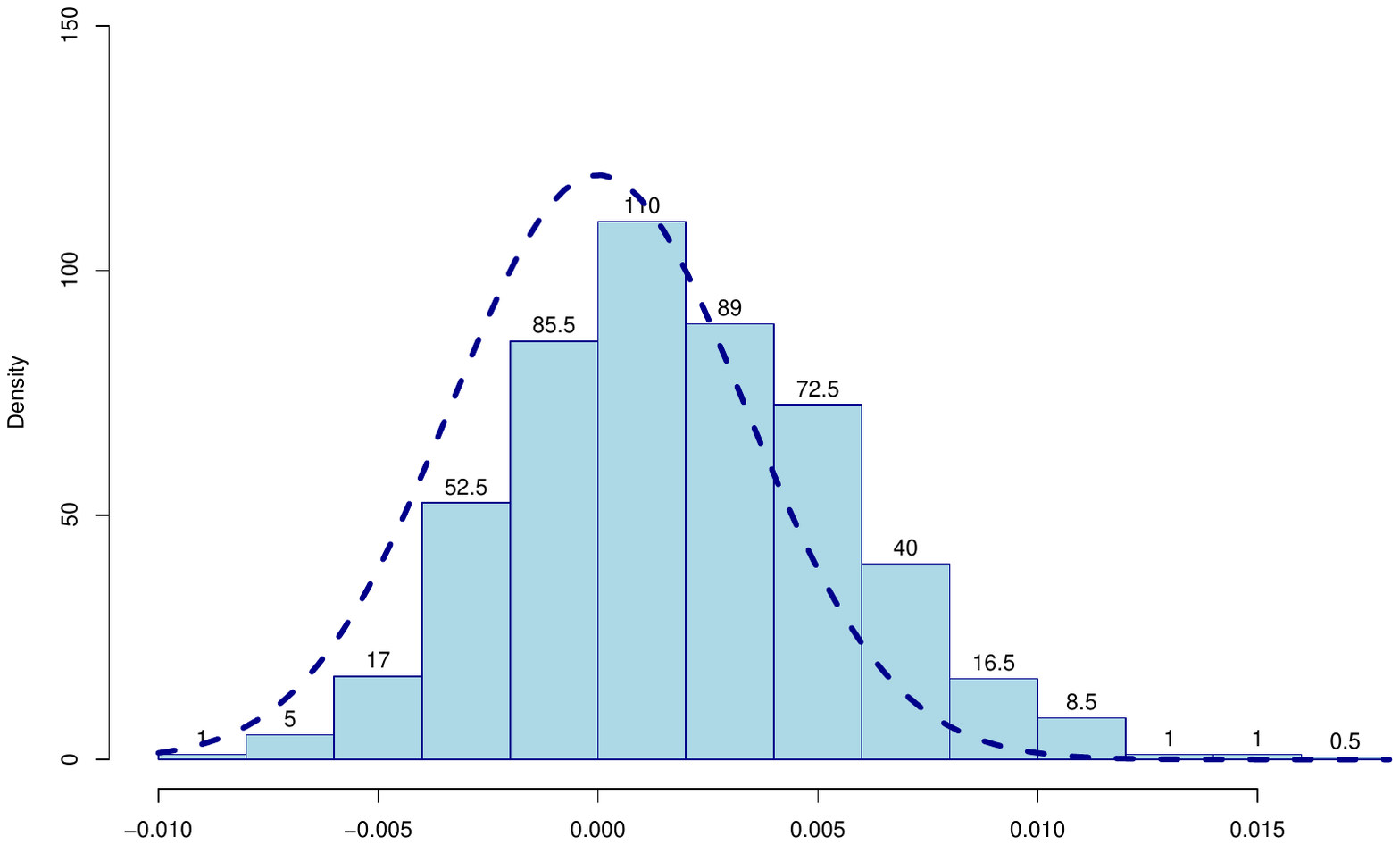}\\
\end{tabular}
\label{sim_hist_V}
\end{figure}

\begin{figure}[h!!]
\caption{\footnotesize Histogram and the asymptotic density (see Theorem \ref{th2}) of $\sqrt{n}(\hat{s}_c-s)$ for the normal
distribution (above), for the $t$-distribution with $3$ degrees of freedom (in the middle), and for the CCC-GARCH process (below). We put $n=1000$.
}
\begin{tabular}{cc}
&\\
$c=0.5$, Normal distribution&$c=0.9$, Normal distribution\\
\includegraphics[scale=0.28]{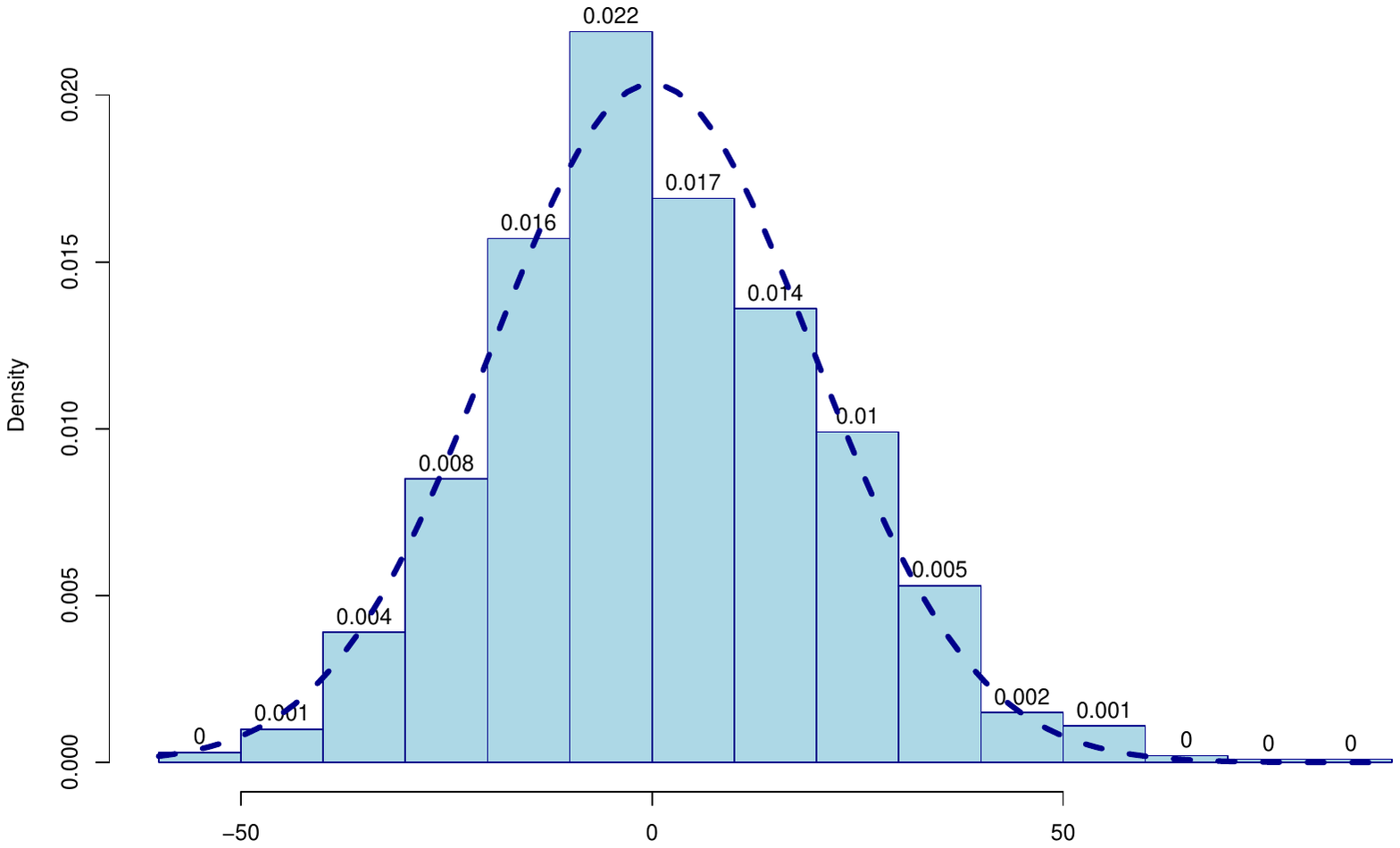}&\includegraphics[scale=0.28]{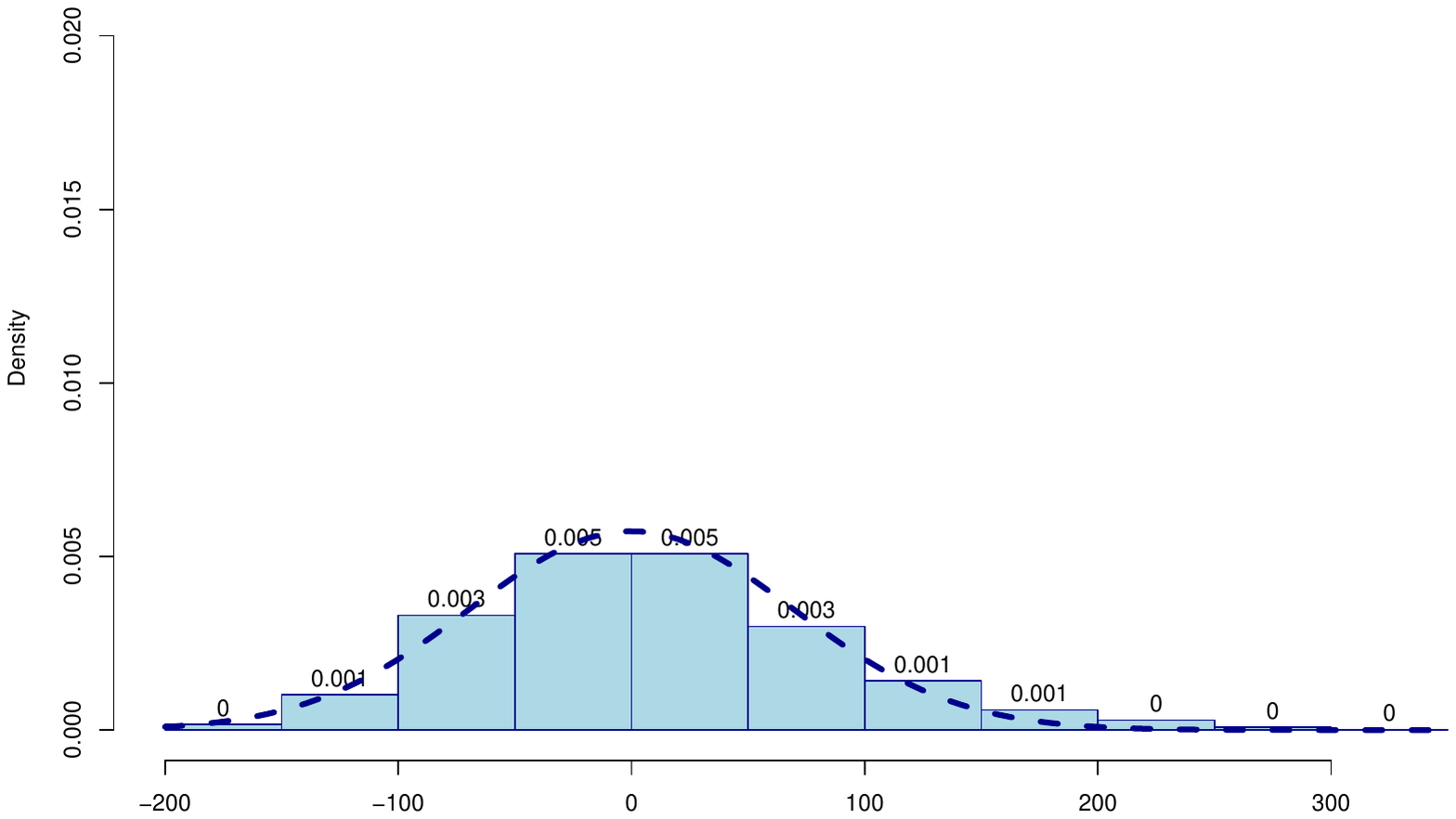}\\
$c=0.5$, $t$ distribution&$c=0.9$, $t$ distribution\\
\includegraphics[scale=0.28]{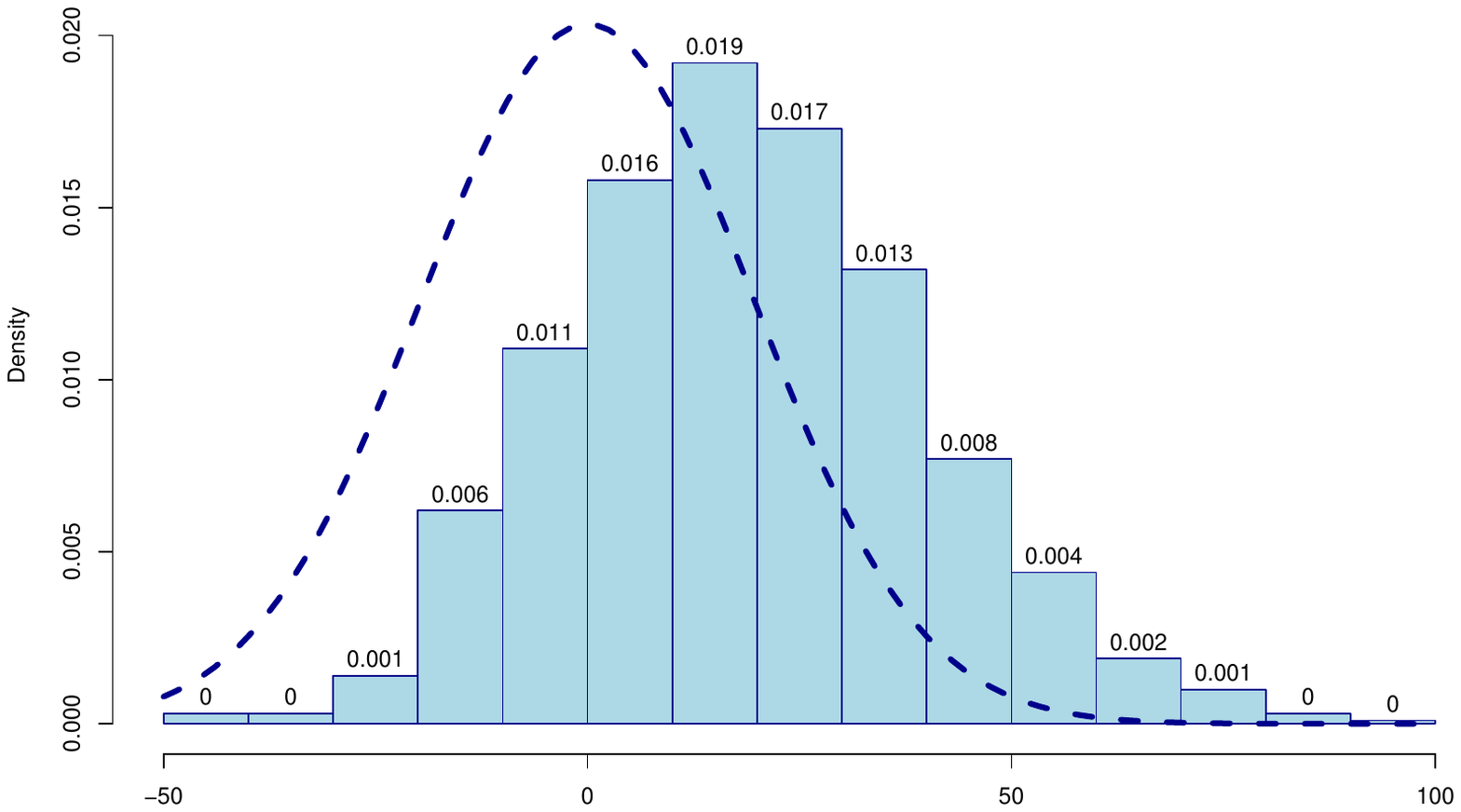}&\includegraphics[scale=0.28]{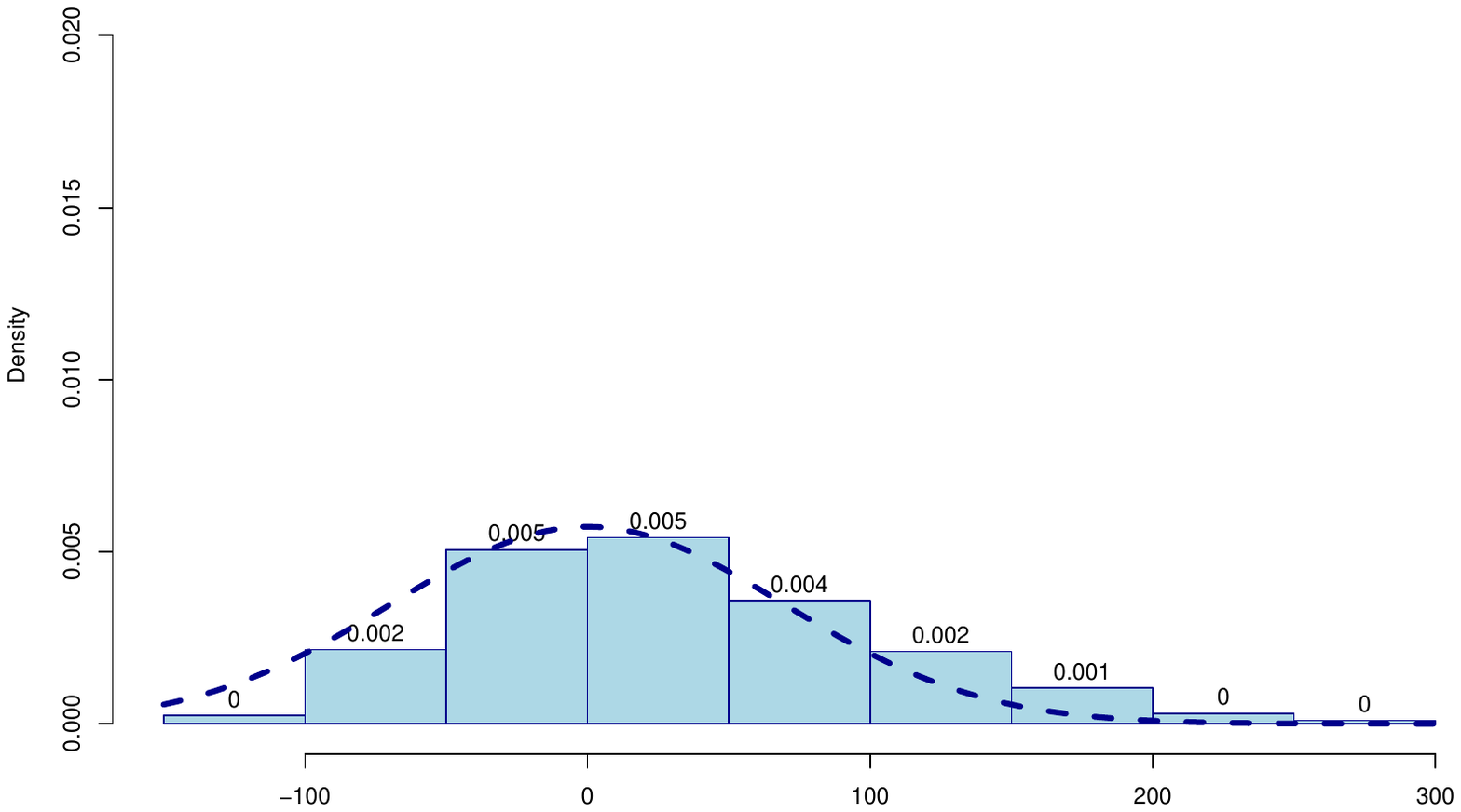}\\
$c=0.5$, CCC-GARCH &$c=0.9$, CCC-GARCH\\
\includegraphics[scale=0.28]{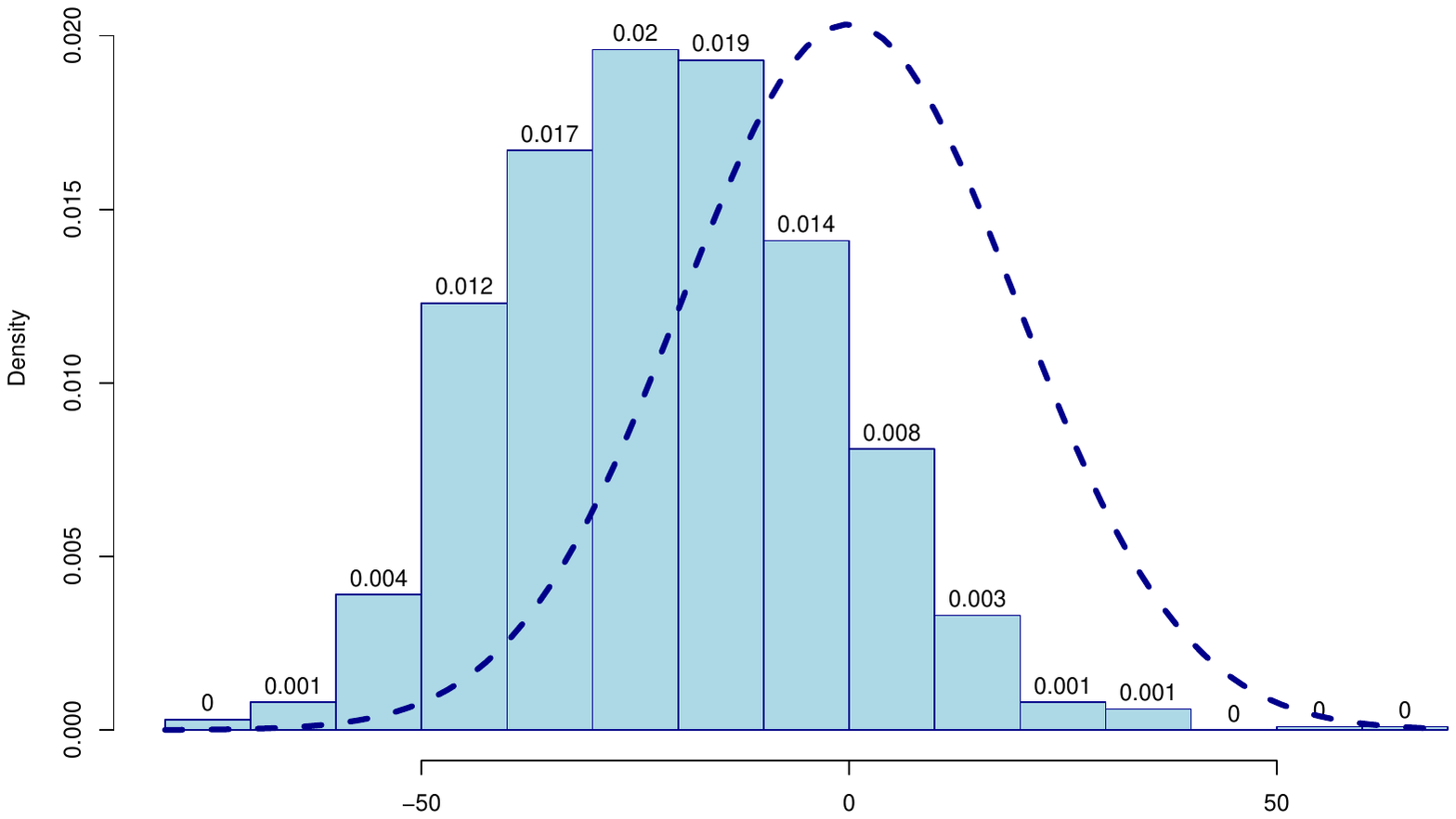}&\includegraphics[scale=0.28]{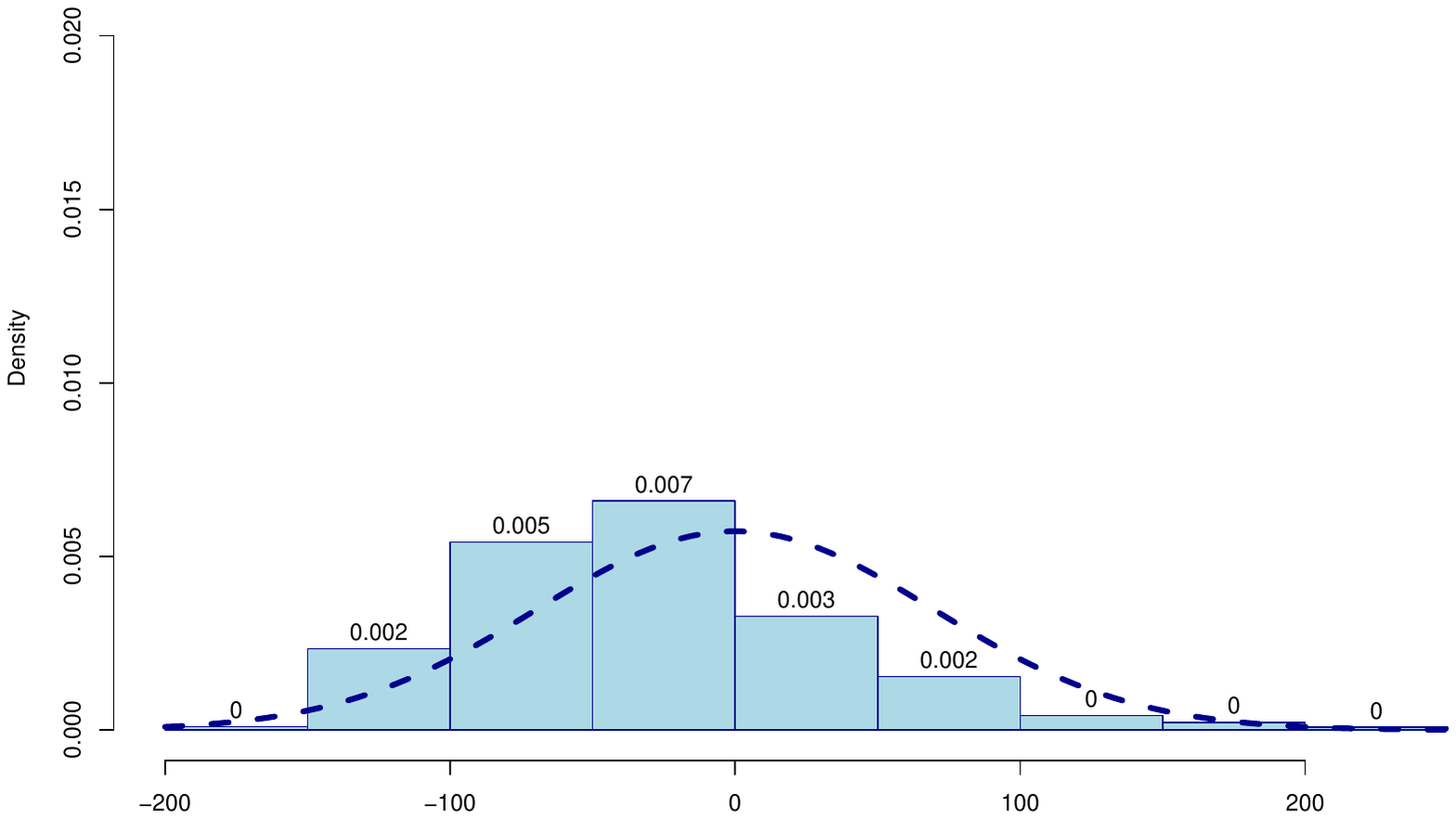}\\
\end{tabular}
\label{sim_hist_s}
\end{figure}

In Figures \ref{sim_hist_R}-\ref{sim_hist_s} we plot the histograms of $\sqrt{n}(\hat{R}_c-R_{GMV})$, $\sqrt{n}(\hat{V}_c-V_{GMV})$, and $\sqrt{n}(\hat{s}_c-s)$ as well as their asymptotic densities presented in Theorem \ref{th2}. Although the results of Theorem \ref{th2} are derived under the assumption of normality, in Figures \ref{sim_hist_R}-\ref{sim_hist_s} we also include the results for the $t$-distribution and the CCC-GARCH process. In Figure \ref{sim_hist_R}, we observe that the densities of the normal distribution provide very good approximations of the corresponding histograms independently of the considered scenario. Furthermore, as expected, the variance in the case of $c=0.9$ is much larger than in the case of $c=0.5$. In contrast to the findings of Figure \ref{sim_hist_R}, a good approximation by the normal distribution is present only in the case of the normal distribution. In the case of $\sqrt{n}(\hat{V}_c-V_{GMV})$, the histograms are shifted slightly to the left for the $t$-distribution and to the right for the CCC-GARCH model (see Figure \ref{sim_hist_s}). Similar results are also present in Figure \ref{sim_hist_s} for $\sqrt{n}(\hat{s}_c-s)$. In particular, a good approximation is observed in the case of a normal distribution, whereas the histogram is shifted to the right in the case of the $t$-distribution and to the left in the case of the CCC-GARCH process.

To summarize the simulation results, we observe fast convergence of the proposed estimators for the three parameters of the efficient frontier in the average quadratic loss. The speed of the convergence is controlled by the existence of the moments of the distribution, i.e., the heavier are its tails the slower is the convergence rate. It is noted that the estimators $\hat{R}_c$ and $\hat{s}_c$ can still be improved by applying a shrinkage estimator for the mean vector. A suitable shrinkage can significantly reduce the noise of the considered estimators for the parameters of the efficient frontier (see, e.g., \cite{Kan2008}).

\subsection{Estimation of the efficient frontier}
To further investigate the performance of the obtained estimators, we compare them with  state-of-art estimators used for the estimation of the inverse covariance matrix, namely the scaled sample estimator (SSE) considered in \cite{Stein1975}, \cite{Mestre2006}, and \cite{Kubokawa2008}, the empirical Bayes estimator of \cite{Efron1976}, and the ridge-type estimator suggested by \cite{Kubokawa2008}.

The scaled sample estimator is given by
\begin{equation}
\widehat{\boldsymbol{\Pi}}_{SSE} = \dfrac{n-p-2}{n-1}\bS_n^{-1},
\end{equation}
whereas the empirical Bayes estimator is obtained by
\begin{equation}
\widehat{\boldsymbol{\Pi}}_{EBE} = \dfrac{n-p-2}{n-1}\bS_n^{-1} + \dfrac{p^2+p-2}{(n-1)\text{tr}(\bS_n)}\bI_p\,.
\end{equation}
Finally, the ridge-type of \cite{Kubokawa2008} is computed as
\begin{equation}
 \widehat{\boldsymbol{\Pi}}_{RTE} = p\left((n-1)\bS_n+\text{tr}(\bS_n)\bI_p\right)^{-1}\,,
\end{equation}
where $\bI_p$ is the $p$-dimensional identity matrix. The estimator of the mean vector is left unchanged and corresponds to the sample. Using the estimators $\widehat{\boldsymbol{\Pi}}_{SSE}$ ($\widehat{\boldsymbol{\Pi}}_{EBE}$ or $\widehat{\boldsymbol{\Pi}}_{RTE}$) and $\bby_n$ in (\ref{rgmv}), (\ref{vgmv}) and (\ref{slope}) instead of $\bSigma_n^{-1}$ and $\bm_n$, respectively, we obtain the scaled sample, the empirical Bayes, and the ridge-type estimators of the parameters of the efficient frontier.

\begin{figure}[h!!]
\caption{\footnotesize Estimated efficient frontier for the normal
distribution (above), for the $t$-distribution with $3$ degrees of freedom (in the middle), and for the CCC-GARCH process (below). We put $n=100$.
}
\begin{tabular}{cc}
&\\
$c=0.5$, Normal distribution&$c=0.9$, Normal distribution\\
\includegraphics[scale=0.28]{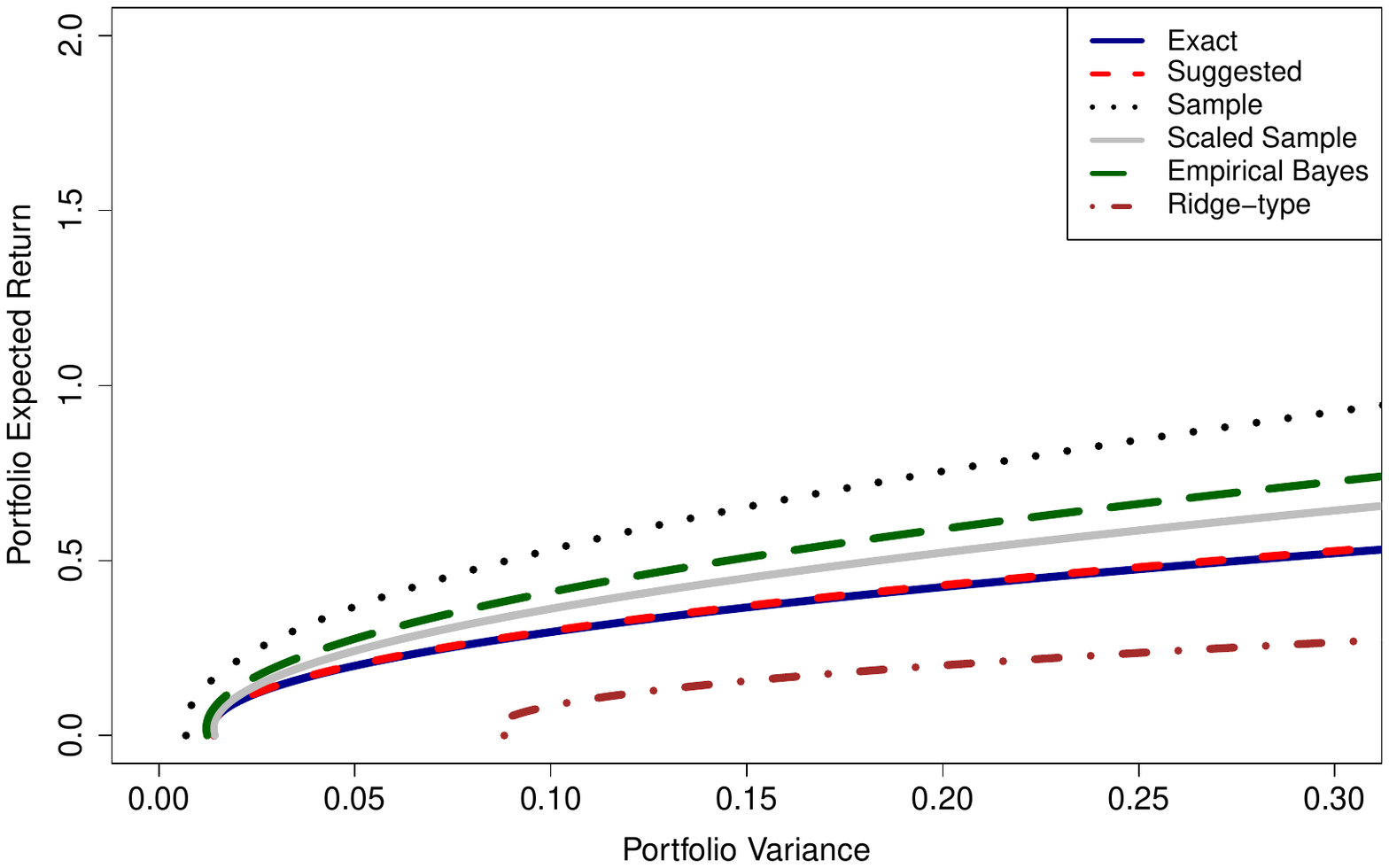}&\includegraphics[scale=0.28]{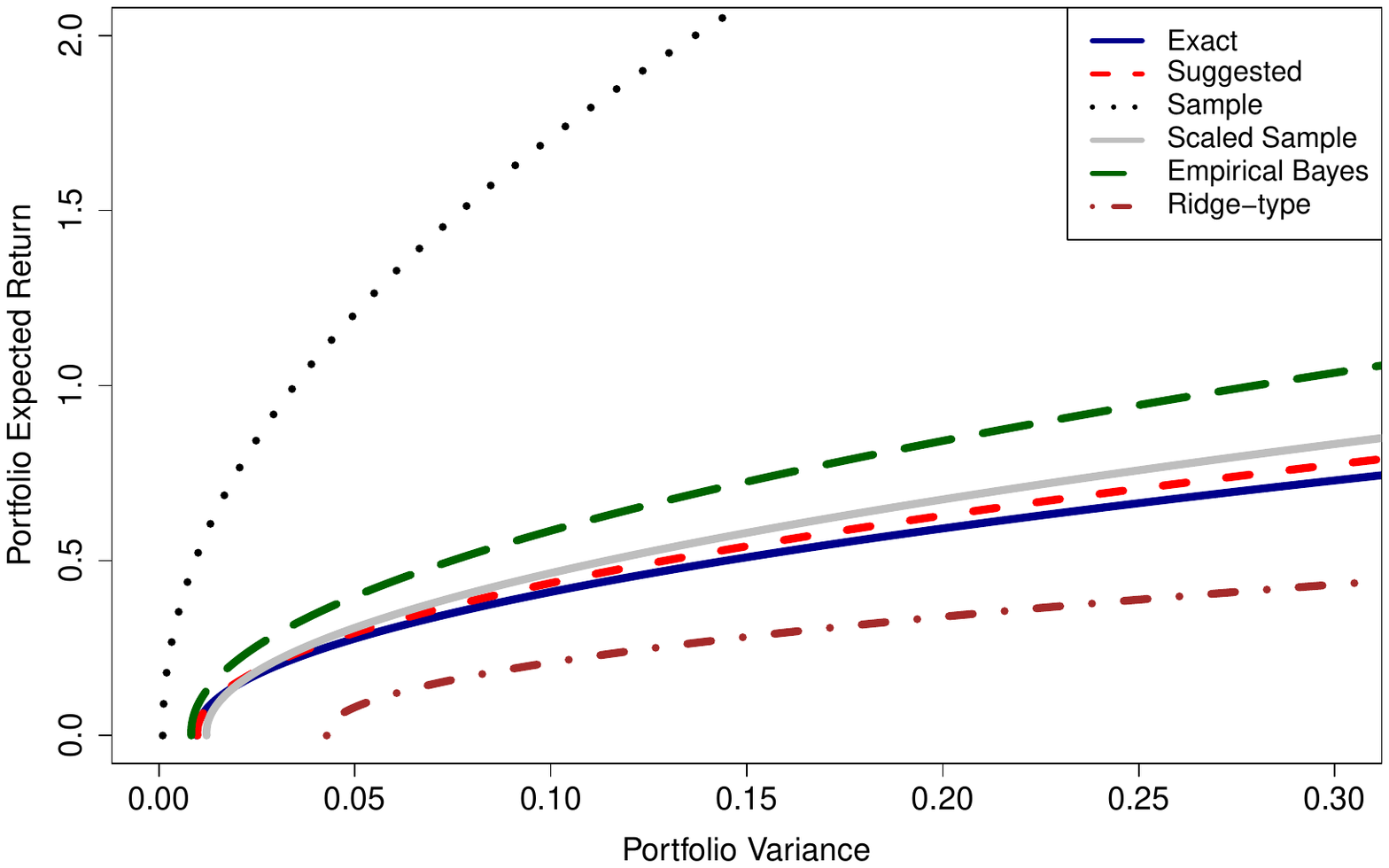}\\
$c=0.5$, $t$ distribution&$c=0.9$, $t$ distribution\\
\includegraphics[scale=0.28]{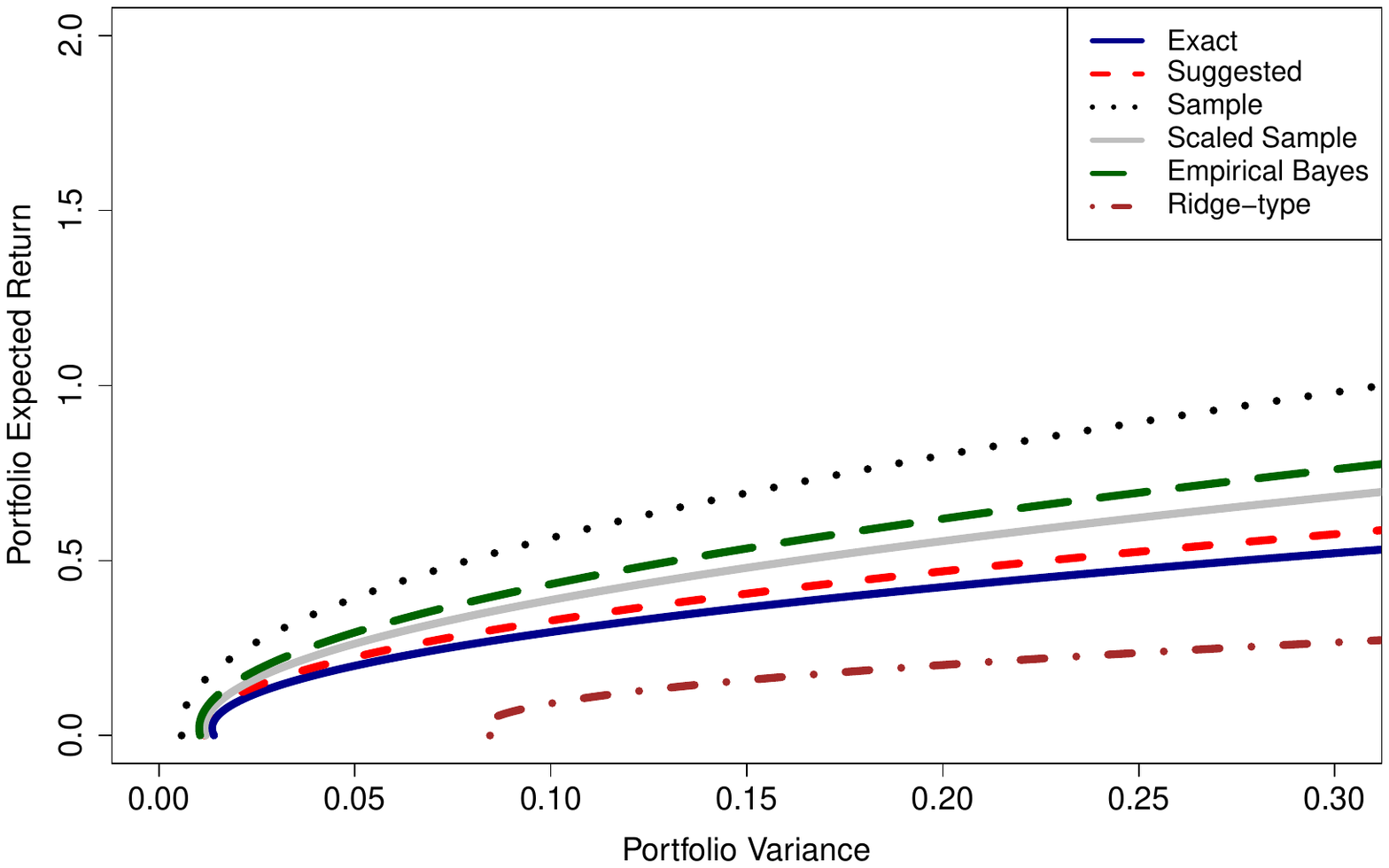}&\includegraphics[scale=0.28]{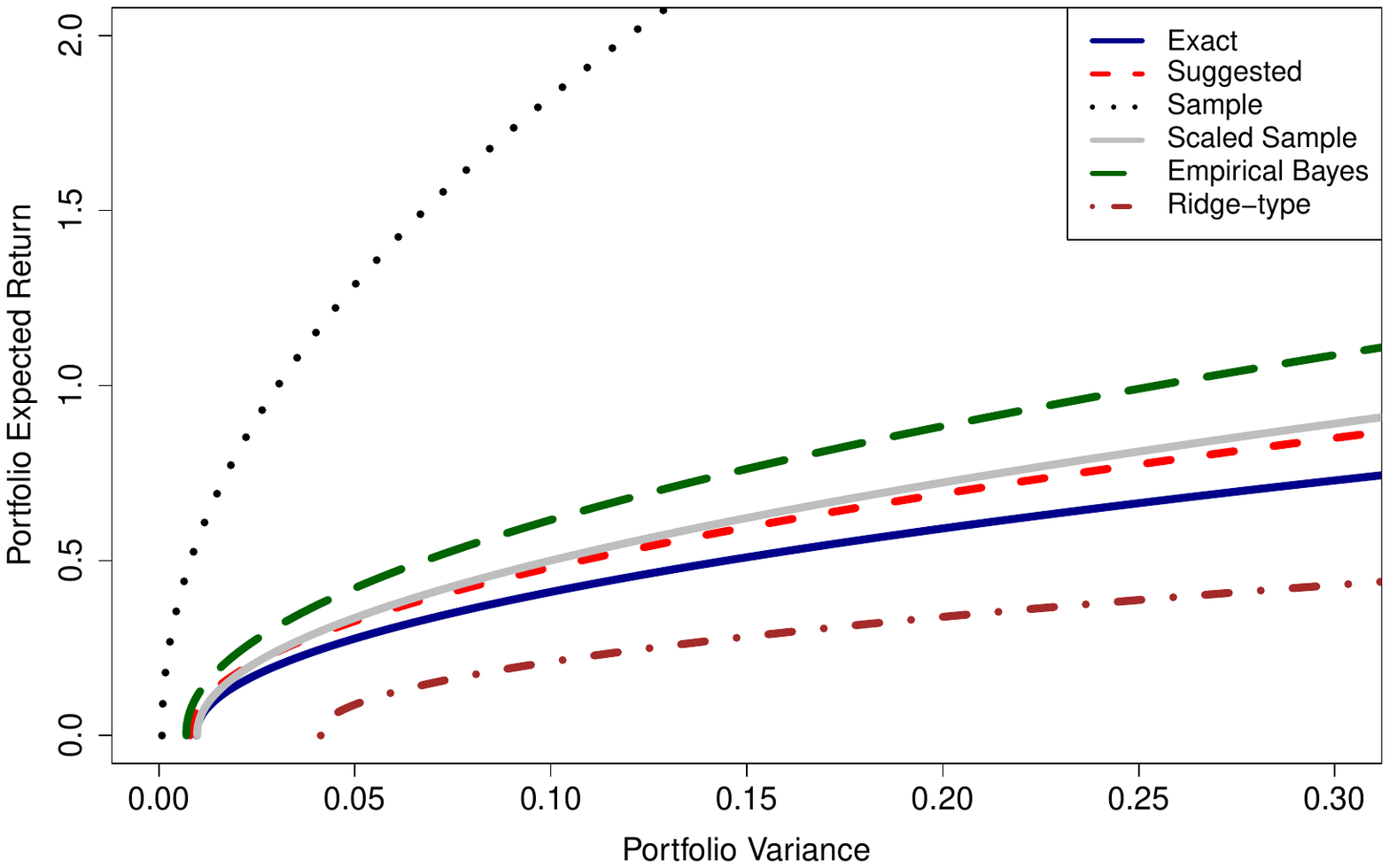}\\
$c=0.5$, CCC-GARCH &$c=0.9$, CCC-GARCH\\
\includegraphics[scale=0.28]{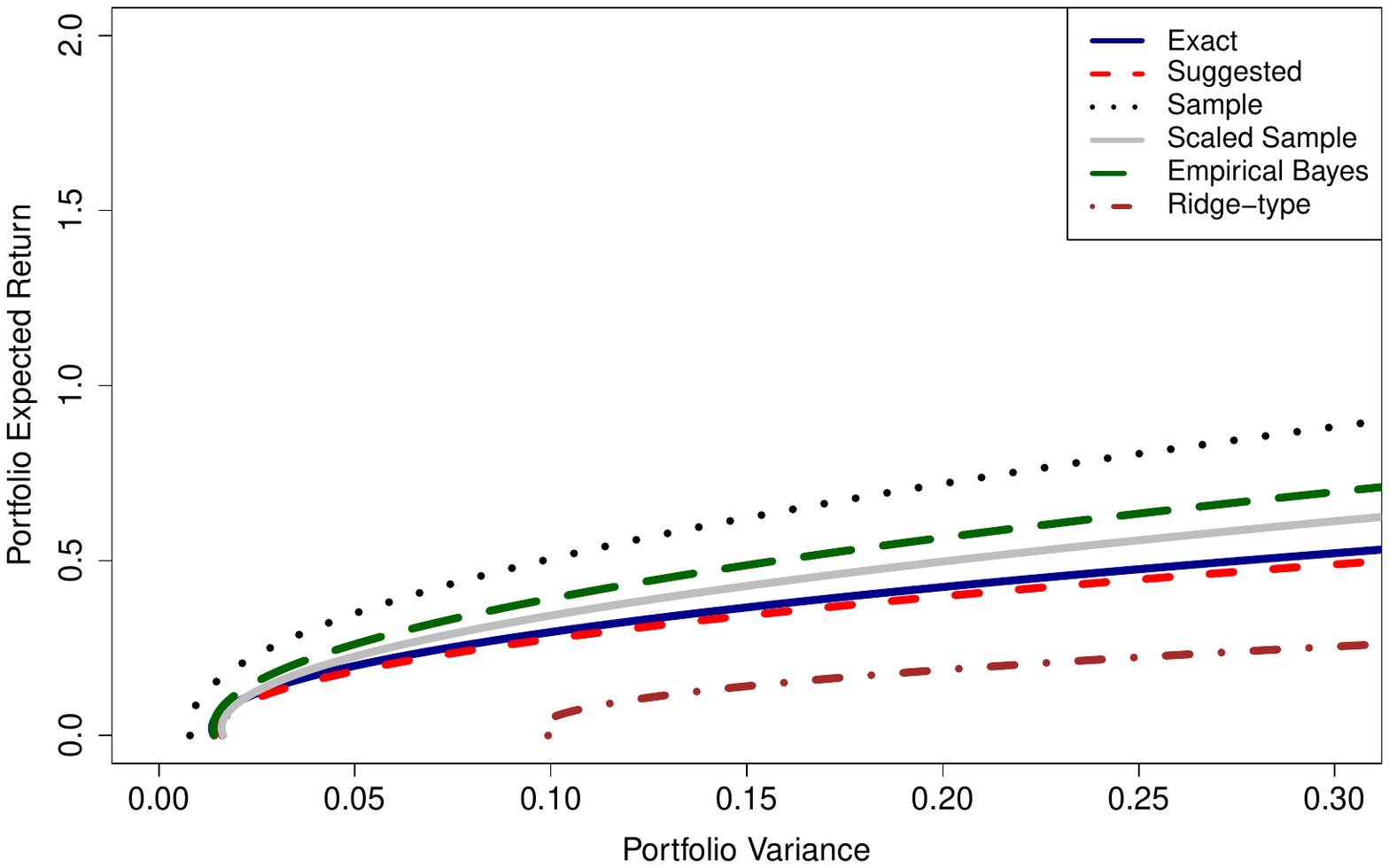}&\includegraphics[scale=0.28]{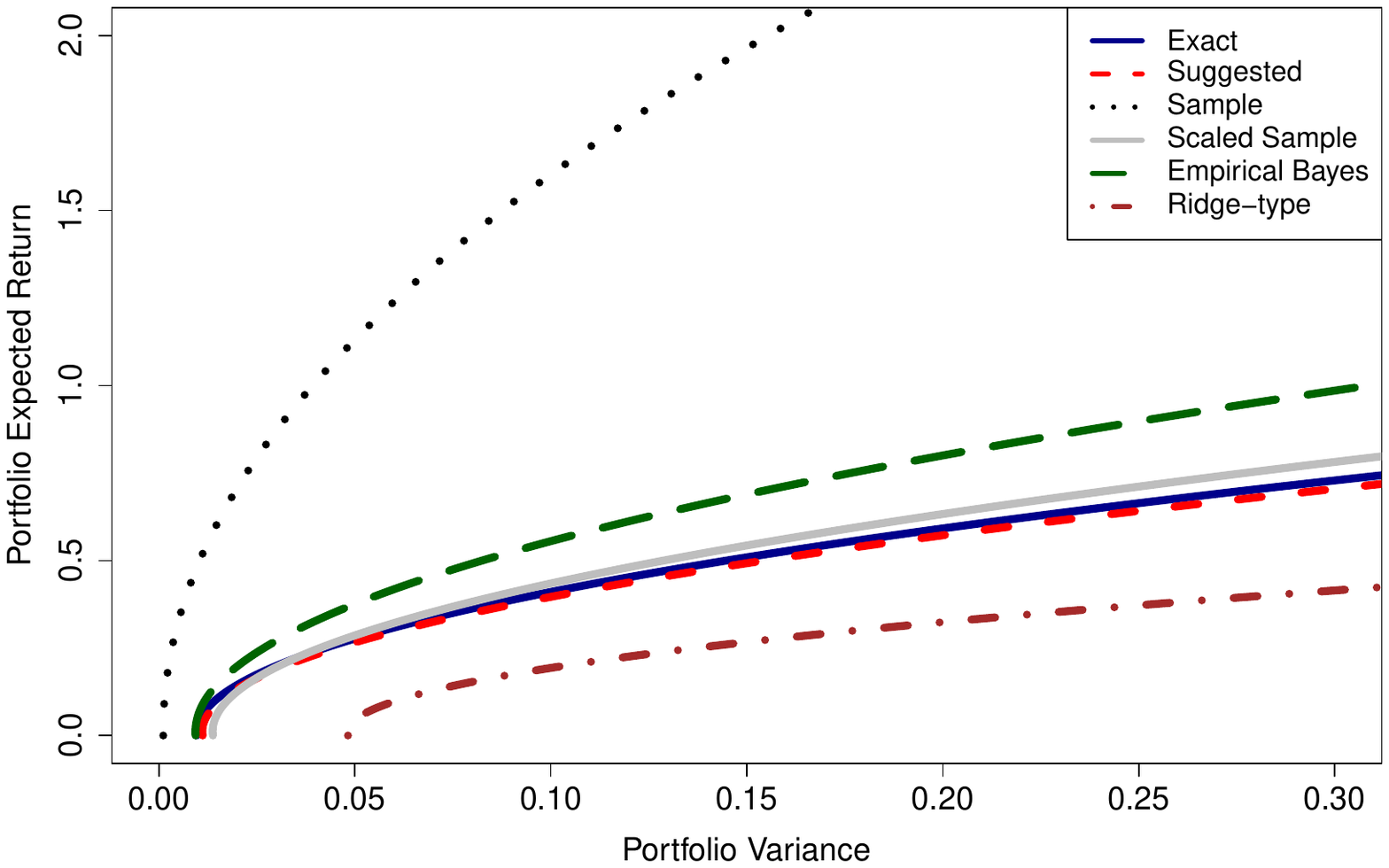}\\
\end{tabular}
\label{sim_ef}
\end{figure}

In Figure \ref{sim_ef}, we consider the behavior of the whole efficient frontier by comparing the proposed estimators  with the three benchmarks presented above. Moreover, we also include the population efficient frontier and its sample estimator. The results are presented for all three scenarios considered in Section 4.1 in case of $c\in\{0.5,0.9\}$ and $n=100$. 

As expected, we observe that the sample estimator performs very poorly in all of the considered cases. It considerably overestimates the location of the true efficient frontier in the mean-variance space. In contrast, the consistent estimator of the efficient frontier shows a very good performance in almost all of the considered cases. It nearly coincides with the population efficient frontier in the case of normally distributed data for both values of $c$. In the case of the $t$-distribution, it slightly overestimates the location of the efficient frontier, while the underestimation is present for the CCC-GARCH process. Remarkably, the deviations from the population efficient frontier are very small in both cases even for $c=0.9$. As second-best option, we rank the scaled sample estimator followed by the empirical Bayes estimator. Finally, the ridge-type estimator underestimates the location of the population efficient frontier in the mean-variance space for all considered cases.

To summarize, we conclude that the proposed estimator for the efficient frontier shows significant improvement over the sample estimator for large dimensional data. In the case of the normal distribution, it is a great alternative to all of the considered benchmarks. Under heavy-tailed data, the proposed estimator is still much better in comparison to the estimators based on the scaled sample estimator, the empirical Bayes estimator and the ridge-type estimator of the inverse covariance matrix.

\section{Empirical Application}

In this section, we apply the theoretical results of the paper to real data. Our goal is to estimate efficient frontiers based on a large cross-section of stocks without relying on data from overly extended periods. Instead, we seek to provide insights into the time-varying nature of efficient frontiers over relatively short intervals. This approach excludes the use of daily data, where estimating efficient frontiers for hundreds of assets would necessitate sample periods spanning several months or years. Consequently, we use one-minute intraday returns of a broad cross-section of S\&P 500 constituents over the period from March 3, 2017, to June 6, 2022. This period is particularly interesting as it encompasses the COVID-19 pandemic. The choice of this time interval highlights the importance of short estimation periods while maintaining a large number of assets. Using daily data at the end of the period would inevitably include observations from before the pandemic, leading to deteriorated estimators of the actual frontier. This further underscores the relevance of high-frequency data and short estimation windows for portfolio selection.

We use a $p=200$-dimensional portfolio constructed based on the most frequently traded assets each day of the sample window. We set $n=375$, corresponding to the number of $1$-minute intervals on a single trading day from 9:45 to 16:00 (leaving out the first and last 15 minutes to avoid market opening and closure effects). To weaken the influence of very large and very small values of asset returns on the estimation procedure, we perform a proper truncation of large (in absolute terms) returns in the considered data sets that is commonly referred to as winsorization. This method is widely used by quantitative portfolio managers \citep[p.180]{Chincarini2006}.

Intraday data at high-frequency, for example, 1-minute data, induce market-microstructure noise. A common practice is to deploy 5-min observations (see \cite{Anderson2001} and \cite{BarndorffNielsen2002}). Using returns at higher frequency causes the difficulty that the underlying data is clearly not i.i.d. due to serial correlation and persistent variances.

Therefore, we alternatively consider $5-$, $10$-, $30$-, and $60$-minute returns, which are shown to be significantly less autocorrelated but are still subject to volatility clustering. The latter effect does not align with the assumptions underlying the proposed estimators. However, our simulation results in Section 3 provide promising evidence that our estimator is sufficiently robust to violations from the underlying assumptions, particularly GARCH-type effects. Moreover, high-frequency data are subject to intraday periodicities challenging the assumption of an underlying deterministic covariance matrix. \cite{Bibinger2014} show that intraday seasonalities in covariances are clearly less pronounced than seasonalities in variances. Since in our high-dimensional portfolio application, the contribution of covariances clearly dominates the role of variances, we suggest that the effects of seasonalities on our estimates should be of minor importance.

To make the estimates based on alternative sampling schemes comparable, we fix the ratio $p/n$ with $p=200$ and $n=375$. This, however, requires utilizing data from neighboring days. In this case, we estimate the parameters using rolling windows, which are moved forward on a day-to-day basis. Hence, efficient frontiers based on $5$ minutes utilize data from rolling $5$-day windows before the corresponding day. The resulting estimated efficient frontier can be seen as the 5-min frontier averaged over the past $5$ trading days. Accordingly, using $10$-, $30$-, and $60$-minute returns result in estimates averaged over periods of $10$-, $30$-, and $60$ trading days. To make the resulting efficient frontiers comparable, the underlying returns are scaled to 60-min holding periods corresponding to the lowest sampling frequency in the given context. The efficient frontier parameters $R_{GMV}$, $V_{GMV}$, and $s$ are alternatively estimated using the sample estimator, the proposed consistent estimator, the empirical Bayes estimator, and the ridge-type estimator.

\begin{figure}
  \begin{tabular}{cc}
    \includegraphics[width=0.5\textwidth]{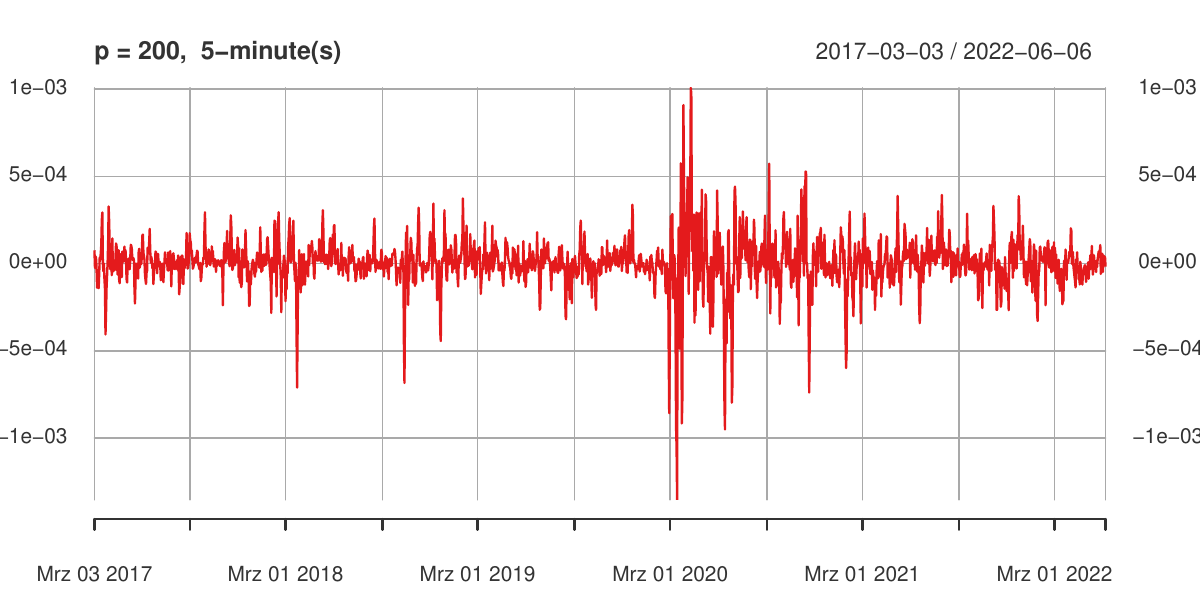}&\includegraphics[width=0.5\textwidth]{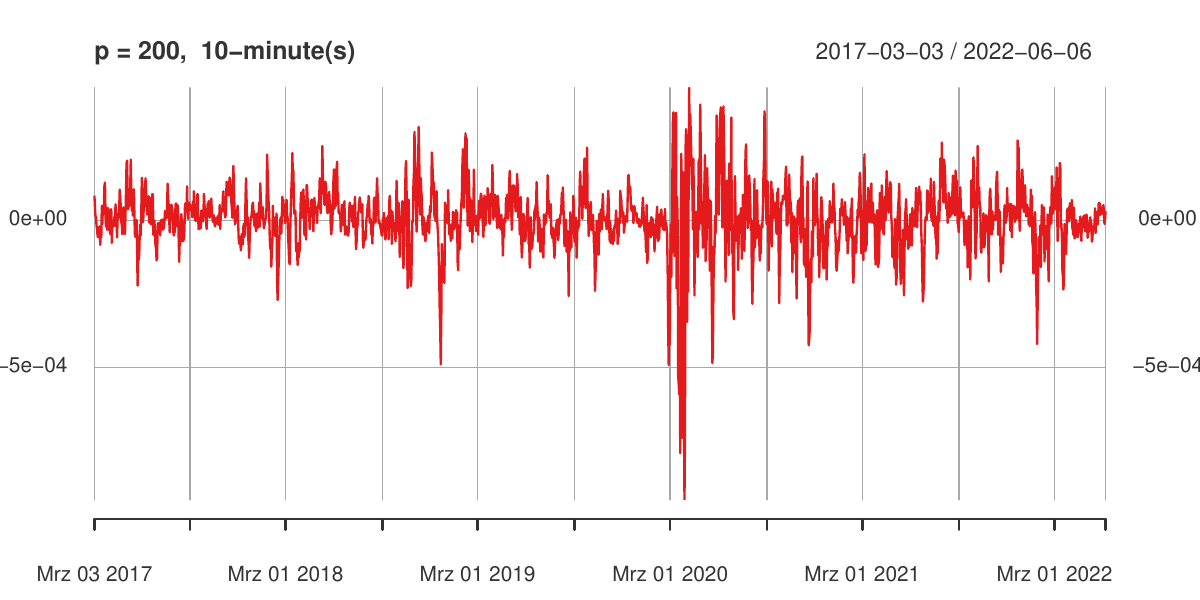}\\
    \includegraphics[width=0.5\textwidth]{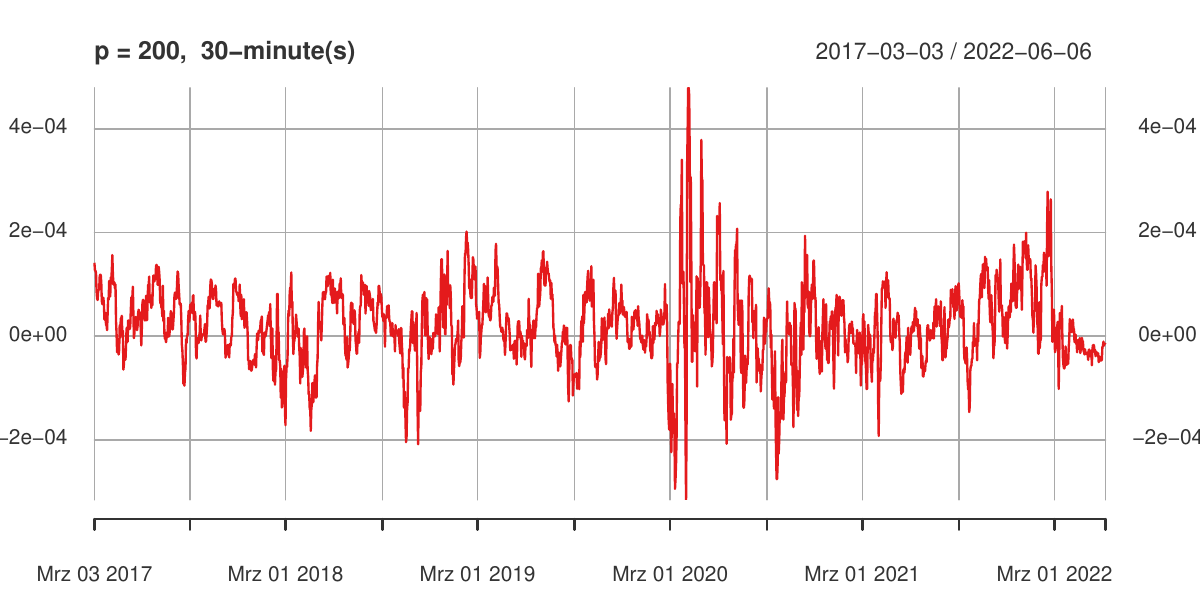}&\includegraphics[width=0.5\textwidth]{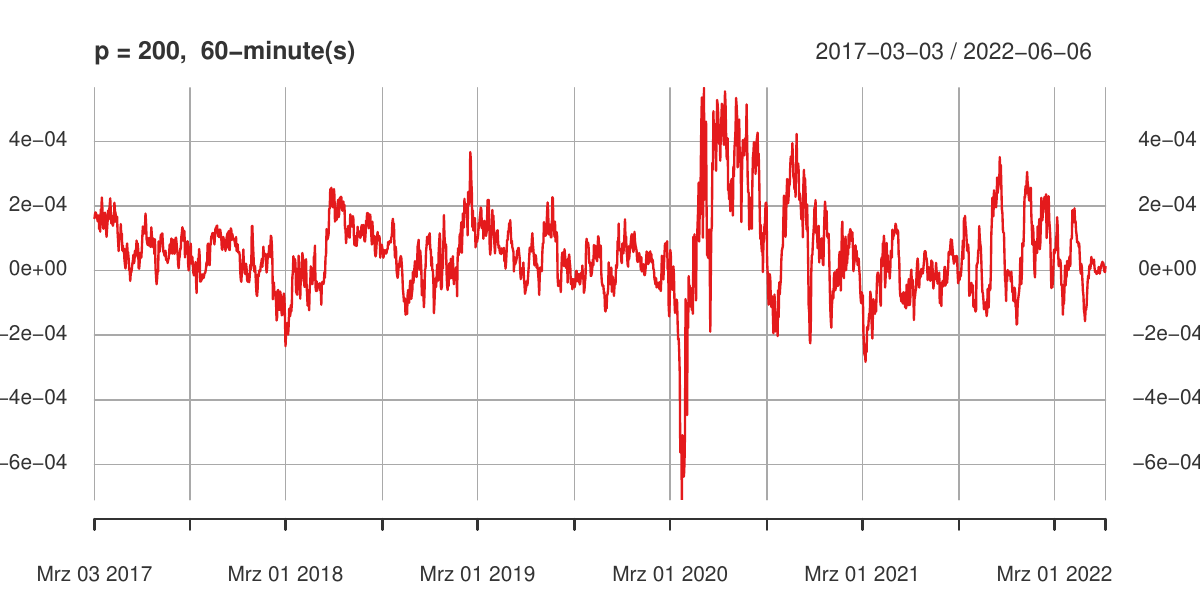}\\
  \end{tabular}
\caption{\footnotesize Estimated expected return of the global minimum variance portfolio based on $5$-, $10$-,  $30$-, and $60$-minute returns on first 200 stocks from S\&P500 from 03 March 2017 to 6 June 2022.
}
\label{emp_R}
  \end{figure}

\begin{figure}
  \begin{tabular}{cc}
    \includegraphics[width=0.5\textwidth]{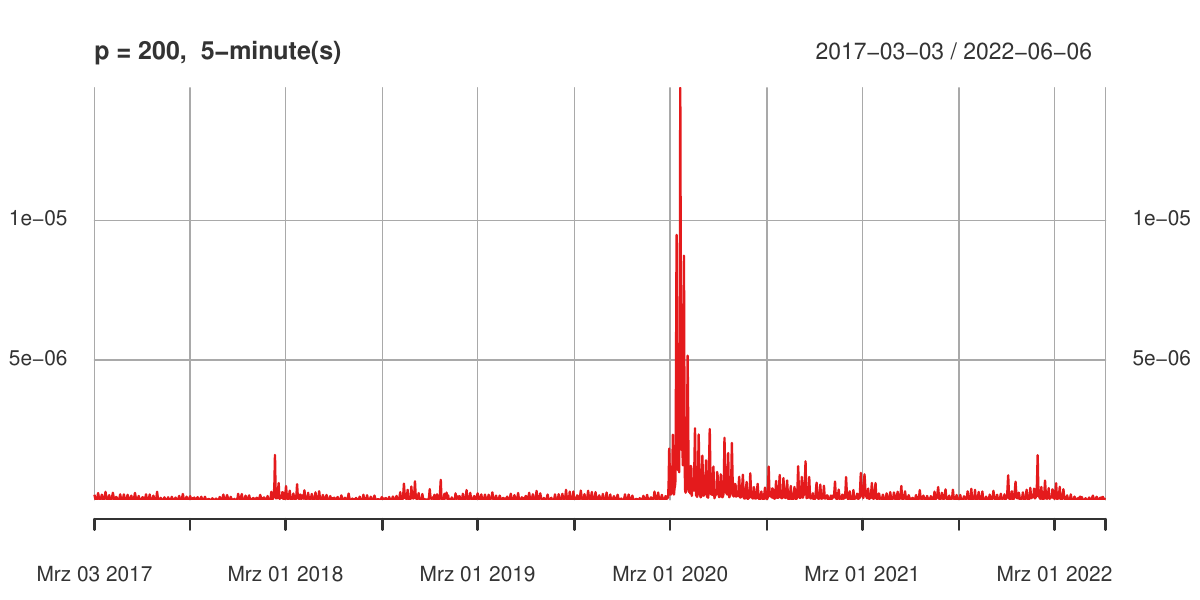} &\includegraphics[width=0.5\textwidth]{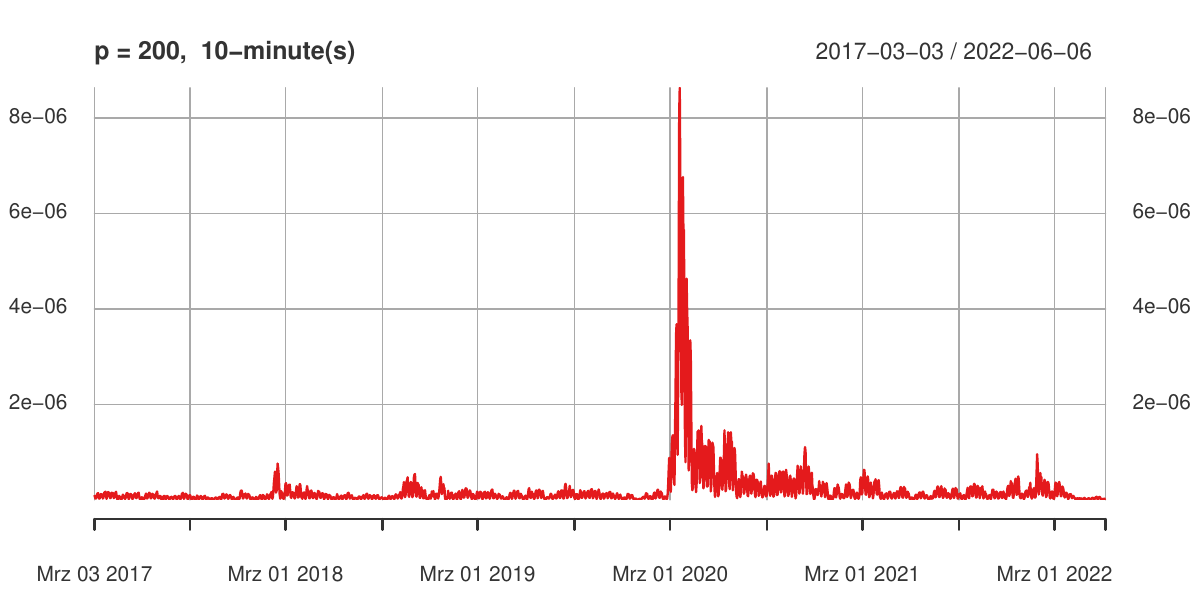}\\
    \includegraphics[width=0.5\textwidth]{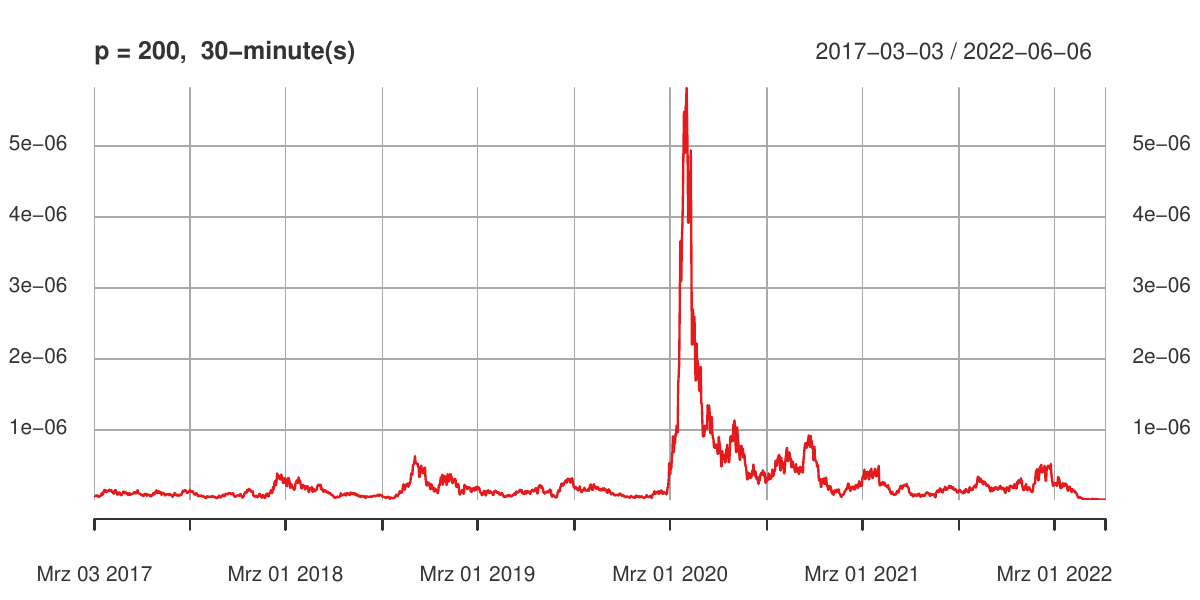}&\includegraphics[width=0.5\textwidth]{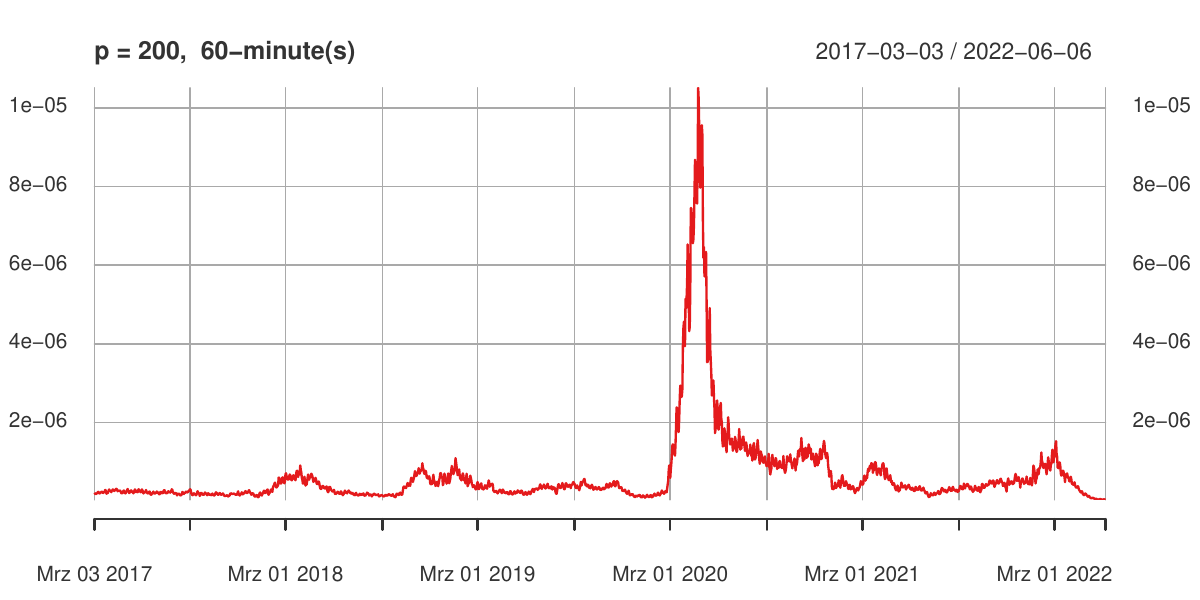}\\
  \end{tabular}
  \caption{\footnotesize Estimated variance of the global minimum variance portfolio based on $5$-, $10$-,  $30$-, and $60$-minute returns on first 200 stocks from S\&P500 from 03 March 2017 to 6 June 2022.}
  \label{emp_V}
  \end{figure}

\begin{figure}
  \begin{tabular}{cc}
    \includegraphics[width=0.5\textwidth]{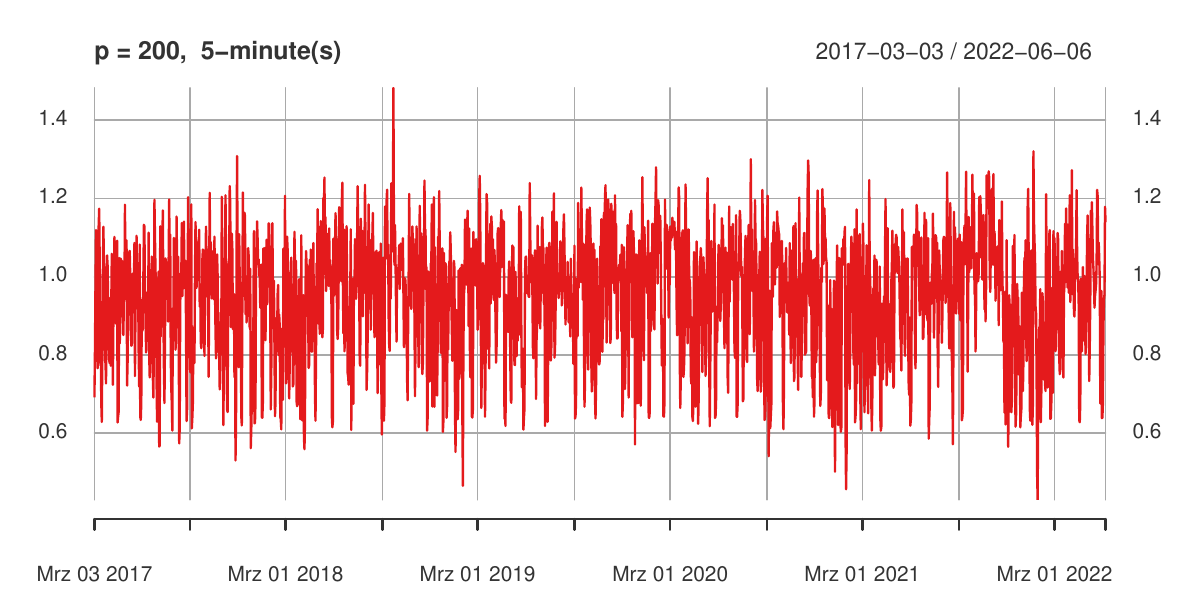} &\includegraphics[width=0.5\textwidth]{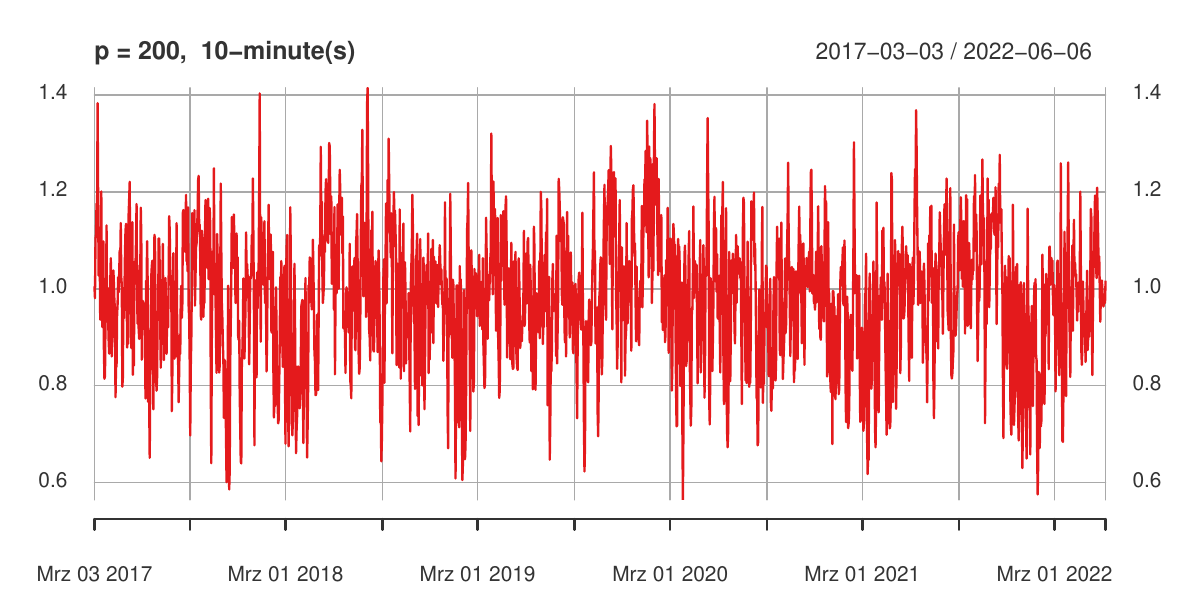}\\
    \includegraphics[width=0.5\textwidth]{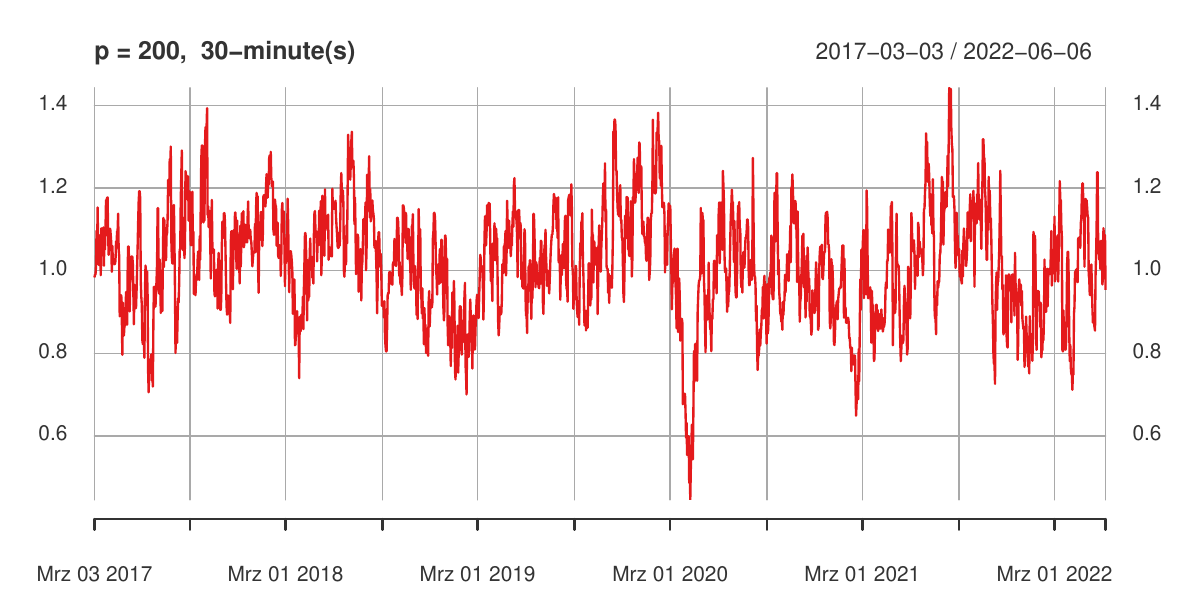}&\includegraphics[width=0.5\textwidth]{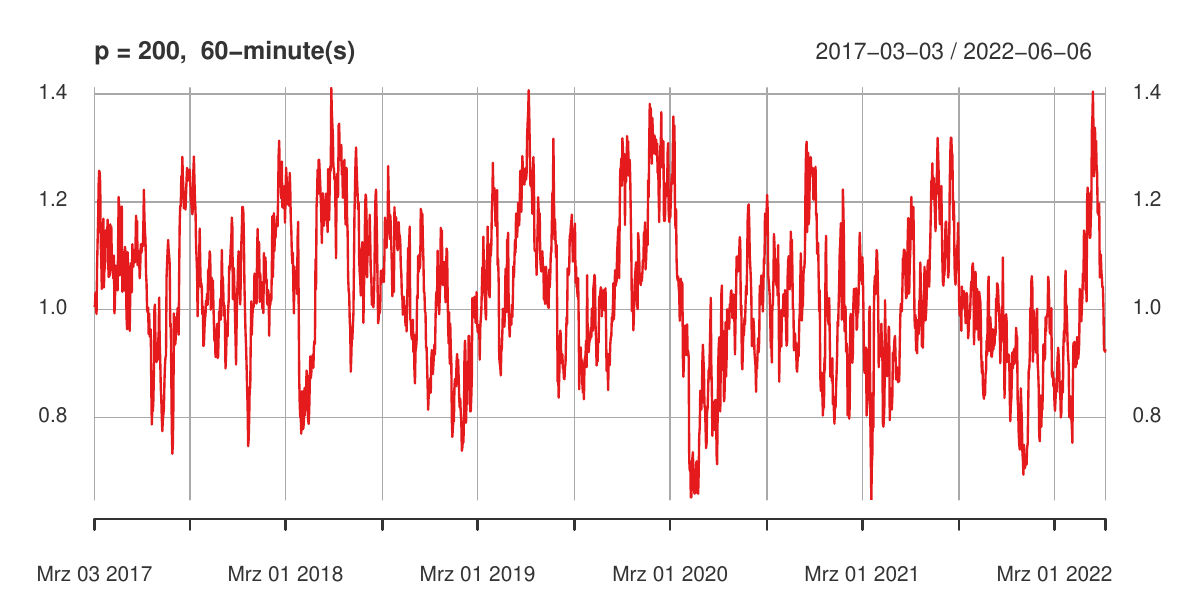}\\
  \end{tabular}
  \caption{\footnotesize Estimated slope parameter of the efficient frontier based on  $5$-, $10$-,  $30$-, and $60$-minute returns on first 200 stocks from S\&P500 from 03 March 2017 to 6 June 2022. }
  \label{emp_s}
  \end{figure}

Figures \ref{emp_R}, \ref{emp_V}, and \ref{emp_s} display the estimates based on our proposed estimator of the expected return, the variance, and the slope of the global minimum variance portfolio for different frequencies. In both figures, we observe more volatile behavior when the returns of higher frequency are used. Remarkably, the parameters of the efficient frontier react to the financial crisis in different ways. While a large increase in the variance of the GMV portfolio is observed, the changes in the other two parameters are not large, especially in the case of the slope parameter of the efficient frontier whose behavior shows only minor changes during this period. Important is the fact that the slope parameter decreases during the crisis. In general, however, the level and pattern of all characteristics are similar for different data frequencies but with more variability for more frequent returns. 

\begin{figure}
  \begin{tabular}{cc}
    \includegraphics[width=0.5\textwidth]{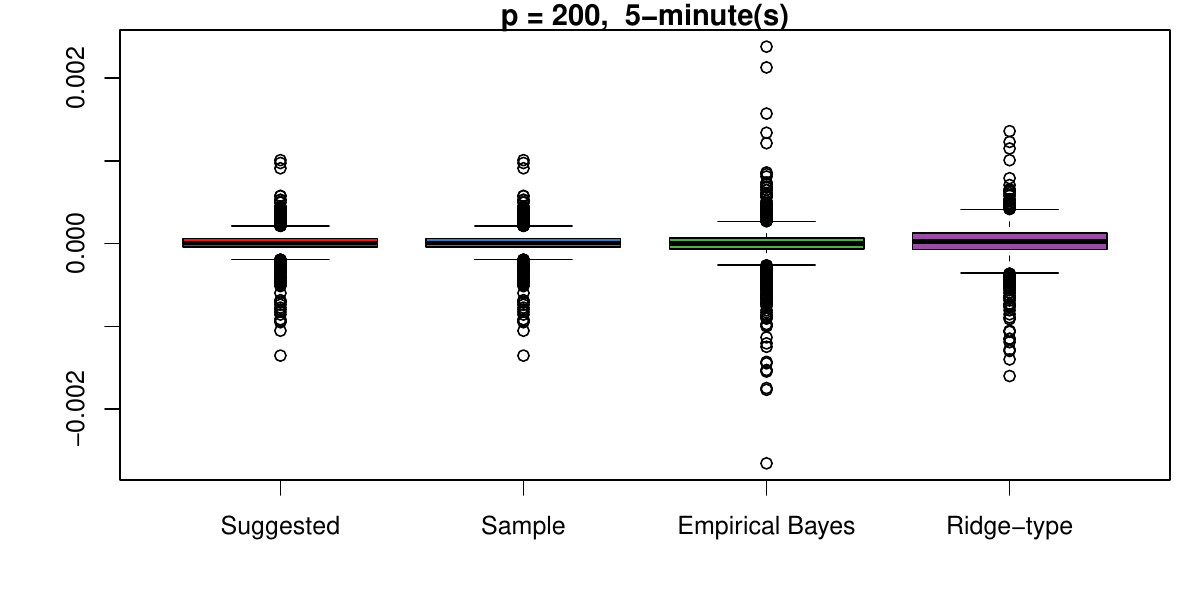} &\includegraphics[width=0.5\textwidth]{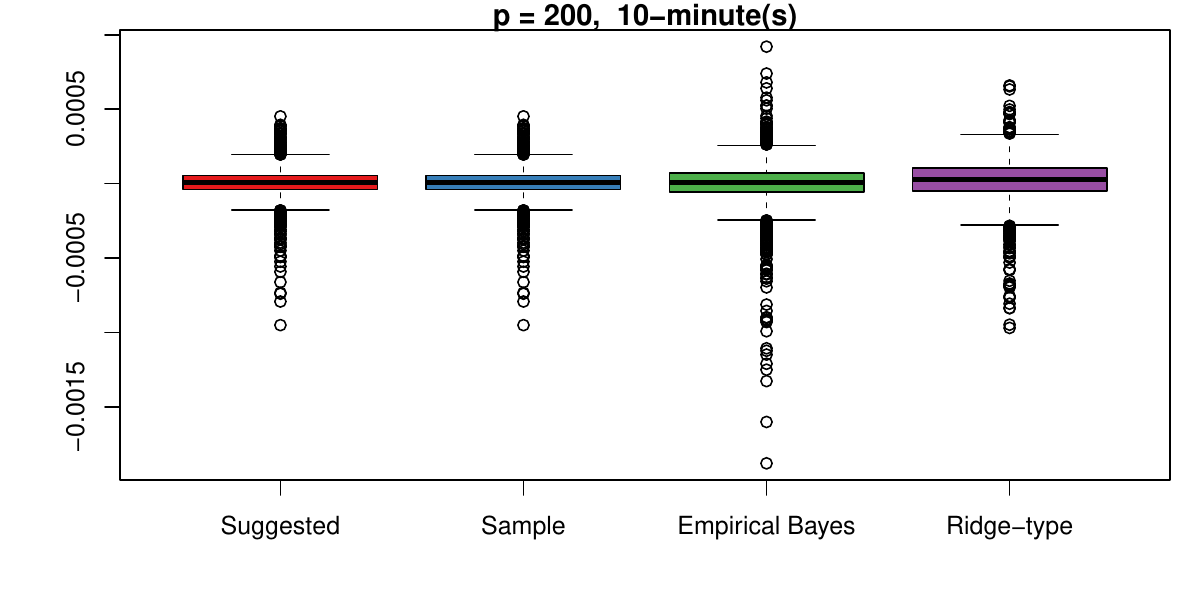}\\
    \includegraphics[width=0.5\textwidth]{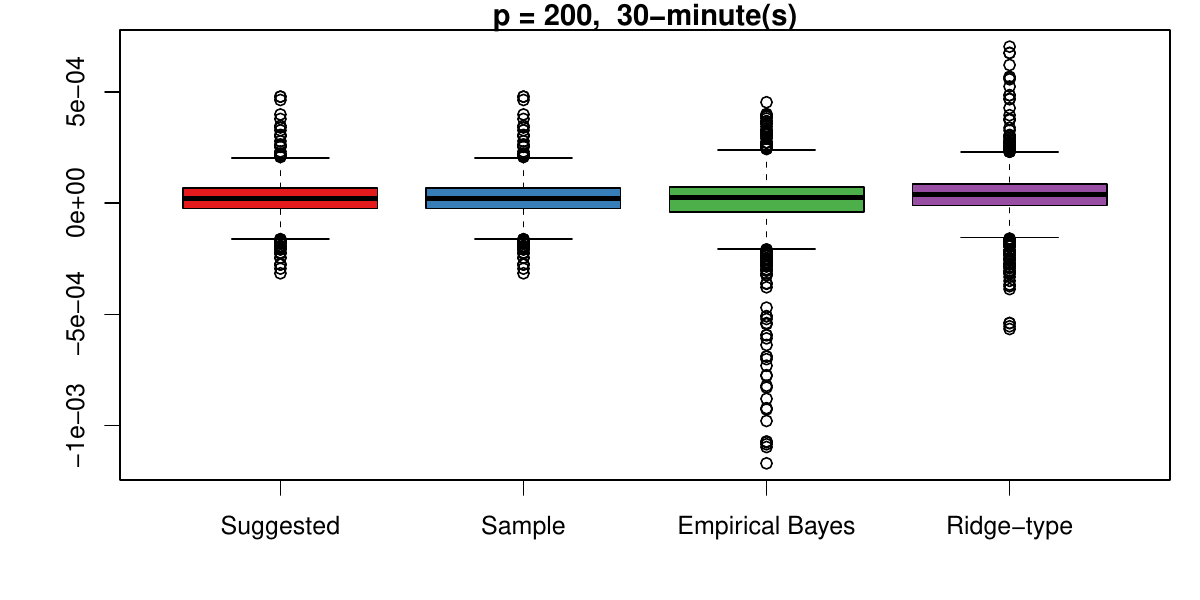}&\includegraphics[width=0.5\textwidth]{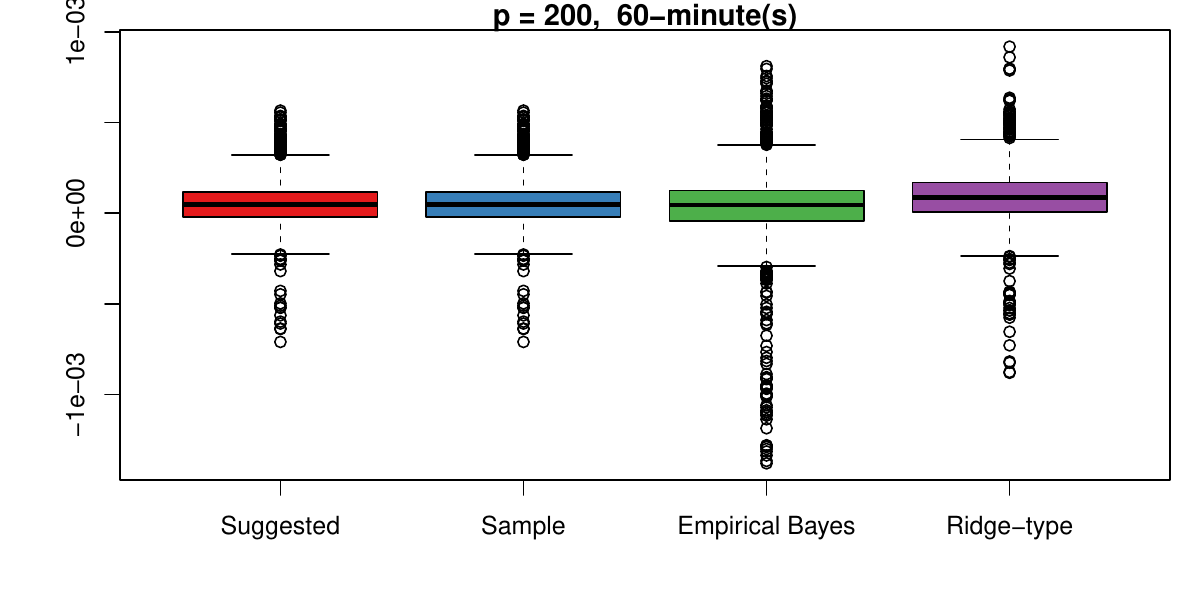}\\
  \end{tabular}
  \caption{Box plot of the estimated expected return of the global minimum variance portfolio based on $5$-, $10$-, $30$-, and $60$-minute returns on first 200 stocks from S\&P500 from 03 March 2017 to 6 June 2022.}
  \label{emp_box_R}
  \end{figure}

\begin{figure}
  \begin{tabular}{cc}
    \includegraphics[width=0.5\textwidth]{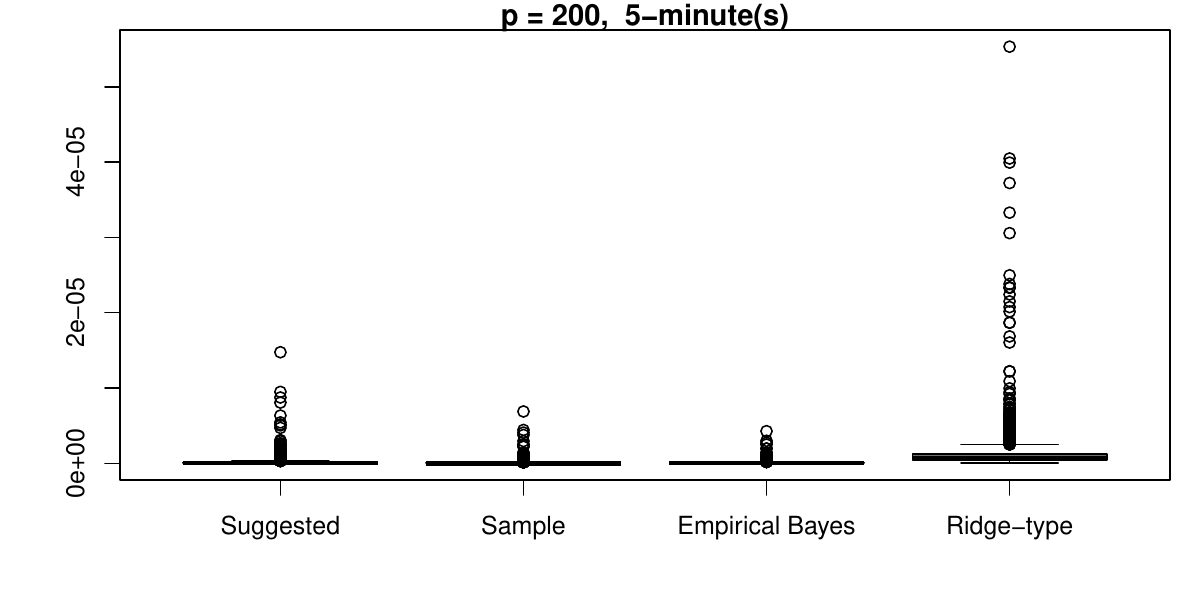} &\includegraphics[width=0.5\textwidth]{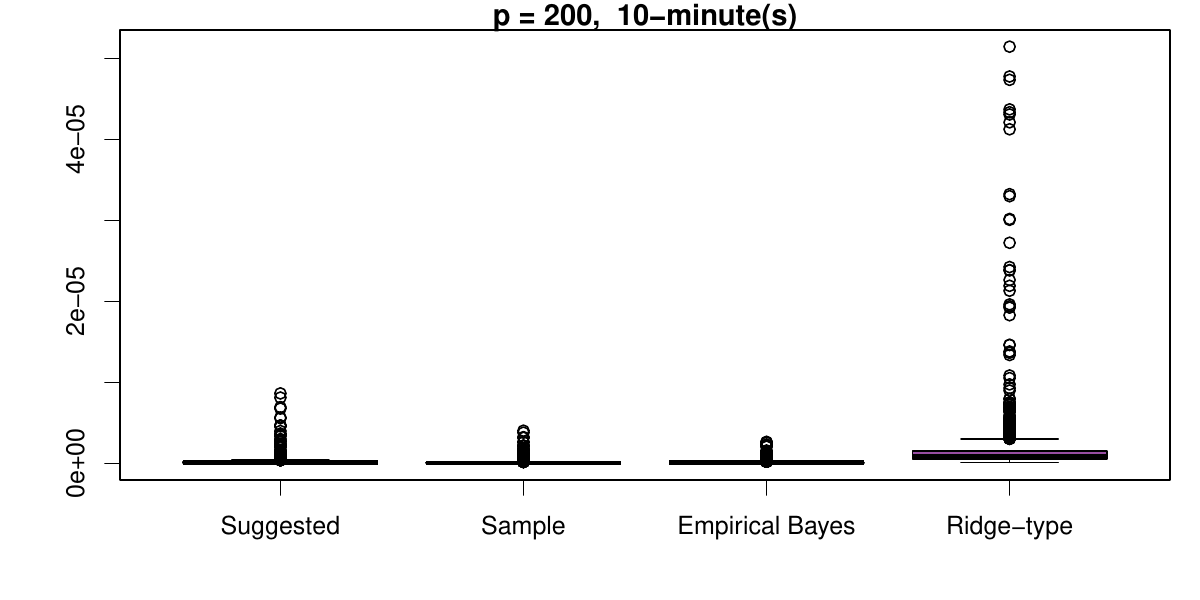}\\
    \includegraphics[width=0.5\textwidth]{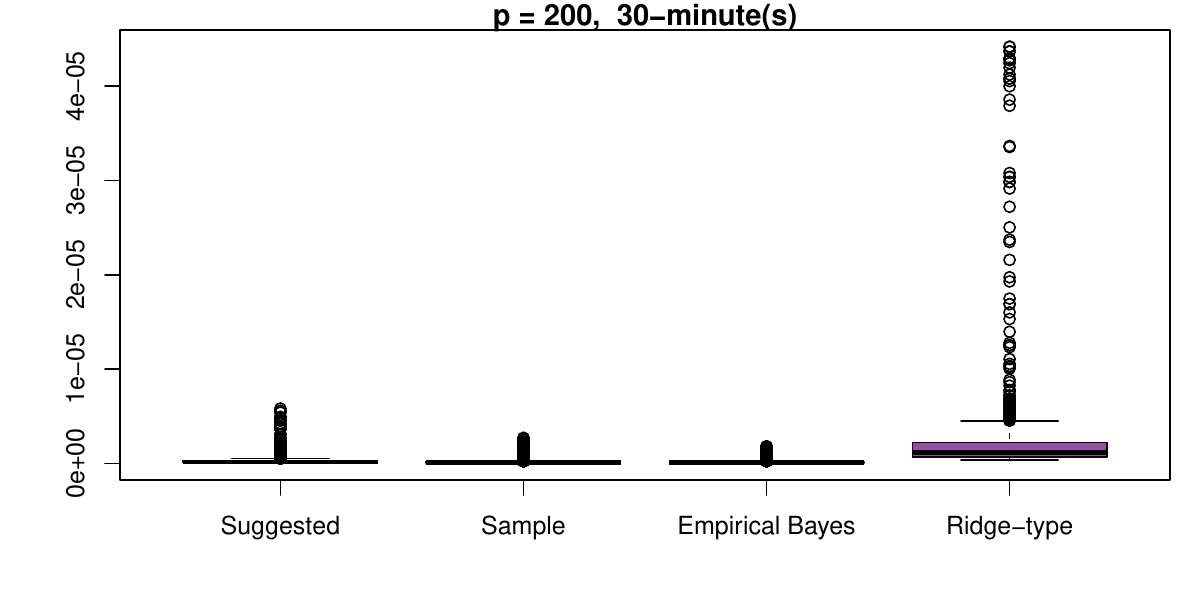}&\includegraphics[width=0.5\textwidth]{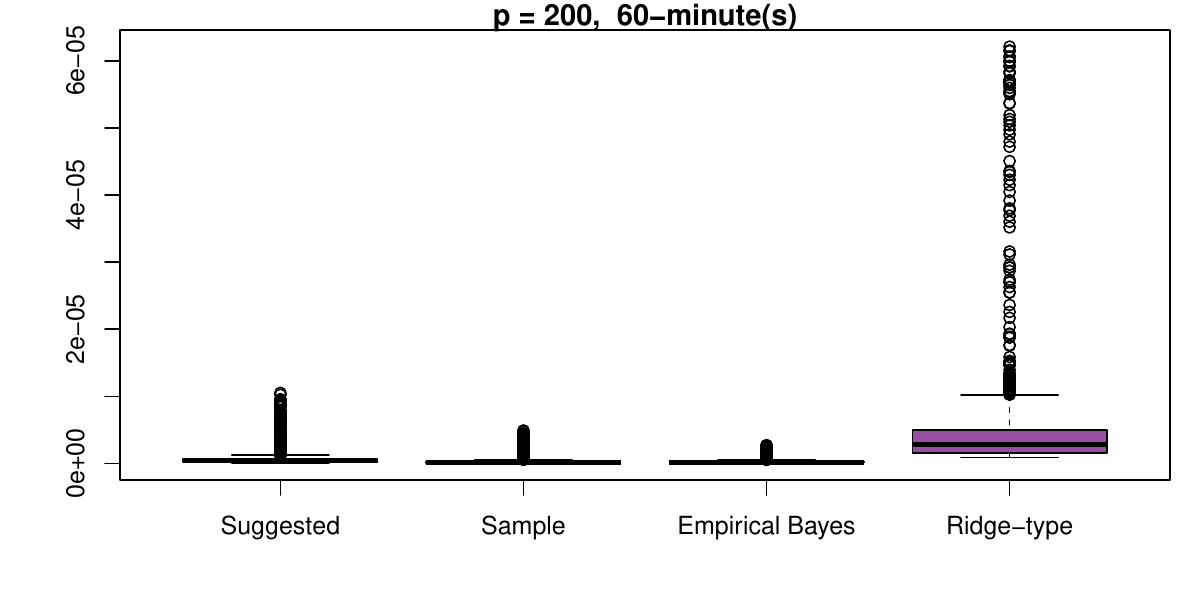}\\
  \end{tabular}
  \caption{Box plot of the estimated variance of the global minimum variance portfolio based on  $5$-, $10$-, $30$-, and $60$-minute returns on first 200 stocks from S\&P500 from 03 March 2017 to 6 June 2022.}
  \label{emp_box_V}
  \end{figure}

\begin{figure}
  \begin{tabular}{cc}
    \includegraphics[width=0.5\textwidth]{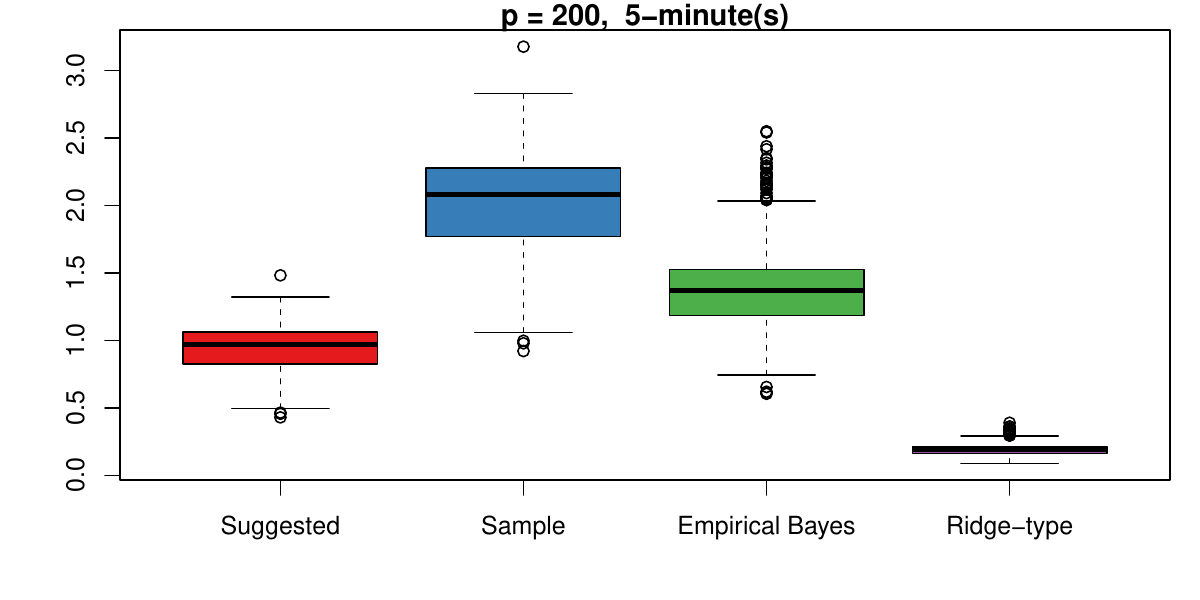} &\includegraphics[width=0.5\textwidth]{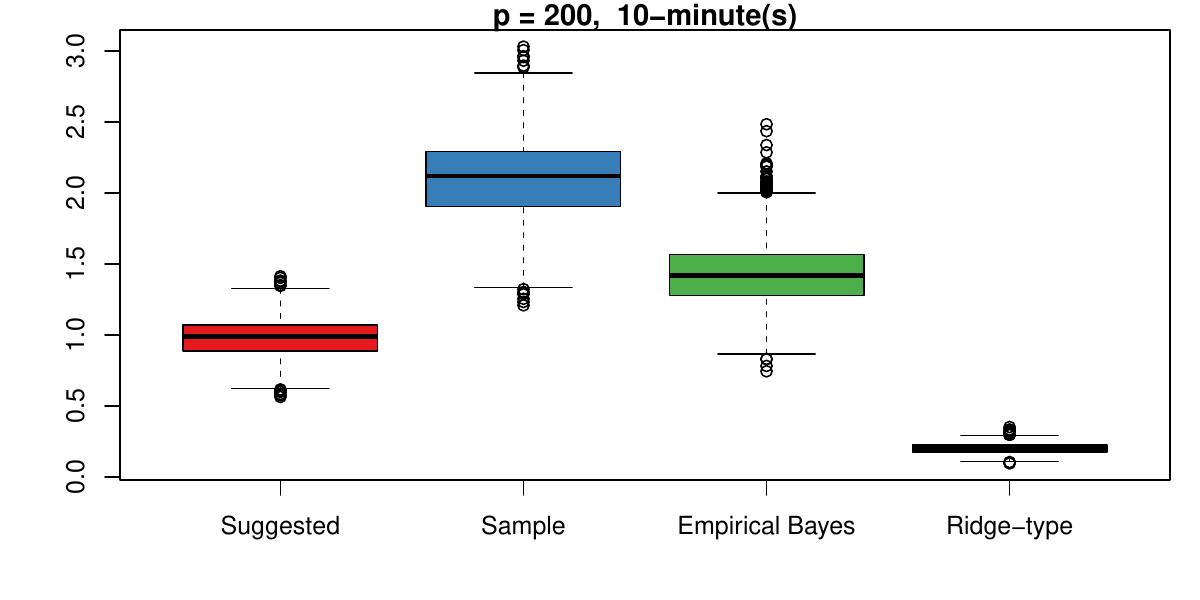}\\
    \includegraphics[width=0.5\textwidth]{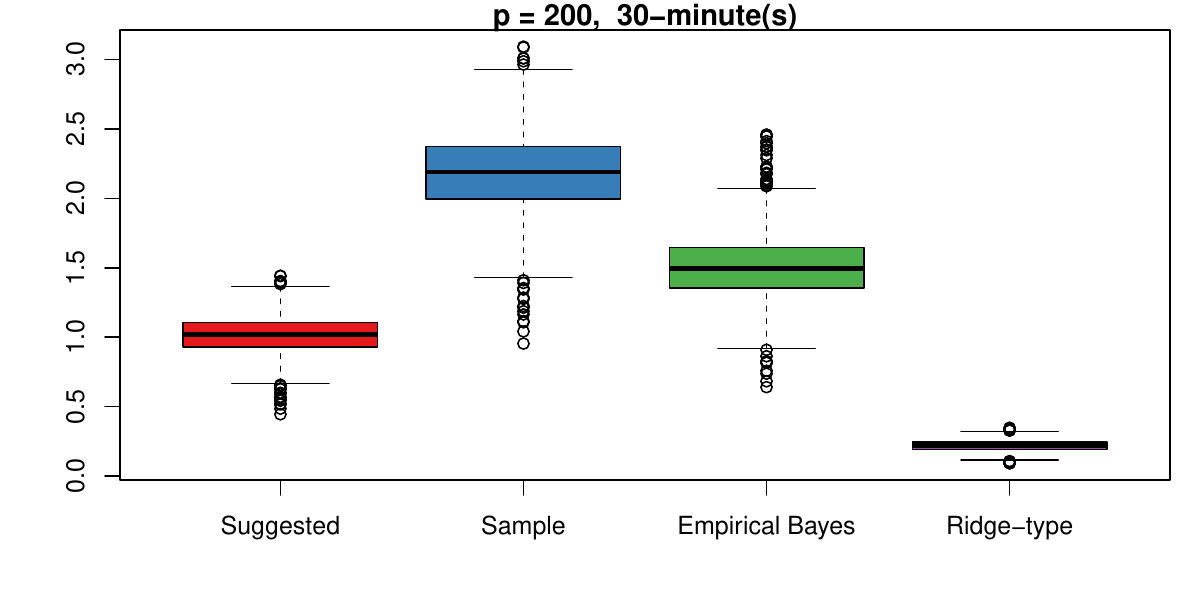}&\includegraphics[width=0.5\textwidth]{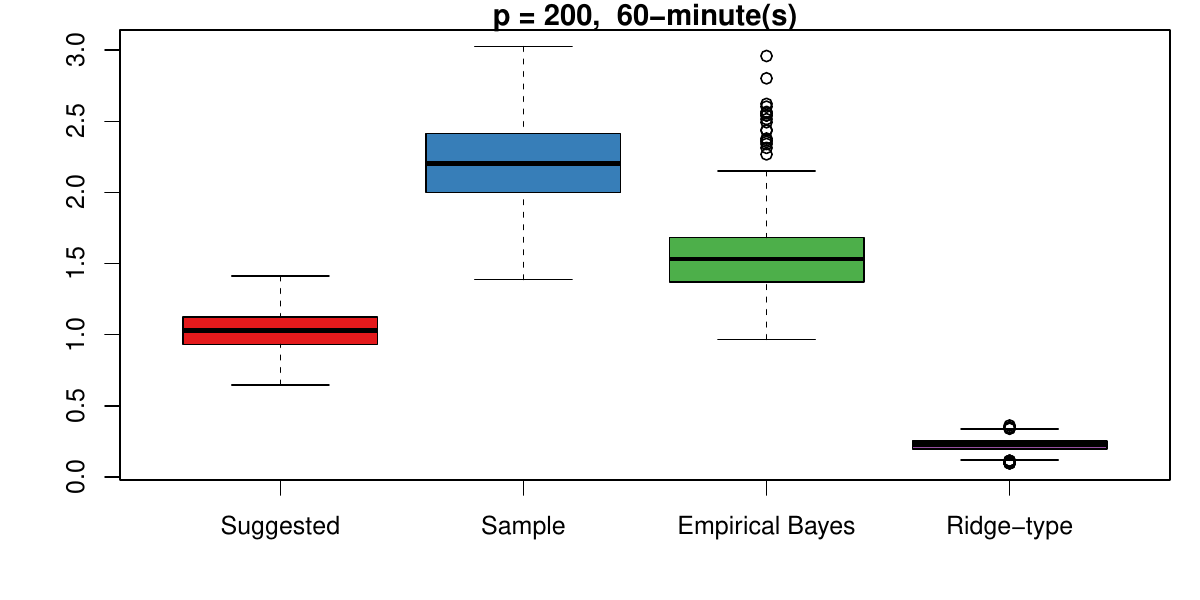}\\
  \end{tabular}
  \caption{Box plot of the estimated slope coefficient of the efficient frontier based on $5$-, $10$-,  $30$-, and $60$-minute returns on first 200 stocks from S\&P500 from 03 March 2017 to 6 June 2022.}
  \label{emp_box_s}
  \end{figure}

In Figures \ref{emp_box_R}-\ref{emp_box_s}, we present the box plots of the considered estimators for the three parameters of the efficient frontier. In the case of the expected return of the global minimum variance portfolio (cf. Figure \ref{emp_box_R}), the box plots are almost identical with the exception of the empirical Bayes estimator with more volatile behavior for all of the considered frequencies. In contrast, the box plots in the case of the other two parameters of the efficient frontier are very different (cf. Figures \ref{emp_box_V} and \ref{emp_box_s}). Furthermore, we observe that the values of the estimated expected return and the estimated variance of the global minimum variance portfolios get closer to zero with increasing frequency. Surprisingly, the ridge-type estimators are the most volatile for the variance and the least volatile for the slope. This effect is not present in the figure for the efficient frontier. Finally, as expected, the sample estimator underestimates the GMV portfolio's variance and overestimates the efficient frontier's slope parameter.

\begin{figure}[h!!]
\begin{tabular}{cc}
&\\
\includegraphics[width=0.5\textwidth]{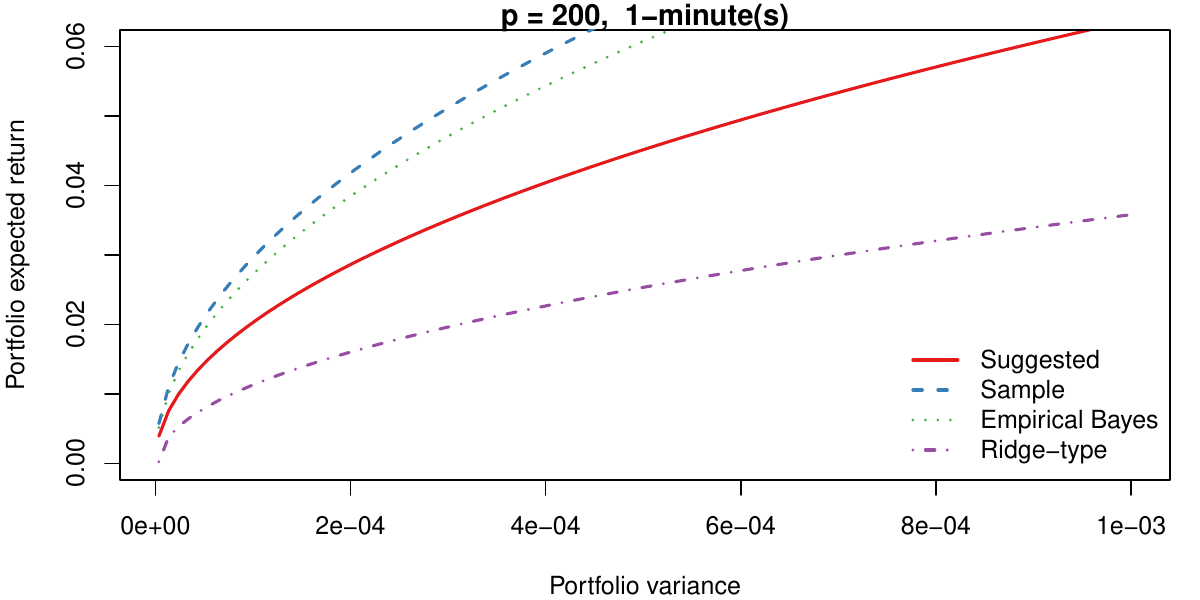}&\includegraphics[width=0.5\textwidth]{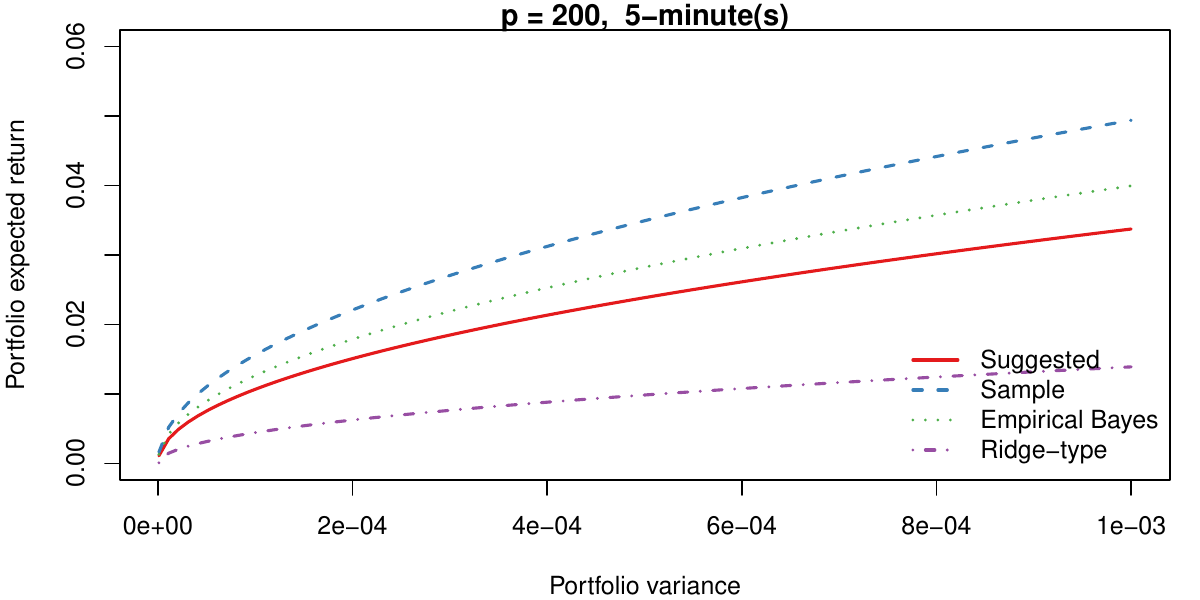}\\
\includegraphics[width=0.5\textwidth]{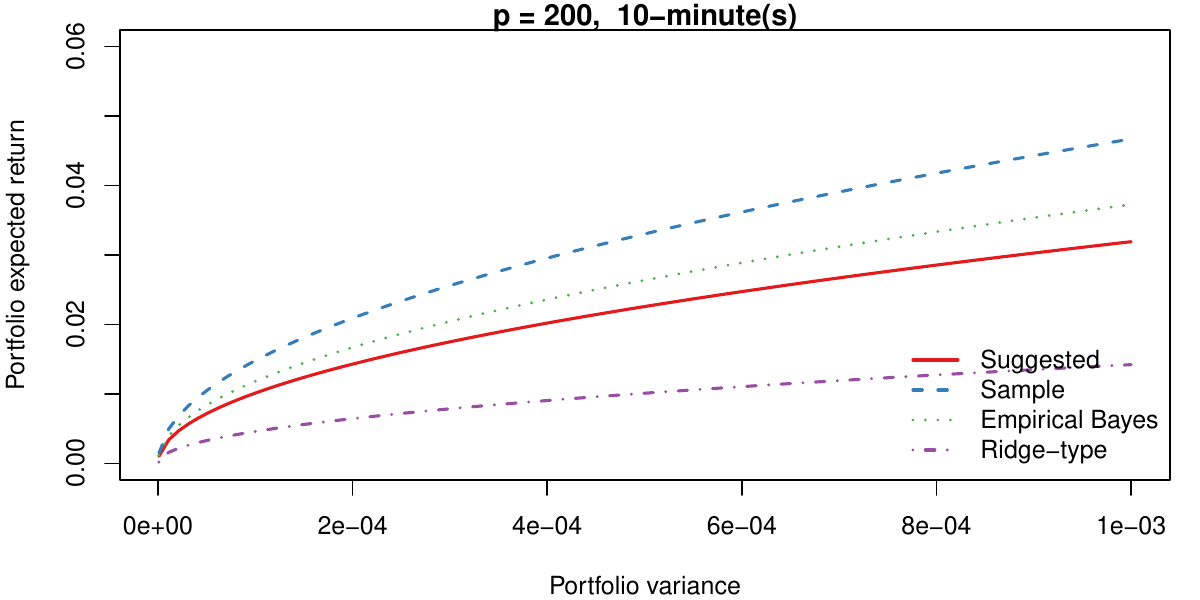}&\includegraphics[width=0.5\textwidth]{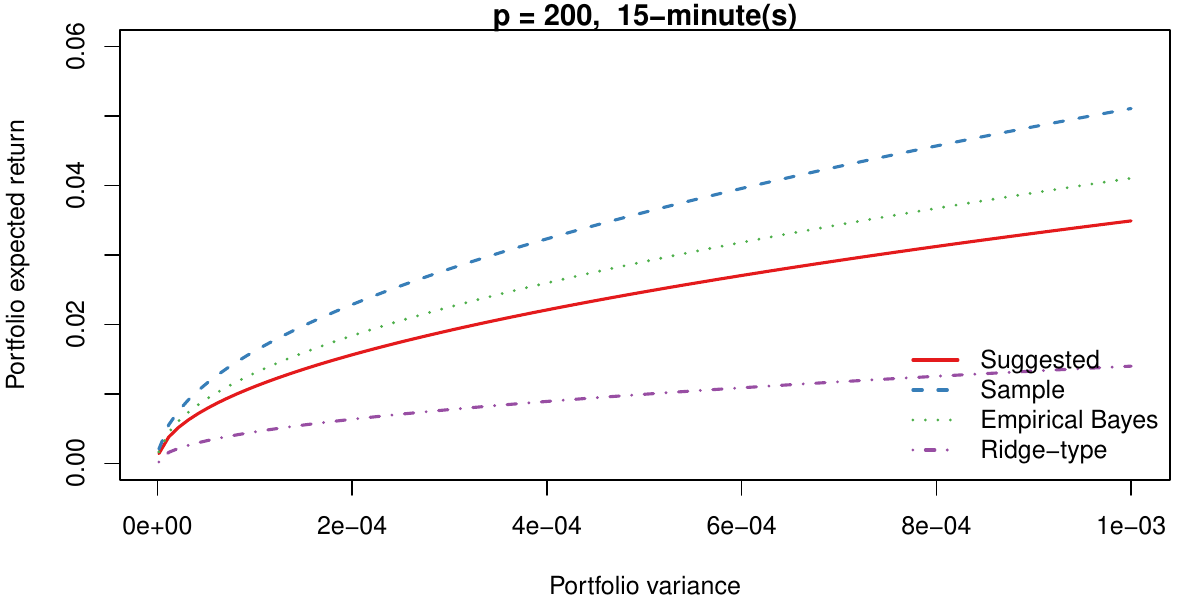}\\
\includegraphics[width=0.5\textwidth]{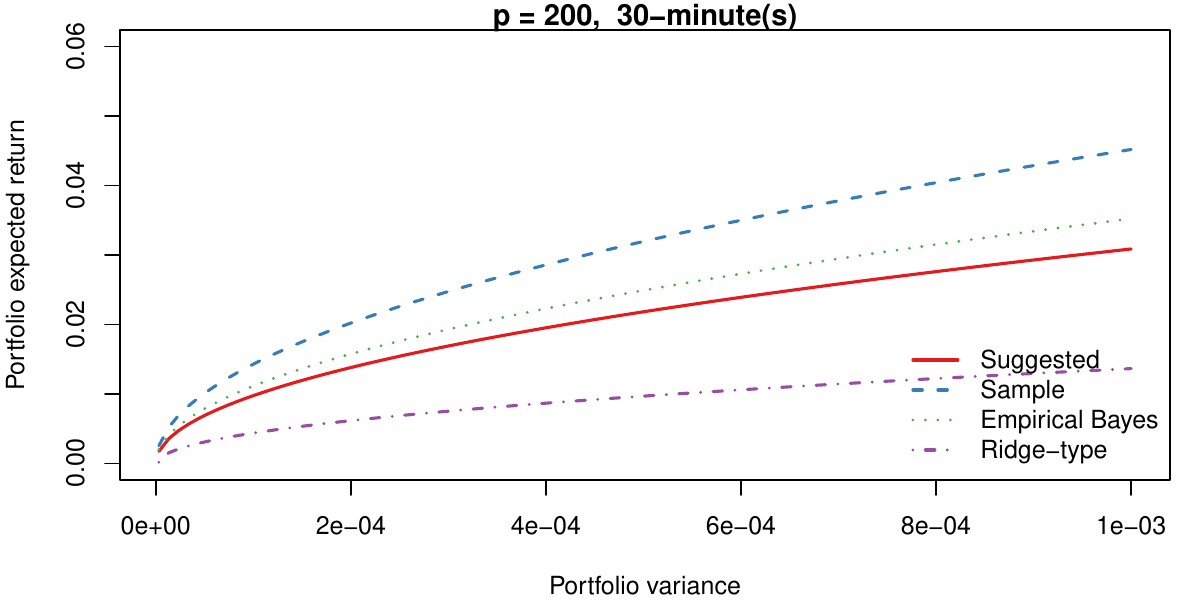}&\includegraphics[width=0.5\textwidth]{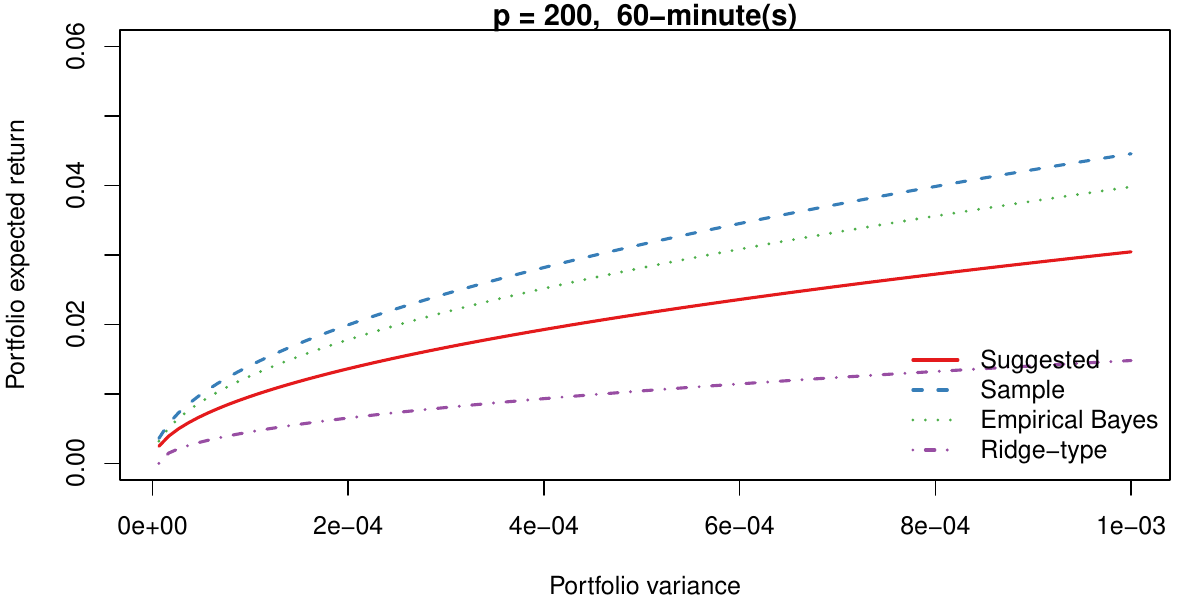}\\
\end{tabular}
\caption{\footnotesize The sample, the consistent, the empirical Bayes, and the ridge-type estimators for the efficient frontier based on $5$-, $10$-,  $30$-, and $60$-minute returns on 200 stocks from S\&P500 from 03 March 2017 to 6 June 2022.
}
\label{emp_ef}
\end{figure}

The estimators for the efficient frontier are presented for several frequencies in Figure \ref{emp_ef} in the case of June 6, 2022. The results of this figure are in line with the finding of the simulation study. Namely, the sample and the empirical Bayes efficient frontiers are overoptimistic and deviate considerably from the consistent estimator of the efficient frontier. On the contrary, the ridge-type estimator is too conservative and reflects the specific behaviour of the variance and the slope in the figures above. 

\section{Conclusions}
The efficient frontier derived by \cite{Markowitz1952} describes a set of portfolios that indicates the optimal asset allocation for the investor. Under uncertainty, the efficient frontier must be estimated from a given data set. In the case of a fixed number of assets \(p\) and a large sample size \(n\), the sample estimator is a good intuitive choice. However, when \(p\) grows together with \(n\) such that their ratio approaches a constant \(c \in (0, 1)\), the sample estimator becomes a poor alternative.

To handle "the curse of dimensionality," we investigate the asymptotic behavior of the sample estimator for the efficient frontier under large-dimensional asymptotics. We identify the asymptotic biases and correct them to develop consistent estimators for the parameters of the efficient frontier. We prove that two of the three parameters of the efficient frontier are highly biased under the large-dimensional asymptotic regime and that their biases are pure functions of the concentration ratio \(c \in (0, 1)\).

Finally, we provide a simulation study to investigate the convergence rate and the performance of the proposed estimator for the efficient frontier. We take state-of-the-art estimators for high-dimensional data and compare them with the suggested one. The proposed estimator, due to its simple form, appears to be a strong alternative in the case of large-dimensional data.

\section{Appendix}
We start with three important lemmas investigating the asymptotic behavior of quadratic forms in $\bx_n$ under the high-dimensional asymptotic regime.\\

\begin{lemma}\label{lem1}
Assume (A1)-(A3). Let a nonrandom $p\times p$-dimensional matrix $\mathbf{\Theta}_p$ and a nonrandom $n\times n$-dimensional matrix $\bTheta_n$ possess a uniformly bounded trace norms (sum of singular values) and let $\bSigma_n=\bI_p$. Then it holds that
\begin{eqnarray}\label{RM2011_id}
&&\left|\text{tr}\left(\mathbf{\Theta}_p(1/n\bx_n\bx_n^\prime-z\bI_p)^{-1}\right)-m(z)\text{tr}\left(\mathbf{\Theta}_p\right)\right|\stackrel{a.s.}{\longrightarrow}0\,, \\
&&\left|\text{tr}\left(\mathbf{\Theta}_n(1/n\bx_n^\prime\bx_n-z\bI_n)^{-1}\right)-\underline{m}(z)\text{tr}\left(\mathbf{\Theta}_n\right)\right|\stackrel{a.s.}{\longrightarrow}0 
\end{eqnarray}
$\text{for}~ p/n\longrightarrow c \in (0, +\infty) ~~ \text{as} ~n\rightarrow\infty$ where
\begin{equation}\label{mm}
 m(z)=(x(z)-z)^{-1}~~\text{and}~~\underline{m}(z)=-\dfrac{1-c}{z}+cm(z)
\end{equation}
with
\begin{equation}\label{RM2011_id_xz}
x(z)=\dfrac{1}{2}\left(1-c+z+\sqrt{(1-c+z)^2-4z}\right)\,.
\end{equation}
\end{lemma}

\noindent\textbf{Proof of Lemma \ref{lem1}:}
The application of Theorem 1 in \cite{Rubio2011} leads to (\ref{RM2011_id}) where $x(z)$ is a unique solution in $\mathbbm{C}^+$ of the following equation
\begin{equation}\label{eq1-Lemma1}
\dfrac{1-x(z)}{x(z)}=\dfrac{c}{x(z)-z}\,.
\end{equation}
The two solutions of (\ref{eq1-Lemma1}) are given by
\begin{equation}\label{solx}
x_{1,2}(z)=\dfrac{1}{2}\left(1-c+z\pm\sqrt{(1-c+z)^2-4z}\right)\,.
\end{equation}
In order to decide which of two solutions is feasible, we note that $x_{1,2}(z)$ is the Stieltjes transform with a positive imaginary part. Thus, without loss of generality, we can take $z=1+c+i2\sqrt{c}$ and get
{\small
\begin{equation}\label{im}
\textbf{Im}\{x_{1,2}(z)\}=\textbf{Im}\left\{\dfrac{1}{2}\left(2+i2\sqrt{c}\pm i2\sqrt{2c}\right)\right\}=\textbf{Im}\left\{1+i\sqrt{c}(1\pm\sqrt{2})\right\}=\sqrt{c}\left(1\pm\sqrt{2}\right)\,,
\end{equation}
}
which is positive only if the sign $"+"$ is chosen. Hence, the solution is given by
\begin{equation}\label{solx_a}
x(z)=\dfrac{1}{2}\left(1-c+z+\sqrt{(1-c+z)^2-4z}\right)\,.
\end{equation}
The second assertion of the lemma follows directly from \cite{Bai2010}.

\vspace{1cm}
\begin{lemma}\label{lem2}
Assume (A1)-(A3). Let $\boldsymbol{\theta}$ and $\boldsymbol{\xi}$ be the universal nonrandom vectors from the set $\mathcal{V}=\left\{p^{-q/2}\bSigma_n^{-1/2}\bi, p^{-q/2}\bSigma_n^{-1/2}\bm_n\right\}$
and let $\tbS=\frac{1}{n}\bx_n\bx_n^\prime$. Then it holds that
\begin{eqnarray}
  \left|\boldsymbol{\xi}^\prime\tbS_n^{-1}\boldsymbol{\theta}- (1-c)^{-1} \boldsymbol{\xi}^\prime \boldsymbol{\theta}\right| &\stackrel{a.s.}{\longrightarrow}& 0 \,,\label{1}\\
   \bbx_n^\prime\tbS_n^{-1}\bbx_n &\stackrel{a.s.}{\longrightarrow}& c  \,,\label{2}\\
 n^{-1/2}\bbx_n^\prime\tbS_n^{-1}\boldsymbol{\theta}&\stackrel{a.s.}{\longrightarrow}&0 \label{3}
\end{eqnarray}
$\text{for}~ p/n\longrightarrow c \in (0, +\infty) ~\text{as} ~n\rightarrow\infty ~~$.
\end{lemma}

\noindent\textbf{Proof of Lemma \ref{lem2}:}

It holds that
\begin{eqnarray*}
\bxi^\prime\tbS_n^{-1}\btheta&=&\lim\limits_{z\rightarrow0^+} \text{tr}\left[(\tbS_n-z\bI)^{-1}\btheta\bxi^\prime\right]\nonumber\\
&=&\lim\limits_{z\rightarrow0^+}\text{tr}\left[\left(\dfrac{1}{n}\bx_n\bx^\prime_n-z\bI\right)^{-1}\btheta\bxi^\prime\right]\,.
\end{eqnarray*}

Assumption (A3) ensures that $\btheta\bxi^\prime$ possesses a uniformly bounded trace norm. Then the application of Lemma \ref{lem1} leads to
\[
\left|\text{tr}\left[\left(\dfrac{1}{n}\bx_n\bx^\prime_n-z\bI\right)^{-1}\btheta\bxi^\prime\right]
- m(z)\boldsymbol{\xi}^\prime \boldsymbol{\theta}\right| \stackrel{a.s.}{\longrightarrow} 0 \,, \]
$\text{for}~ p/n\longrightarrow c \in (0, +\infty)~~\text{as} ~n\rightarrow\infty$ where $m(z)=(x(z)-z)^{-1}$ with $x(z)$ given in \eqref{RM2011_id_xz}. Finally, noting that $m(0)=(1-c)^{-1}$ leads to \eqref{1}.

Next, we deal with $\bbx_n^\prime\tbS_n^{-1}\bbx_n$. It holds that
\begin{eqnarray*}
\bbx_n^\prime \tbS_n^{-1} \bbx_n &=& \frac{1}{n^2} \bi^\prime_n\bx^\prime_n\tbS^{-1}_n\bx_n\bi_n= \lim\limits_{z\rightarrow0^+}\text{tr}\left[\frac{1}{\sqrt{n}}\bx^\prime_n\left(\dfrac{1}{n}\bx_n\bx_n^\prime-z\bI\right)^{-1}
\frac{1}{\sqrt{n}}\bx_n\left(\dfrac{\bi_n\bi^\prime_n}{n}\right)\right]\\
&=& \lim\limits_{z\rightarrow0^+} \text{tr}\left[\left(\dfrac{\bi_n\bi^\prime_n}{n}\right)\right]+z\text{tr}\left[\left(\dfrac{1}{n}\bx_n^\prime\bx_n-z\bI_n\right)^{-1}\left(\dfrac{\bi_n\bi^\prime_n}{n}\right)\right]\,,
\end{eqnarray*}
where the last equality follows from the Woodbury formula (matrix inversion lemma, see, e.g., \cite{Horn1985}). The application of Lemma \ref{lem1} leads to
\begin{equation*}
\text{tr}\left[\left(\dfrac{\bi_n\bi^\prime_n}{n}\right)\right]+z\text{tr}\left[\left(\dfrac{1}{n}\bx_n^\prime\bx_n-z\bI_n\right)^{-1}\left(\dfrac{\bi_n\bi^\prime_n}{n}\right)\right]
\stackrel{a.s.}{\longrightarrow}\left[1+(c-1)+czm(z)\right]\text{tr}\left[\left(\dfrac{\bi_n\bi^\prime_n}{n}\right)\right]
\end{equation*}
for $p/n\longrightarrow c<1$ as $n\rightarrow \infty$ where $m(z)$ is given by (\ref{mm}). Setting $z\rightarrow0^+$ and taking into account $\lim\limits_{z\rightarrow0^+}m(z)=\dfrac{1}{1-c}$ we get
\begin{equation*}
\bbx_n^\prime \tbS_n^{-1} \bbx_n\stackrel{a.s.}{\longrightarrow} 1+c-1 = c~\text{for}~\dfrac{p}{n}\longrightarrow c\in(0,1)~\text{as}~n\rightarrow\infty\,.
\end{equation*}

In case of $\bbx_n^\prime\tbS_n^{-1}\tbtheta$, we first note that $\boldsymbol{\theta}$ possesses bounded Euclidean norm due to assumption (A3). The application of Sherman-Morrison formula leads to
\begin{equation}
\tbS_n^{-1} = \tbS_{(k)}^{-1} - \dfrac{1/n\tbS_{(k)}^{-1}\bX_k\bX_k^\prime\tbS_{(k)}^{-1}}{1+1/n\bX_k^\prime\tbS_{(k)}^{-1}\bX_k} ~\text{for}~ k=1,...,n\,,
\end{equation}
where $\tbS_{(k)}^{-1}=(\tbS_n - 1/n\bX_k\bX_k^\prime)^{-1}$ and $\bX_k$ stands for the $k$th column of the matrix $\bx_n$. Hence,
\begin{eqnarray*}
\left|n^{-1/2}\bbx_n^\prime\tbS_n^{-1}\btheta\right| &=& n^{-3/2}\bi_n^\prime\bx_n^\prime \tbS_n^{-1}\btheta\nonumber\\
 &=& \left|n^{-3/2}\sum\limits_{k=1}^n\bX_k^\prime\left(\tbS_{(k)}^{-1} - \dfrac{1/n\tbS_{(k)}^{-1}\bX_k\bX_k^\prime\tbS_{(k)}^{-1}}{1+1/n\bX_k^\prime\tbS_{(k)}^{-1}\bX_k}\right)\btheta\right|\nonumber\\
 &=& \left|n^{-3/2} \sum\limits_{k=1}^n\dfrac{1}{1+1/n\bX_k^\prime\tbS_{(k)}^{-1}\bX_k}\bX_k^\prime\tbS_{(k)}^{-1}\btheta\right|\\
 &\le& \sqrt{\dfrac{1}{n}\sum\limits_{k=1}^n\dfrac{1}{1+1/n\bX_k^\prime\tbS_{(k)}^{-1}\bX_k}}
 \sqrt{\dfrac{1}{n^2}\sum\limits_{k=1}^n\dfrac{(\bX_k^\prime\tbS_{(k)}^{-1}\btheta)^2}{1+1/n\bX_k^\prime\tbS_{(k)}^{-1}\bX_k}}\\
 &\le&\sqrt{\dfrac{1}{n^2}\sum\limits_{k=1}^n\dfrac{(\bX_k^\prime\tbS_{(k)}^{-1}\btheta)^2}{1+1/n\bX_k^\prime\tbS_{(k)}^{-1}\bX_k}}\nonumber\,.
\end{eqnarray*}

From the other side, we get
\begin{eqnarray}
 \btheta^\prime \tbS_n^{-1}\btheta &=& \dfrac{1}{n}\sum\limits_{k=1}^n  \btheta^\prime\tbS_n^{-1}\btheta\\
&=& \dfrac{1}{n}\sum\limits_{k=1}^n  \btheta^\prime \left(\tbS_{(k)}^{-1} - \dfrac{1/n\tbS_{(k)}^{-1}\bX_k\bX_k^\prime\tbS_{(k)}^{-1}}{1+1/n\bX_k^\prime\tbS_{(k)}^{-1}\bX_k}\right)\btheta\nonumber\\
 &=& \dfrac{1}{n}\sum\limits_{k=1}^n  \btheta^\prime\tbS_{(k)}^{-1} \btheta -\dfrac{1}{n^2}\sum\limits_{k=1}^n\dfrac{(\bX_k^\prime\tbS_{(k)}^{-1}\btheta)^2}{1+1/n\bX_k^\prime\tbS_{(k)}^{-1}\bX_k}\nonumber\,.
\end{eqnarray}
The application of \eqref{1} leads to $\btheta^\prime \tbS_n^{-1}\btheta \stackrel{a.s.}{\longrightarrow} (1-c)^{-1} \btheta^\prime \btheta$ and
\[\btheta^\prime \tbS_{(k)}^{-1}\btheta \stackrel{a.s.}{\longrightarrow} (1-c)^{-1} \btheta^\prime \btheta ~~ \text{for} ~~ k=1,...,n \]
which proofs that
\[\dfrac{1}{n^2}\sum\limits_{k=1}^n\dfrac{(\bX_k^\prime\tbS_{(k)}^{-1}\btheta)^2}{1+1/n\bX_k^\prime\tbS_{(k)}^{-1}\bX_k} \stackrel{a.s.}{\longrightarrow}0 \]
and, hence, $n^{-1/2}\bbx_n^\prime\tbS_n^{-1}\btheta\stackrel{a.s.}{\longrightarrow}0$ $\text{for}~ p/n\longrightarrow c \in (0, +\infty) ~\text{as} ~n\rightarrow\infty ~~$.

\vspace{1cm}
\begin{lemma}\label{lem3}
Assume (A1)-(A3). Let $\boldsymbol{\theta}$ and $\boldsymbol{\xi}$ be the universal nonrandom vectors from the set $\mathcal{V}=\left\{\bi, \bm_n\right\}$. Let $q\ge 1$. Then it holds that
\begin{eqnarray}
   p^{-q}\left|\boldsymbol{\xi}^\prime\bS_n^{-1}\boldsymbol{\theta}- (1-c)^{-1} \boldsymbol{\xi}^\prime \bSigma^{-1}_n\boldsymbol{\theta}\right| &\stackrel{a.s.}{\longrightarrow}& 0 \,,\label{1a}\\
   \bbx_n^\prime\bSigma_n^{1/2}\bS_n^{-1}\bSigma_n^{1/2}\bbx_n &\stackrel{a.s.}{\longrightarrow}& \frac{c}{1-c}  \,,\label{2a}\\
 p^{-q}\bbx_n^\prime\bSigma_n^{1/2}\bS_n^{-1}\boldsymbol{\theta}&\stackrel{a.s.}{\longrightarrow}&0 \label{3a}
\end{eqnarray}
$\text{for}~ p/n\longrightarrow c \in (0, +\infty)~~\text{as} ~n\rightarrow\infty$.
\end{lemma}

\noindent\textbf{Proof of Lemma \ref{lem3}:}

From the Sherman-Morrison formula we get
\[\bS_n^{-1}=(\bSigma_n^{1/2}\tbS_n\bSigma_n^{1/2}-\bSigma_n^{1/2}\bbx_n\bbx_n^\prime\bSigma_n^{1/2})^{-1}
=\bSigma_n^{-1/2}\tbS_n^{-1}\bSigma_n^{-1/2}+\frac{\bSigma_n^{-1/2}\tbS_n^{-1}\bbx_n\bbx_n^\prime\tbS_n^{-1}\bSigma_n^{-1/2}}{1-\bbx_n^\prime\tbS_n^{-1}\bbx_n}\,.\]

Hence, using the results of Lemma \ref{lem2} we get
\begin{eqnarray*}
p^{-q}\left|\boldsymbol{\xi}^\prime\bS_n^{-1}\boldsymbol{\theta}- (1-c)^{-1} \boldsymbol{\xi}^\prime \bSigma^{-1}_n\boldsymbol{\theta}\right|
&\le& p^{-q}\left|\boldsymbol{\xi}^\prime\bSigma_n^{-1/2}\tbS_n^{-1}\bSigma_n^{-1/2}\boldsymbol{\theta}-(1-c)^{-1}\boldsymbol{\xi}^\prime \bSigma^{-1}_n\boldsymbol{\theta}\right|\\
&+&\left|\frac{ p^{-q} \bxi^\prime\bSigma_n^{-1/2}\tbS_n^{-1}\bbx_n\bbx_n^\prime\tbS_n^{-1}\bSigma_n^{-1/2}\btheta}{1-\bbx_n^\prime\tbS_n^{-1}\bbx_n}\right|\stackrel{a.s.}{\longrightarrow}0\,,\\
 \bbx_n^\prime\bSigma_n^{1/2}\bS_n^{-1}\bSigma_n^{1/2}\bbx_n &=& \bbx_n^\prime\tbS_n^{-1}\bbx_n+\frac{ (\bbx_n^\prime\tbS_n^{-1}\bbx_n)^2}{1-\bbx_n^\prime\tbS_n^{-1}\bbx_n}\stackrel{a.s.}{\longrightarrow}
 c+\frac{c^2}{1-c}=\frac{c}{1-c}\,,\\
p^{-q}\bbx_n^\prime\bSigma_n^{1/2}\bS_n^{-1}\boldsymbol{\theta}&=&
p^{-(q-1)/2}\left(n/p\right)^{1/2}n^{-1/2}p^{-q/2}\bbx_n^\prime\tbS_n^{-1}\bSigma_n^{-1/2}\boldsymbol{\theta}\\
&+&p^{-(q-1)/2}\left(n/p\right)^{1/2}\frac{n^{-1/2}p^{-q/2} \bbx_n^\prime\tbS_n^{-1}\bbx_n\bbx_n^\prime\tbS_n^{-1}\bSigma_n^{-1/2}\btheta}{1-\bbx_n^\prime\tbS_n^{-1}\bbx_n}\stackrel{a.s.}{\longrightarrow}0
\end{eqnarray*}
$\text{for}~ p/n\longrightarrow c \in (0, +\infty) ~~ \text{as} ~n\rightarrow\infty$.

\vspace{1cm}
\noindent\textbf{Proof of Theorem \ref{th1}:}
The statement \eqref{Vgmv} follows immediately from \eqref{1a} with $\bxi=\btheta=\bi_p$.

In order to prove \eqref{Rgmv}, we consider
\begin{eqnarray*}
\hat{R}_{GMV}=\frac{p^{-q} \bi_p \bS_n^{-1} \bby_n}{p^{-q} \bi_p \bS_n^{-1} \bi_n}
=\frac{p^{-q} \bi_p \bS_n^{-1} \bm_n}{p^{-q} \bi_p \bS_n^{-1} \bi_p}+\frac{p^{-q} \bi_p \bS_n^{-1} \bSigma_n^{1/2}\bbx_n}{p^{-q} \bi_p \bS_n^{-1} \bi_p}\,.
\end{eqnarray*}
The application of \eqref{1a} and \eqref{3a} leads to
\[\frac{p^{-q} \bi_p \bS_n^{-1} \bm_n}{p^{-q} \bi_p \bS_n^{-1} \bi_p}\stackrel{a.s.}{\longrightarrow} \frac{p^{-q} (1-c)^{-1} \bi_p \bSigma_n^{-1} \bm_n}{p^{-q} (1-c)^{-1}\bi_p \bSigma_n^{-1} \bi_p}
= \frac{\bi_p \bSigma_n^{-1} \bm_n}{\bi_p \bSigma_n^{-1} \bi_p}= R_{GMV}\]
and
\[ \frac{p^{-q} \bi_p \bS_n^{-1} \bSigma_n^{1/2}\bbx_n}{p^{-q} \bi_p \bS_n^{-1} \bi_p} \stackrel{a.s.}{\longrightarrow} 0\,,\]
$\text{for}~ p/n\longrightarrow c \in (0, +\infty) ~~ \text{as} ~n\rightarrow\infty$, from which \eqref{Rgmv} follows.

Finally, we get
\begin{eqnarray*}
\hat{s}&=&\bby_n^\prime \hat{\mathbf{Q}} \bby_n=(\bm_n+\bSigma_n^{1/2}\bbx_n)^\prime \bS_n^{-1} (\bm_n+\bSigma_n^{1/2}\bbx_n)-\frac{((\bm_n+\bSigma_n^{1/2}\bbx_n)^\prime \bS_n^{-1}\bi_p)^2}
{\bi_p^\prime \bS_n^{-1}\bi_p}\\
&=& \bm_n^\prime \bS_n^{-1}\bm_n+2\bm_n^\prime\bS_n^{-1}\bSigma_n^{1/2}\bbx_n+\bbx_n^\prime\bSigma_n^{1/2}\bS_n^{-1}\bSigma_n^{1/2}\bbx_n\\
&-&\frac{(\bm_n^\prime \bS_n^{-1}\bi_p)^2}{\bi_p^\prime \bS_n^{-1}\bi_p}
-\frac{(\bbx_n^\prime \bSigma_n^{1/2}\bS_n^{-1}\bi_p)^2}{\bi_p^\prime \bS_n^{-1}\bi_p}
-2\frac{(\bm_n^\prime \bS_n^{-1}\bi_p)(\bbx_n^\prime \bSigma_n^{1/2}\bS_n^{-1}\bi_p)}{\bi_p^\prime \bS_n^{-1}\bi_p}\,.
\end{eqnarray*}
The application of the results of Lemma \ref{lem3} leads to \eqref{s}. The theorem is proved.

\vspace{1cm}
\begin{lemma}\label{lem4}
\begin{enumerate}[a)]
\item Let $Z_{p,n}\sim \chi^2_{n-p}$ ($\chi^2$-distribution with $n-p$ degrees of freedom). Then
\begin{equation}
\sqrt{n} \left(\frac{Z_{p,n}}{n-p}-1\right) \stackrel{d.}{\longrightarrow} \mathcal{N}\left(0,\frac{2}{1-c}\right) ~~ \text{for} ~~ \frac{p}{n} \longrightarrow c \in (0,1) ~\text{as}~ n \rightarrow \infty \,.
\end{equation}
\item Let $F_{p,n}\sim F_{p,n-p,n\,\lambda}$ (non-central $F$-distribution with $p$ and $n-p$ degrees of freedom and non-centrality parameter $n \lambda$) with $\lambda \in (0,\infty)$. Then
\begin{equation}
\sqrt{n} \left(\frac{n-p}{p}F_{p,n}-1-\frac{n}{p}\lambda\right) \stackrel{d.}{\longrightarrow} \mathcal{N}\left(0,\frac{2}{c}\left(1+2\dfrac{\lambda}{c}\right)+\frac{2}{1-c}\left(1+\dfrac{\lambda}{c}\right)^2\right)
\end{equation}
$\text{for} ~~ \frac{p}{n} \longrightarrow c \in (0,1) ~\text{as}~ n \rightarrow \infty$.
\end{enumerate}
\end{lemma}

\noindent\textbf{Proof of Lemma \ref{lem4}:}
\begin{enumerate}[a)]
\item The statement follows directly from the properties of $\chi^2$-distribution and the fact that $c<1$ which ensures that $n-p \longrightarrow \infty$ for $\frac{p}{n} \longrightarrow c \in (0,1)$ as $n \rightarrow \infty$.

\item From the properties of the non-central $F$-distribution, we get
\[\frac{n-p}{p}F_{p,n} \stackrel{d.}{=}\frac{f_1/p}{f_2/(n-p)}\,,\]
where $f_1\sim \chi^2_{p;n\lambda}$ (non-central $\chi^2$-distribution with $p$ degrees of freedom and non-centrality parameter $n\lambda$), $f_2\sim \chi^2_{n-p}$, and $f_1$, $f_2$ are independently distributed.

It holds that
\begin{eqnarray*}
\sqrt{n} \left(\frac{n-p}{p}F_{p,n}-1-\frac{n}{p}\lambda\right)&=&\dfrac{1}{f_2/(n-p)}\left(\sqrt{n} \left(\dfrac{f_1}{p}-1-\frac{n}{p}\lambda\right)\right.\\
&-&\left.\sqrt{n} \left(\dfrac{f_2}{n-p}-1\right)\left(1+\frac{n}{p}\lambda\right)\right)\,.
\end{eqnarray*}

From Lemma \ref{lem4}.a, we get
\[\sqrt{n} \left(\dfrac{f_2}{n-p}-1\right)\stackrel{d.}{\longrightarrow} \mathcal{N}\left(0,\frac{2}{1-c}\right) ~~ \text{for} ~~ \frac{p}{n} \longrightarrow c \in (0,1) ~\text{as}~ n \rightarrow \infty \,.\]
Furthermore, the application of Lemma 3 in \cite{bodnar2016exact} leads to $f_2/(n-p) \stackrel{a.s.}{\longrightarrow} 1 $ and
\[\sqrt{n}  \left(\dfrac{f_1}{p}-1-\frac{n}{p}\lambda\right) \stackrel{d.}{\longrightarrow} \mathcal{N}\left(0,\frac{2}{c}\left(1+2\dfrac{\lambda}{c}\right)\right) \]
for $\frac{p}{n} \longrightarrow c \in (0,1)$ as $n \rightarrow \infty$. Putting these results together and applying Slutsky's lemma (see, e.g., Theorem 1.5 in \cite{dasgupta2008asymptotic} we obtain the statement of Lemma \ref{lem4}.b.
\end{enumerate}

\vspace{1cm}
The following result is Lemma A1 in \cite{BodnarSchmid2009}\\[0.2cm]

\begin{lemma}\label{lem5}
Let $\mathbf{x}_1,\ldots,\mathbf{x}_n$ be a random sample of independent vectors such that $\mathbf{x}_i\sim\mathcal{N}_p(\bm_n,\bSigma_n)$ for $i=1,\ldots,n$. Then for any $p$ and $n$ with $n>p$ it holds that
\begin{enumerate}[a)]
\item $\hat{V}_{GMV}$ is independent of $(\hat{R}_{GMV},\hat{s})$.
\item $(n-1){\hat{V}_{GMV}}/{V_{GMV}} \sim \chi^2_{n-p}$.
\item $\frac{n(n-p+1)}{(n-1)(p-1)} \hat{s}\sim F_{p-1,n-p+1,n\,s}$.
\item $\hat{R}_{GMV}| \hat{s}=y \sim \mathcal{N}\left(R_{GMV},\frac{1+\frac{n}{n-1}y}{n}V_{GMV}\right)$.
\end{enumerate}
\end{lemma}

\vspace{1cm}
\noindent\textbf{Proof of Theorem \ref{th2}:}

From Lemma \ref{lem5}.a, we get that $\hat{V}_{GMV}$ is independent of $(\hat{R}_{GMV},\hat{s})$ for any $p$ and $n$ and, hence, $\hat{V}_{GMV}$ is independent of $(\hat{R}_{GMV},\hat{s})$  also under high-dimensional asymptotics. Moreover, from Lemma \ref{lem4}.a we get
\begin{eqnarray*}
\sqrt{n}\left(\hat{V}_{c}-V_{GMV}\right)=\sqrt{n}\left(\frac{1}{1-p/n}\hat{V}_{GMV}-V_{GMV}\right)=\sqrt{n}\left(\frac{n}{n-p}\dfrac{\hat{V}_{GMV}}{V_{GMV}}-1\right)V_{GMV}\,.
\end{eqnarray*}
The application of Lemma \ref{lem5}.b leads to $\sqrt{n}\left(\hat{V}_{c}-V_{GMV}\right)\stackrel{d.}{\longrightarrow} \mathcal{N}\left(0,\frac{2}{1-c}V_{GMV}^2\right)$.

Similarly, we obtain
\begin{eqnarray*}
\sqrt{n}\left(\hat{s}_{c} -\frac{p}{n}-s\right)=\sqrt{n}\left(\left(1 -\dfrac{p}{n}\right)\hat{s}- \dfrac{p}{n}-s\right)
=\sqrt{n}\left(\frac{n-p}{p}\hat{s}-1-\dfrac{n}{p}s\right)\dfrac{p}{n}\,.
\end{eqnarray*}
The application of Lemma \ref{lem4}.b and Lemma \ref{lem5}.c leads to
\[\sqrt{n}\left(\hat{s}_{c}-s-\frac{p}{n}\right)\stackrel{d.}{\longrightarrow} \mathcal{N}\left(0,2\left(c+2s\right)+\frac{2}{1-c}\left(c+s\right)^2\right)\,.\]

Finally, the application of Lemma \ref{lem5}.d leads to
\begin{eqnarray*}
\sqrt{n}\left(
          \begin{array}{c}
            \hat{R}_{c}-R_{GMV} \\
            \hat{s}_{c}-s-\frac{p}{n} \\
          \end{array}
        \right)\stackrel{d.}{=}
\left(
          \begin{array}{c}
            \sqrt{1+\frac{n}{n-1}\frac{\hat{s}_{c}+p/n}{1-p/n}}\sqrt{V_{GMV}}\sqrt{n}z_0\\
            \sqrt{n}(\hat{s}_{c}-s) \\
          \end{array}
        \right) \,,
\end{eqnarray*}
where
\begin{eqnarray*}
\sqrt{n}\left(\begin{array}{c}
z_0\\
            \hat{s}_{c}-s -\frac{p}{n} \\
          \end{array}
        \right)\stackrel{d.}{\longrightarrow} \mathcal{N}\left(\left(
                                                   \begin{array}{c}
                                                     0 \\
                                                     0 \\
                                                   \end{array}
                                                 \right),
                                                 \left(
                                                   \begin{array}{cc}
                                                     1 & 0 \\
                                                     0 & \sigma_s^2 \\
                                                   \end{array}
                                                 \right)
        \right) \,,
\end{eqnarray*}
with $\sigma_s^2$ as in (\ref{sig_s2}). Now, the application of the $\delta$-method (see, e.g., Theorem 3.7 in \cite{dasgupta2008asymptotic}) and the fact that $\hat{V}_{GMV}$ is independent of $(\hat{R}_{GMV},\hat{s})$ leads to the statement of the theorem.

\bibliography{EFF}

\newpage

\newpage 

\newpage

\end{document}